\documentclass[iop,revtext4-1]{emulateapj}
\usepackage[colorlinks=true,citecolor=blue,urlcolor=red,linkcolor=blue]{hyperref}
\usepackage{mathtools}
\usepackage{amsmath}

\long\def\symbolfootnote[#1]#2{\begingroup\def\thefootnote{\fnsymbol{footnote}}\footnote[#1]{#2}\endgroup}

\def\HII{\hbox{H\,{\sc ii}}}

\def\OIVsub{\hbox{[O\,{\sc iv}]$_{25.9}$}}

\def\NeIIsub{\hbox{[Ne\,{\sc ii}]$_{12.8}$}}

\def\NeIIIsub{\hbox{[Ne\,{\sc iii}]$_{15.6}$}}

\def\NeVsub{\hbox{[Ne\,{\sc v}]$_{14.3}$}}
\def\SIIIa{\hbox{[S\,{\sc iii}]18.7$\,\mu$m}}
\def\SIIIasub{\hbox{[S\,{\sc iii}]$_{18.7}$}}
\def\SIIIb{\hbox{[S\,{\sc iii}]33.5$\,\mu$m}}
\def\SIIIbsub{\hbox{[S\,{\sc iii}]$_{33.5}$}}

\def\CII{\hbox{[C\,{\sc ii}]157.7$\,\mu$m}}
\def\CIIno{\hbox{[C\,{\sc ii}]}}
\def\CIIsub{\hbox{[C\,{\sc ii}]$_{158}$}}
\def\CIIpdr{\hbox{[C\,{\sc ii}]$^{\rm PDR}_{158}$}}
\def\CIIion{\hbox{[C\,{\sc ii}]$^{\rm ion}_{158}$}}
\def\CIIionhii{\hbox{[C\,{\sc ii}]$^{\rm ion, {\sc HII}}_{158}$}}
\def\CIIiondiff{\hbox{[C\,{\sc ii}]$^{\rm diff, {\sc HII}}_{158}$}}

\def\OIa{\hbox{[O\,{\sc i}]63.2$\,\mu$m}}
\def\OIasub{\hbox{[O\,{\sc i}]$_{63}$}}
\def\OIb{\hbox{[O\,{\sc i}]145.5$\,\mu$m}}
\def\OIbsub{\hbox{[O\,{\sc i}]$_{145}$}}
\def\OIno{\hbox{[O\,{\sc i}]}}

\def\OIIIasub{\hbox{[O\,{\sc iii}]$_{52}$}}
\def\OIIIb{\hbox{[O\,{\sc iii}]88.4$\,\mu$m}}

\def\OIIIbsub{\hbox{[O\,{\sc iii}]$_{88}$}}
\def\NIIa{\hbox{[N\,{\sc ii}]121.9$\,\mu$m}}
\def\NIIasub{\hbox{[N\,{\sc ii}]$_{122}$}}
\def\NIIb{\hbox{[N\,{\sc ii}]205.2$\,\mu$m}}
\def\NIIbsub{\hbox{[N\,{\sc ii}]$_{205}$}}
\def\NIIno{\hbox{[N\,{\sc ii}]}}
\def\NIII{\hbox{[N\,{\sc iii}]57.3$\,\mu$m}}
\def\NIIIsub{\hbox{[N\,{\sc iii}]$_{57}$}}

\def\OHbsub{\hbox{OH$_{119}$}}

\def\SSi{\hbox{$S_{\rm 9.7\mu m}$}}
\def\H2{\hbox{H$_{2}$}}

\def\PAHasub{\hbox{PAH$_{11.3}$}}

\def\PAHcsub{\hbox{PAH$_{7.7}$}}
\def\PAHd{\hbox{6.2$\,\mu$m\,PAH}}

\def\PAHdsub{\hbox{PAH$_{6.2}$}}

\def\Sa{\hbox{$S_{\rm 63}$}}
\def\Sb{\hbox{$S_{\rm 158}$}}
\def\Sc{\hbox{$S_{\rm 30}$}}
\def\Sd{\hbox{$S_{\rm 15}$}}

\def\kB{$k_{\rm B}$}
\def\kms{${\rm km~s}^{-1}$}

\def\AV{\hbox{$A_V$}}

\def\nhvol{$n_{\rm H}$}
\def\nhcol{$N_{\rm H}$}
\def\nhcr{$n_{\rm H}^{cr}$}
\def\nhhcr{$n_{\rm H_2}^{cr}$}
\def\Go{$G_{\rm 0}$}
\def\G{$G$}
\def\ne{$n_{\rm e}$}
\def\necr{$n_{\rm e}^{\rm cr}$}

\def\Mgas{\hbox{$M_{\rm gas}$}}

\def\Lsun{\hbox{$L_\odot$}}

\def\Mstar{\hbox{$M_{\star}$}}
\def\LIR{\hbox{$L_{\rm IR}$}}
\def\LIReff{\hbox{$L_{\rm IR, eff}$}}

\def\SigmaIR{\hbox{$\Sigma_{\rm IR}$}}
\def\SigmaIRstar{\hbox{$\Sigma_{\rm IR}^*$}}

\def\LIRwave{\hbox{$L_{\rm IR\,[8-1000\,\mu m]}$}}

\def\LFIR{\hbox{$L_{\rm FIR}$}}
\def\LFIReff{\hbox{$L_{\rm FIR, eff}$}}
\def\IFIR{\hbox{$I_{\rm FIR}$}}

\def\Areaeff{\hbox{$\pi \rm R^2_{\rm 70\mu m, eff}$}}

\def\LFIRwave{\hbox{$L_{\rm FIR\,[42.5-122.5\,\mu m]}$}}
\def\FIRwave{\hbox{$\rm FIR_{\rm [42.5-122.5\,\mu m]}$}}

\def\Teff{\hbox{$T_{\rm eff}$}}
\def\Tdust{\hbox{$T_{\rm dust}$}}
\def\Tkin{\hbox{$T^{\rm kin}_{\rm gas}$}}
\def\Tpdr{\hbox{$T_{\rm PDR}$}}
\def\Mdust{\hbox{$M_{\rm dust}$}}
\def\Zsun{\hbox{$Z_{\odot}$}}
\def\alphaAGN{\hbox{$\alpha_{\rm AGN}$}}
\def\alphabolAGN{\hbox{$\alpha_{\rm AGN}^{\rm bol}$}}
\def\alphamirAGN{\hbox{$\alpha_{\rm AGN}^{\rm MIR}$}}
\def\Aff{\hbox{$\phi_{\rm A}$}}
\def\Vff{\hbox{$\phi_{\rm V}$}}
\def\chisq{\hbox{$\chi^2$}}
\def\chinusq{\hbox{$\chi_\nu^2$}}

\def\cnmm{\hbox{cm$^{-2}$}}
\def\cnmmm{\hbox{cm$^{-3}$}}
\def\cmmm{\hbox{cm$^{3}$}}
\def\kms{\hbox{km$\,$s$^{-1}$}}

\def\cmns{\hbox{cm$\,$s$^{-1}$}}
\def\wnmm{\hbox{W$\,$m$^{-2}$}}

\def\lsd{\hbox{L${_\odot}\,$kpc$^{-2}$}}

\shorttitle{GOALS-Herschel: FIR Line Emission from Local LIRGs}

\shortauthors{D\'iaz-Santos et al.}

\begin{document}

\title{A \textit{Herschel}/PACS Far Infrared Line Emission Survey of Local Luminous Infrared Galaxies}


\author{T.~D\'{\i}az-Santos\altaffilmark{1,\dag},
L.~Armus\altaffilmark{2},
V.~Charmandaris\altaffilmark{3,4},
N.~Lu\altaffilmark{15,16},
S.~Stierwalt\altaffilmark{5,6},
G.~Stacey\altaffilmark{12},
S.~Malhotra\altaffilmark{9},
P.~P.~van~der~Werf\altaffilmark{11},
J.~H.~Howell\altaffilmark{2},
G.~C.~Privon\altaffilmark{10,14},
J.~M.~Mazzarella\altaffilmark{7},
P.~F.~Goldsmith\altaffilmark{13},
E.~J.~Murphy\altaffilmark{6},
L.~Barcos-Mu\~noz\altaffilmark{6,22},
S.~T.~Linden\altaffilmark{5,6},
H.~Inami\altaffilmark{8},
K.~L.~Larson\altaffilmark{2},
A.~S.~Evans\altaffilmark{5,6},
P.~Appleton\altaffilmark{7,17},
K.~Iwasawa\altaffilmark{18,21}
S.~Lord\altaffilmark{20},
D.~B.~Sanders\altaffilmark{19},
and J.~A.~Surace\altaffilmark{2}
}

\altaffiltext{\dag}{Contact email: tanio.diaz@mail.udp.cl}
\affil{$^1$N\'ucleo de Astronom\'ia de la Facultad de Ingenier\'ia, Universidad Diego Portales, Av. Ej\'ercito Libertador 441, Santiago, Chile}
\affil{$^2$Infrared Processing and Analysis Center, MC 314-6, Caltech, 1200 E. California Blvd., Pasadena, CA 91125}
\affil{$^3$Institute for Astronomy, Astrophysics, Space Applications \& Remote Sensing, National Observatory of Athens, GR-15236, Penteli, Greece}
\affil{$^4$University of Crete, Department of Physics, GR-71003, Heraklion}
\affil{$^5$Department of Astronomy, University of Virginia, P.O. Box 400325, Charlottesville, VA 22904}
\affil{$^6$National Radio Astronomy Observatory, 520 Edgemont Road, Charlottesville, VA 22903}
\affil{$^7$Infrared Processing and Analysis Center, MC 100-22, Caltech, 1200 E. California Blvd., Pasadena, CA 91125}
\affil{$^8$Centre de Recherche Astrophysique de Lyon, Universite de Lyon, Universite Lyon 1, CNRS, Observatoire de Lyon, 9 avenue Charles Andre, Saint-Genis Laval Cedex F-69561, France}
\affil{$^9$Astrophysics Science Division, Goddard Space Flight Center, 8800 Greenbelt Road, Greenbelt, MD 20771, USA}
\affil{$^{10}$Departamento de Astronom\'ia, Universidad de Concepci\'on, Casilla 160-C, Concepci\'on, Chile}
\affil{$^{11}$Leiden Observatory, Leiden University, P.O. Box 9513, NL-2300 RA Leiden, The Netherlands}
\affil{$^{12}$Department of Astronomy, Cornell University, Ithaca, NY 14853, USA}
\affil{$^{13}$Jet Propulsion Laboratory, California Institute of Technology, 4800 Oak Grove Drive, Pasadena, CA 91109, USA}
\affil{$^{14}$Instituto de Astrof\'isica, Facultad de F\'isica, Pontificia Universidad Cat\'olica de Chile, Casilla 306, Santiago 22, Chile}
\affil{$^{15}$China-Chile Joint Center for Astronomy (CCJCA), Camino El Observatorio 1515, Las Condes, Santiago, Chile}
\affil{$^{16}$National Astronomical Observatories, Chinese Academy of Sciences (CAS), Beijing 100012, China}
\affil{$^{17}$NASA Herschel Science Center, IPAC, California Institute of Technology, MS 100-22, Cech, Pasadena, CA 91125}
\affil{$^{18}$Institut de Cincies del Cosmos (ICCUB), Universitat de Barcelona (IEEC-UB), Marti i Franques 1, 08028 Barcelona, Spain}
\affil{$^{19}$Institute for Astronomy, University of Hawaii, 2680 Woodlawn Drive, Honolulu, HI 96822}
\affil{$^{20}$The SETI Institute, 189 Bernardo Avenue Suite 100, Mountain View, CA, 94043, USA}
\affil{$^{21}$ICREA, Pg. Llu\'is Companys 23, E-08010 Barcelona, Spain}
\affil{$^{22}$Joint ALMA Observatory, Alonso de C\'{o}rdova 3107, Vitacura, Santiago, Chile
\\
}

\begin{abstract}

We present an analysis of \OIasub, \OIIIbsub, \NIIasub, and \CIIsub\, far-infrared (FIR) fine-structure line observations obtained with \textit{Herschel}/PACS, for $\sim$\,240 local luminous infrared galaxies (LIRGs) in the Great Observatories All-sky LIRG Survey (GOALS). We find pronounced declines --deficits-- of line-to-FIR-continuum emission for \NIIasub, \OIasub\, and \CIIsub\, as a function of FIR color and infrared luminosity surface density, \SigmaIR. The median electron density of the ionized gas in LIRGs, based on the \NIIasub/\NIIbsub\, ratio, is \ne\,=\,41\,\cnmmm. We find that the dispersion in the \CIIsub\, deficit of LIRGs is attributed to a varying fractional contribution of photo-dissociation-regions (PDRs) to the observed \CIIsub\, emission, $f(\CIIpdr)$\,=\,\CIIpdr/\CIIsub, which increase from $\sim$\,60\,\% to $\sim$\,95\% in the warmest LIRGs. The \OIasub/\CIIpdr\, ratio is tightly correlated with the PDR gas kinetic temperature in sources where \OIasub\, is not optically-thick or self-absorbed. For each galaxy, we derive the average PDR hydrogen density, \nhvol, and intensity of the interstellar radiation field, \G, in units of \Go, and find \G/\nhvol\, ratios $\sim$\,0.1--50\,\Go\,\cmmm, with ULIRGs populating the upper end of the distribution. There is a relation between \G/\nhvol\, and \SigmaIR, showing a critical break at \SigmaIRstar\,$\simeq$\,5\,$\times$\,10$^{10}$\,\lsd. Below \SigmaIRstar, \G/\nhvol\, remains constant, $\simeq$\,0.32\,\Go\,\cmmm, and variations in \SigmaIR\, are driven by the number density of star-forming regions within a galaxy, with no change in their PDR properties. Above \SigmaIRstar, \G/\nhvol\, increases rapidly with \SigmaIR, signaling a departure from the typical PDR conditions found in normal star-forming galaxies towards more intense/harder radiation fields and compact geometries typical of starbursting sources.
\\

\end{abstract}


\section{Introduction}\label{s:intro}

One of the most fundamental processes studied in virtually any field of physics is the dissipation of energy \citep{Thomson1874}. In particular, in extra-galactic astrophysics, investigating how interstellar gas cools down is crucial to our understanding of galaxy formation and evolution, since gravity is only able to collapse structures when they are sufficiently cold.

Thirty years ago, data obtained with the \textit{Kuiper Airborne Observatory} (\textit{KAO}) revealed that the far-infrared (FIR) spectra of nearby galaxies were populated with some of the most intense emission lines observed across the electromagnetic spectrum, indicating that they are very efficient cooling channels of the interstellar medium (ISM) \citep{Watson1984, Stacey1991, Lord1996b}. A decade later, the \textit{Infrared Space Observatory} \citep[\textit{ISO};][]{Kessler1996} increased the number of galaxies with FIR fine-structure line detections to the dozens \citep{Malhotra1997, Malhotra2001, Luhman1998, Luhman2003}. But twenty more years needed to pass until the \textit{Herschel Space Observatory} \citep[\textit{Herschel} hereafter;][]{Pilbratt2010} re-opened a new window into the FIR universe providing spectroscopic data for significantly larger samples of nearby, intermediate and high-redshift galaxies \citep[e.g.,][]{GC2011, DS2013, DS2014, Rigopoulou2014, Magdis2014, Brisbin2015, Cormier2015, Rosenberg2015, Malhotra2017}.

The most important fine-structure lines emitted by atomic species in the $\sim$\,50--200\,$\mu$m wavelength range are: \NIII\, (\NIIIsub), \OIa\, (\OIasub), \OIIIb\, (\OIIIbsub), \NIIa\, (\NIIasub), \OIb\, (\OIbsub), \CII\, (\CIIsub) and \NIIb\, (\NIIbsub). Each of them originates from a different phase of the ISM, with C$^+$ and O emission mostly arising from the neutral and molecular medium within dense photo-dissociation regions (PDRs) surrounding newly formed massive stars \citep[][and references therein]{Tielens1985, Hollenbach1997}, and N$^{+,++}$ and O$^{++}$ emission being produced by warm ionized gas, both in the diffuse medium and dense (\HII) regions.

Of particular importance among star-forming galaxies are the so-called luminous IR galaxies (LIRGs; \LIR\,=\,10$^{11-12}$\,\Lsun). LIRGs cover the entire evolutionary merger sequence, ranging from isolated galaxies, to early interacting systems, to advanced mergers. They exhibit enhanced star formation rates (SFRs) and specific SFRs (SSFRs\,=\,SFR/\Mstar), consequence of the funneling of large amounts of gas and dust towards their nuclei due to the loss of angular momentum during the dynamical interaction. And while the presence of active galactic nuclei (AGN) in LIRGs is frequent \citep{Petric2011, AH2012}, their contribution to the bolometric luminosity of the hosts is still very limited in comparison to ultra-luminous IR galaxies \citep[ULIRGs; \LIR\,$\ge$\,10$^{12}$\,\Lsun][]{Veilleux2009}. Therefore, nearby LIRGs are a key galaxy population bridging the gap between normal, Milky-Way (MW) type star-forming galaxies and the most extreme, AGN-dominated (quasar-like) systems \citep{Sanders1996}. This diversity is also reflected on the fact that they cover the entire transition between main-sequence (MS) galaxies and starbursts \citep{DS2013}.

At high redshfit, LIRGs dominate the obscured star formation activity between \textit{z}\,$\sim$\,1--3 \citep[e.g.,][]{Berta2011, Murphy2011, Magnelli2011}, and a number of works have already shown that local LIRGs (and not ULIRGs) are probably the closest local analogs of this high-\textit{z}, IR-bright galaxy population \citep{Desai2007,Pope2008, MD2009, DS2010b, Stacey2010}. Thus, a comprehensive study of the physical properties of low-redshift LIRGs, and specifically of their ISM, is critical for our understanding of the evolution of galaxies and AGN across cosmic time.

To this end, we have performed a systematic study of the most important FIR cooling lines of the ISM in a complete, flux-limited sample of nearby LIRGs, the Great Observatories All-sky LIRG Survey \citep[GOALS;][]{Armus2009}, using \textit{Herschel} and its Photodetector Array Camera and Spectrometer \citep[PACS;][]{Poglitsch2010} as well as the Fourier-transform spectrometer (FTS) of the Spectral and Photometric Imaging Receiver \citep[SPIRE;][]{Griffin2010}. We combine these observations with mid-IR (MIR) spectroscopy previously obtained with the Infrared Spectrograph \citep[IRS;][]{Houck2004} on board the \textit{Sptizer Space Telescope} \citep[\textit{Spitzer} hereafter;][]{Werner2004} to provide a panchromatic view of the heating and cooling of the ISM in LIRGs across a wide range of integrated properties such as IR luminosity, compactness, dust temperature, AGN activity, and merger stage.

The paper is organized as follows: In \S\ref{s:sample} we present the LIRG sample and the new \textit{Herschel} spectroscopy, as well as the ancillary observations used in this work. In \S\ref{s:datared} we describe the processing and analysis of the data. The basic results are presented in \S\ref{s:results}. The ISM properties derived for the LIRG sample are discussed throughout \S\ref{s:discussion} in relation to specific emission line ratios. In particular, in \S\ref{ss:ciipdr} and \S\ref{ss:ciidef} we describe how the fractional contribution of PDRs to the \CIIsub\, emission shapes the observed trend between the \CIIsub\, deficit and the FIR color of LIRGs. Section \S\ref{ss:ciiion} presents the electron densities found for the ionized gas phase of the ISM derived from the nitrogen lines and discusses the implications. In \S\ref{ss:oicii} we present a link between the \OIasub/\CIIpdr\, and the kinetic temperature of the PDR gas. We explore PDR covering factors and metallicity variations in \S\ref{ss:oiiioi}, and confront FIR emission line ratios involving oxygen and nitrogen emission lines to photo-ionization models of \HII\, regions in \S\ref{ss:oiiinii}. We derive the average PDR properties for each galaxy in the sample in \S\ref{ss:pdrmodel} and show the existence of a critical break in the PDR conditions as a function of luminosity surface density in \S\ref{ss:g0nHSigmaIR}. We discuss the physical implications of this results in \S\ref{ss:dustyreg}.
The summary of the results and conclusions are given in \S\ref{s:summary}.

\section{Sample and Observations}\label{s:sample}

\subsection{The GOALS Sample}\label{ss:sample}

The Great Observatories All-sky LIRG Survey (GOALS; \citealt{Armus2009}) encompasses the complete sample of 202 LIRG and ULIRG systems contained in the \textit{IRAS} Revised Bright Galaxy Sample \citep[RBGS;][]{Sanders2003}, which in turn is also a complete, flux-limited sample of 629 galaxies with \textit{IRAS} $S_{60\,\mu m}\,>\,5.24$\,Jy and Galactic latitudes $|b|\,>\,5\,^\circ$. There are 180 LIRGs and 22 ULIRGs in GOALS and their median redshift is \textit{z} = 0.0215 (or $\sim\,95.2$\,Mpc), with the closest galaxy being at \textit{z} = 0.0030 (15.9\,Mpc; NGC~2146) and the farthest at \textit{z} = 0.0918 (400\,Mpc; IRAS~07251-0248). To date, there are many published works that have already exploited the science content of multi-wavelength data obtained mostly from space-born facilities including \textit{GALEX} UV \citep{Howell2010}, \textit{HST} optical and near-IR imaging \citep{Haan2011,Kim2013}, along with \textit{Chandra} X-ray \citep{Iwasawa2011}, as well as \textit{Spitzer}/IRS \citep{DS2010b, DS2011, Petric2011, Stierwalt2013, Stierwalt2014, Inami2013} and \textit{Herschel}/PACS and SPIRE spectroscopy \citep{DS2013, DS2014, Zhao2013, Lu2014, Lu2015, Zhao2016b}. Moreover, a number of ground-based observatories such as VLA, CARMA and ALMA have also been used to observe the GOALS sample \citep[e.g.,][among others]{Murphy2013a, Murphy2013b, Xu2014, Xu2015, Zhao2016a}.

The RBGS, and therefore the GOALS sample, were defined based on \textit{IRAS} observations. However, the higher angular resolution achieved by \textit{Spitzer} allowed us to spatially disentangle galaxies within the same LIRG system into separate components. From the more than 290 \textit{individual} galaxies in GOALS, not all were observed by \textit{Herschel}. In systems with two or more galactic nuclei, minor companions with MIPS\,24$\,\mu$m flux density ratios smaller than 1:5 with respect to the brightest galaxy were not targeted due to their small contribution to the total IR luminosity of the system. The method used to calculate the IR luminosities of individual galaxies (\LIRwave, as defined in \citealt{Sanders1996})\footnote{$F_{\rm IR\,[8-1000\,\mu m]}\,=\,1.8\,\times\,10^{-14}\,(13.48\,S_{\rm 12\mu m}\,+5.16\,S_{\rm 25\mu m}\,+2.58\,S_{\rm 60\mu m}\,+\,S_{\rm 100\mu m})$ [\wnmm], with $S_\nu$ in [Jy]. \LIR\,=\,$4\pi\,D_{\rm L}^2\,F_{\rm IR}$. The luminosity distances, $D_{\rm L}$, are taken from \cite{Armus2009}.} is described in \cite{DS2013} (see end of section~\ref{ss:dataanalysis} for further details on the scaling of the IR luminosities used in this work).

\subsection{Herschel/PACS Observations}\label{ss:pacsobs}

We have obtained FIR spectroscopic observations for 200 LIRG systems from GOALS using the Integral Field Spectrometer (IFS) of the PACS instrument on board \textit{Herschel} (IRASF08339+6517 and IRASF09111--1007 were not observed).  Since some targets contain multiple components, there are 241 individual galaxies with available spectra in at least one emission line. PACS/IFS range spectroscopy of the \OIasub, \OIIIbsub, \NIIasub\, and \CIIsub\, fine-structure emission lines was obtained for 239, 161, 75 and 239 individual sources, respectively. Most of the data were collected as part of our OT1 and OT2 programs (OT1\_larmus\_1, OT2\_larmus\_1; P.I.: L. Armus) accounting more than 200 hours of observing time in total. Additional observations that are publicly available in the \textit{Herschel} archive were included from various projects. The main programs from where these complementary data were gathered are: KPGT\_esturm\_1 (P.I.: E. Sturm), KPOT\_pvanderw\_1 (P.I.: P. van der Werf) and OT1\_dweedman\_1 (P.I.: D. Weedman).

The IFS on PACS is able to perform simultaneous spectroscopy in the $51-73$ or $70-105\,\mu$m (3rd and 2nd orders, respectively; ``blue'' camera) and the $102-210\,\mu$m (1st order; ``red'' camera) ranges. The IFU is composed of a 5\,$\times$\,5 array of individual detectors (spaxels) each with a field of view (FoV) of $\sim\,$9.4\arcsec\, on a side, for a total of 47\arcsec\,$\times$\,47\arcsec. The physical size of the PACS FoV at the median distance of our LIRG sample is $\sim$\,20\,kpc on a side. The number of spectral elements in each pixel is 16, which are rearranged together via an image slicer over two 16\,$\times$\,25 Ge:Ga detector arrays (blue and red cameras). The spectral range selected for the observations was scanned several times, increasing the spectral resolution up to at least Nyquist sampling.

While we requested line maps for some LIRGs of the sample (from two to a few raster positions depending on the target), pointed (one single raster) chop-nod observations were taken for the majority of galaxies. For those galaxies with maps, only one raster position was used to obtain the line fluxes presented in this work. For a more detailed discussion on how the observations were set up, we refer the reader to \cite{DS2013}.

\subsection{Herschel/SPIRE Observations}\label{ss:spireobs}

In addition to the PACS/IFS spectra, we obtained observations of the \NIIbsub\, emission line using the SPIRE Fourier Transform Spectrometer (FTS) for 121 galaxies in the GOALS sample \citep[OT1\_nlu\_1; P.I.: N. Lu;]{Lu2017}. These observations were part of a broader project whose primarily aim is to study the dense and warm molecular gas in LIRGs \citep[see][]{Lu2014,Lu2015}. Details about the data processing as well as the results concerning the \NIIbsub\, line emission are presented in \cite{Zhao2013,Zhao2016b}. See section~\ref{ss:dataanalysis} for further details about how these data are used.

\subsection{Spitzer/IRS Observations}\label{ss:irsobs}

As part of the \textit{Spitzer} GOALS legacy program, all galaxies observed with \textit{Herschel}/PACS have available \textit{Spitzer}/IRS low resolution, R\,$\sim\,60-120$ (SL module: $5.2-14.5\,\mu$m, and LL module: $14-38\,\mu$m), as well as medium resolution, R\,$\sim\,600$ (SH module: $9.9-19.6\,\mu$m, and LH module: $18.7-37.2\,\mu$m), slit spectroscopy. The IRS spectra were extracted with the \textit{Spitzer} IRS Custom Extraction (SPICE) software\footnote{http://irsa.ipac.caltech.edu/data/SPITZER/docs/dataanalysistools/tools/spice/}, using the standard extraction aperture and a point source calibration mode. The projected angular sizes of the apertures on the sky are 3.7\arcsec\,$\times$\,12\arcsec\, at 9.8$\,\mu$m in SL, 10.6\arcsec\,$\times$\,35\arcsec\, at 26$\,\mu$m in LL, 4.7\arcsec\,$\times$\,15.5\arcsec\, at 14.8$\,\mu$m in SH, and 11.1\arcsec\,$\times$\,36.6\arcsec\, at 28$\,\mu$m in LH. Thus, the area covered by the SL and LL apertures is approximately equivalent (within a factor of $\sim\,2$) to that of an individual spaxel and a 3\,$\times$\,3 spaxel box of the IFS in PACS, respectively. Likewise, the SH and LH apertures are approximately equivalent to one spaxel and a 3\,$\times$\,3 aperture box in PACS, respectively. Aside from the line fluxes, which were presented in \cite{Stierwalt2014}, other observables derived from the IRS data that we use in this work are: the strength of the 9.7$\,\mu$m silicate feature, \SSi\footnote{The silicate strength is defined as \SSi\,=\,ln\,($f_{\rm peak}/f_{\rm cont}$), where $f_{\rm peak}$ is the flux density at the peak absorption (or emission) close to 9.7\,$\mu$m, and $f_{\rm cont}$ is the flux density of the continuum emission measured outside of the feature, interpolated at the wavelength of the peak. Thus, negative values indicate the feature appears in absorption while positive values indicate it is emission.}, and the equivalent width (EW) of the \PAHd, both of which were presented in \cite{Stierwalt2013}. We refer the reader to these works for further details about the reduction, extraction and calibration of the IRS spectra as well as for the main results derived from the analysis.

\section{Herschel/PACS Data Reduction and Analysis}\label{s:datared}

\subsection{Data Processing}\label{ss:dataproc}

The \textit{Herschel} Interactive Processing Environment (HIPE; v13.0) application was used to retrieve the raw data from the \textit{Herschel} Science Archive (HSA\footnote{http://herschel.esac.esa.int/Science\_Archive.shtml}) as well as to process them. We used the script for ``LineScan'' observations (also valid for ``range'' mode) to reduce our spectra. We processed the data from level 0 up to level 2 using the following steps: Flag and reject saturated data, perform initial calibrations, flag and reject ``glitches'', compute the differential signal of each on-off pair of data-points for each chopper cycle, divide by the relative spectral response function, divide by the response, convert frames to PACS cubes, and correct for flat-fielding. Next, for each camera (red or blue), HIPE builds the wavelength grid, for which we chose a final rebinning with an \textit{oversample}\,=\,2, and an \textit{upsample}\,=\,1 that corresponds to a Nyquist sampling. The spectral resolution achieved for each line was derived directly from the data and is $\sim$\,82\,\kms\, for \OIasub, $\sim$\,120\,\kms\, for \OIIIbsub, $\sim$\,287\,\kms\, for \NIIasub, and $\sim$\,235\,\kms\, for \CIIsub. The final processing steps were: flag and reject remaining outliers, rebin all selected cubes on consistent wavelength grids and, finally, average the nod-A and nod-B rebinned cubes (all cubes at the same raster position are averaged). This is the final science-grade product currently possible for single raster observations. From this point on, the analysis of the spectra was performed using in-house developed {\sc idl} routines.

\subsection{Data Analysis}\label{ss:dataanalysis}

To obtain the line flux of a source we use an iterative procedure described in detail in \cite{DS2013}. The only difference with respect to the method explained in that work is that, due to the highly variable profile of the lines and underlying continuum, for some sources we had to slightly modify the spectral range over which the final continuum-subtracted spectrum is integrated. This was done on a case by case basis over the entire galaxy sample and for each line \textit{and} spaxel, to ensure that the selected range was correct. The uncertainty associated with the line flux is calculated as the standard deviation of the final fitted underlying continuum integrated over the same wavelength range as the line. That is, uncertainties for all quantities used across this analysis are based on the individual spectrum of each line, therefore reflecting the uncertainties associated with --and measured directly from-- the data. Absolute photometric uncertainties, which can be as high as $\sim\,$11--12\%\footnote{http://herschel.esac.esa.int/Docs/PACS/html/ch04s10.html}, are not taken into account except for the analysis performed in section~\ref{ss:pdrmodel} (the version of the calibration files used in this work was PACS\_CAL\_69\_0)\footnote{http://herschel.esac.esa.int/twiki/bin/view/Public/PacsCalTreeHistory}.

We extracted the line fluxes for our LIRGs from the spectra using three different apertures: (a) the spaxel at which the 63$\,\mu$m continuum emission of the galaxy peaks (hereafter referred as ``central'' spaxel); (b) in a 3\,$\times$\,3 spaxel box centered on the central spaxel as defined in (a), limited by the PACS FoV, and (c) the total FoV (5\,$\times$\,5 spaxel box). In order to recover the total flux of the source from the spectrum extracted from the central spaxel (method (a)), we apply a point-source aperture correction (which varies as a function of wavelength). We note that this correction will only recover the total flux in sources that are unresolved by PACS. For extended sources the fluxes obtained in this manner will be lower limits to the integrated flux of the galaxy.

Table~\ref{t:fluxes} presents the measurements made available in electronic form for each line (\OIasub, \OIIIbsub, \NIIasub, and \CIIsub). The line and continuum fluxes provided are those obtained using the (a) method as well as the ``best'' spatially-integrated value for each individual galaxy. This is defined as the highest flux value that maximizes the S/N ratio of the measurement (line or continuum). In other words, we select, among the three methods described above, the one that has the lowest noise in the measurement while still accounting for the entire flux of the galaxy enclosed by the PACS FoV. As an additional constraint we require that no other source is contained within the aperture used to represent the integrated galaxy flux. Note, however, that some galaxies have companions at distances $\lesssim$\,9.4\arcsec\, (a PACS spaxel). These objects are marked in the tables and figures.
As mentioned in section \ref{ss:irsobs}, the angular size of a PACS spaxel is roughly similar to that of the aperture used to extract the \textit{Spitzer}/IRS spectra. We note that the \textit{Spitzer} and \textit{Herschel} pointings usually coincide within $\lesssim$\,2\arcsec. For further details regarding the pointing accuracy and centering of a source within a given spaxel we refer the reader to \cite{DS2013}.

\begin{deluxetable}{lcc}
\tabletypesize{\scriptsize}
\tablecaption{\scriptsize Table Content of Emission Line and Continuum Data}
\tablehead{
\colhead{Column \#} &
\colhead{Quantity} & 
\colhead{Units}}
\startdata 
\colhead{(1)} &
\colhead{Id.} & 
\colhead{\dots} \\ 
\colhead{(2)} &
\colhead{Galaxy name} & 
\colhead{\dots} \\ 
\colhead{(3,4)} &
\colhead{Raster (x,y)} & 
\colhead{[pixel]} \\ 
\colhead{(5,6)} &
\colhead{Central Spaxel (x,y)} & 
\colhead{[pixel]} \\ 
\colhead{(7,8)} &
\colhead{R.A. and Dec of (4,5)} & 
\colhead{\dots} \\ 
\colhead{(9)} &
\colhead{Dist to \OIasub\, cont.} & 
\colhead{[\arcsec]} \\ 
\colhead{(10)} &
\colhead{Central line flux} & 
\colhead{$\times$\,10$^{-17}$ [\wnmm]} \\ 
\colhead{(11)} &
\colhead{Central cont. flux dens.} & 
\colhead{[Jy]} \\ 
\colhead{(12)} &
\colhead{Companion Galaxy?} & 
\colhead{\dots} \\ 
\colhead{(13,14,15)} &
\colhead{R-C, R-L, C-L} & 
\colhead{\dots} \\ 
\colhead{(16)} &
\colhead{Best line flux} & 
\colhead{$\times$\,10$^{-17}$ [\wnmm]} \\ 
\colhead{(17)} &
\colhead{Best cont. flux dens.} & 
\colhead{[Jy]} \\ 
\colhead{(18)} &
\colhead{Best measurement} & 
\colhead{\dots} \\ 
\colhead{(19)} &
\colhead{AOR} & 
\colhead{\dots} \\
\colhead{(20)} &
\colhead{Program} & 
\colhead{\dots}
\enddata
\tablecomments{\scriptsize \textbf{The data table for each FIR emission line is available in the electronic form of this paper.} The columns include the line and continuum fluxes observed with \textit{Herschel}/PACS as well as a number of measurements: (1) Idientification number; (2) Name of the galaxy; (3,4) The raster used to obtain the galaxy measurements; (5,6) Reference central spaxel, defined as the closest spaxel to the 63\,$\mu$m continuum peak of the galaxy within the 5\,$\times$\,5 PACS FoV in the raster (cols. 3-4) of the AOR (col. 19); (7,8) Right ascension and declination of the reference spaxel; (9) Distance from the 63\,$\mu$m continuum peak to the reference spaxel; (10) Flux and uncertainty of the line measured from the reference spaxel ((a) method. Uncertainty does not account for absolute photometric calibration uncertainty in cols. (10,11,16,17)). Negative values in cols. (10-11) indicate upper limits; (11) Flux density and uncertainty of the continuum under the line measured from the reference spaxel ((a) method); (12) A companion galaxy exist within the reference spaxel (1\,=\,yes; 0\,=\,no); (13,14,15) Do the spaxels of the reference (R), continuum (C) and line (L) emission peaks coincide among each other? (1\,=\,yes; 0\,=\,no); (16) Galaxy-integrated flux and uncertainty of the line measured from the best aperture (see col. (18)). Negative values in cols. (16-17) indicate upper limits; (17) Galaxy-integrated flux density and uncertainty of the continuum under the line measured from the best aperture (see col. (18)); (18) Best measurement type: methods (a), (b) or (c) (see text for details); (19) AOR Id of the dataset used; (20) Program Id of the AOR. The IR luminosities and luminosity surface densities of the galaxies, \LIR\, and \SigmaIR\, --the latter defined as (\LIR/2)/$\pi$R$^{2}_{\rm 70\mu m,eff}$--, can be found at: \url{http://goals.ipac.caltech.edu}}\label{t:fluxes}
\end{deluxetable}

In order to estimate the fractional contribution of the PDR component to the total \CIIsub\, emission based on the \CIIsub/\NIIbsub\, ratio in our LIRGs in section~\ref{ss:ciipdr}, we extracted an additional set of \CIIsub\, spectra using a circular aperture with a diameter equal to the angular size of the SPIRE beam at 205\,$\mu$m ($\approx$\,17\arcsec), to which we applied an aperture correction based on the PACS spectrometer beam efficiency maps (v6) provided in the PACS instrument and calibration webpages\footnote{http://herschel.esac.esa.int/twiki/bin/view/Public/PacsCalibrationWeb}, after rebinning them to the regular 5\,$\times$\,5 spaxel FoV. We note that we do not apply the correction factors based on the FIR color of galaxies provided in \cite{Lu2017} to the \NIIbsub\, emission. We use instead the original, point-source calibrated fluxes since we do not use those corrections in our \textit{Herschel}/PACS data.

The individual FIR luminosities of galaxies belonging to a LIRG system formed by of two or more components were derived in a similar manner to the method used to calculate their \LIR\, (section \ref{ss:sample}). In order to obtain the \LIR\, and \LFIR\, of a given galaxy for the different extraction appertures described above, we scaled the integrated \textit{IRAS} FIR luminosity of the system (\LFIRwave, as defined in \citealt{Helou1985})\footnote{$F_{\rm FIR\,[42.5-122.5\,\mu m]}\,=\,1.26\,\times\,10^{-14}\,(2.58\,S_{\rm 60\mu m}\,+\,S_{\rm 100\mu m})$ [\wnmm], with $S_\nu$ in [Jy]. \LFIR\,=\,$4\pi\,D_{\rm L}^2\,F_{\rm FIR}$. The luminosity distances, $D_{\rm L}$, are taken from \cite{Armus2009}.} and \LIR\, with the ratio between the continuum flux density of each individual galaxy evaluated at 63$\,\mu$m in the PACS spectrum (measured in the same aperture as the line) and the total \textit{IRAS} 60$\,\mu$m flux density.

\section{Results}\label{s:results}

\subsection{The Contribution of AGN}\label{ss:bolagnfrac}

Following the formulation described in \cite{Veilleux2009}, we calculate the potential contribution of an AGN to the MIR and bolometric luminosities of each galaxy in GOALS \citep[see also][]{Petric2011} employing up to five \textit{Spitzer}/IRS diagnostics, depending on the availability of data: the \NeVsub/\NeIIsub\, and \OIVsub/\NeIIsub\, emission line ratios, the \PAHd\, EW \citep[see also][]{Armus2007}, the \Sc/\Sd\, dust continuum slope\footnote{We modified the reference value for the \Sc/\Sd\, ratio of a pure starburst/\HII\, source from 22.4 \citep[see Table~9 in][]{Veilleux2009} to 10 in order to reflect more accurately the actual distribution of \Sc/\Sd\, ratios seen in the GOALS sample. We also adapt the PAH EW diagnostic, which in \citep{Veilleux2009} is developed for the 7.7\,$\mu$m feature, to be used with the \PAHd, and modify the reference value for pure starbursts and its bolometric correction factor accordingly. We further assume there is no PAH destruction due to the AGN when calculating its fractional contribution using this method \citep{DS2008, DS2010a, Esquej2014}.}, and the Laurent diagram, which is based on a decomposition of the MIR spectrum of galaxies in three individual components: AGN, PAH and \HII\, emissions \citep[see][for a detailed explanation of this diagnostic]{Laurent2000}. These indicators are based on how the central AGN modifies the MIR line and continuum spectrum of a normal star-forming galaxy --either through the ionization of the surrounding gas to higher states and/or via the heating of dust at higher temperatures than star-formation--, and provide an estimate of its fractional contribution to the MIR emission, \alphamirAGN. Once these values are known, corrections based on SED templates of pure starbursting and AGN-powered sources can be applied to derive the fractional contribution of AGN to the bolometric luminosity of galaxies, \alphabolAGN\, \cite[][their Table~10, and discussion therein]{Veilleux2009}.

While individually each of these diagnostics has its own particular limitations, the combination of all of them allows for a reasonable quantification of the AGN's average fractional luminosity contribution, $<$\alphamirAGN$>$ and $<$\alphabolAGN$>$. Because there is a significant number of non-detections in one or more of the relevant MIR features employed to calculate $<$\alphamirAGN$>$ or $<$\alphabolAGN$>$, we used the Astronomy SURVival analysis package, {\sc asurv} \citep[v1.3]{Lavalley1992}, which adopts the maximum-likelihood Kaplan-Meier (KM) estimator to compute the mean value of univariate distributions containing censored data \citep{Feigelson1985}. Table~\ref{t:agnfrac} presents the \alphamirAGN\, values for each MIR diagnostic, as well as $<$\alphamirAGN$>$ and $<$\alphabolAGN$>$.

\begin{deluxetable*}{lrrrrrrr}
\tabletypesize{\scriptsize}
\tablewidth{0pc}
\tablecaption{\scriptsize Table Content of AGN Fractions}
\tablehead{\colhead{Id.} & \colhead{\NeVsub/} & \colhead{\OIVsub/} & \colhead{\PAHd} & \colhead{\Sc/\Sd} & \colhead{Laurent} & \multicolumn{2}{c}{$<$\alphaAGN$>$} \\
\colhead{} & \colhead{\NeIIsub} & \colhead{\NeIIsub} & \colhead{EW} & \colhead{ratio} & \colhead{diagram} & \multicolumn{2}{c}{KM estimator} \\
\colhead{} & \colhead{MIR} & \colhead{MIR} & \colhead{MIR} & \colhead{MIR} & \colhead{MIR} & \colhead{MIR} & \colhead{Bol.} \\
\colhead{(1)} & \colhead{(2)} & \colhead{(3)} & \colhead{(4)} & \colhead{(5)} & \colhead{(6)} & \colhead{(7)} & \colhead{(8)}}
\tablecomments{\scriptsize \textbf{The data table is available in the electronic form of this paper.} The columns include the fractional AGN contributions to the emission of LIRGs and the associated uncertainties, based on different \textit{Spitzer}/IRS diagnostics. (1) Identification number to match with Table~\ref{t:fluxes}; (2-6): MIR AGN fractios based on each individual diagnostic (see text for details). Negative values indicate upper limits in the case of the \NeVsub/\NeIIsub\, and \OIVsub/\NeIIsub\, ratios (cols. 2-3), and lower limits in the case of the \PAHd\, EW, \Sc/\Sd\, ratio and the Laurent diagram diagnostics (cols. 4-6); (7,8): Average MIR and bolometric AGN fractions based on all diagnostics. The bolometric fractions based on individual diagnostics can be found at: \url{http://goals.ipac.caltech.edu}}\label{t:agnfrac}
\end{deluxetable*}

\subsection{FIR Line Deficits}\label{ss:linedef}

\begin{figure*}[t!]
\vspace{.25cm}
\epsscale{.54}
\plotone{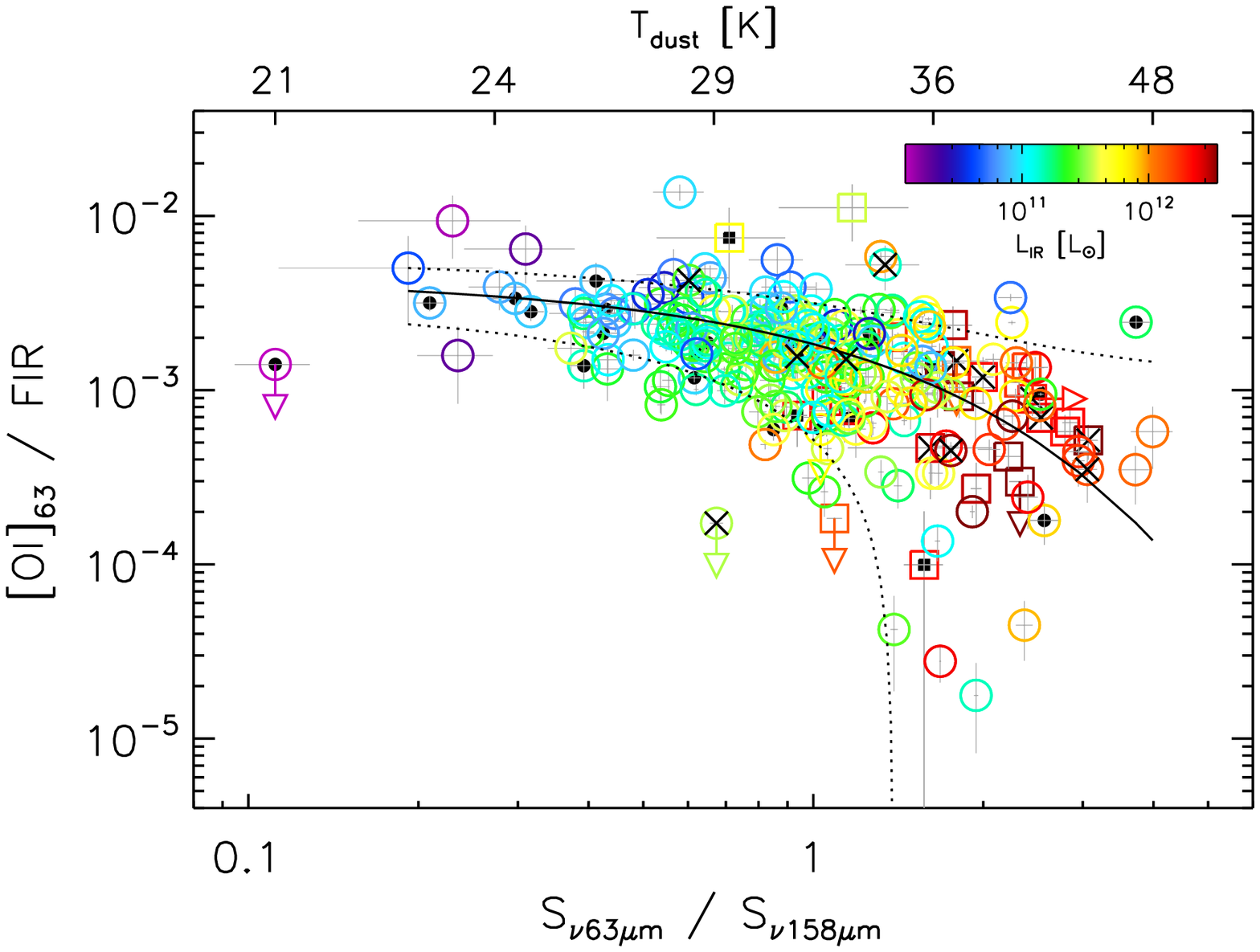}
\plotone{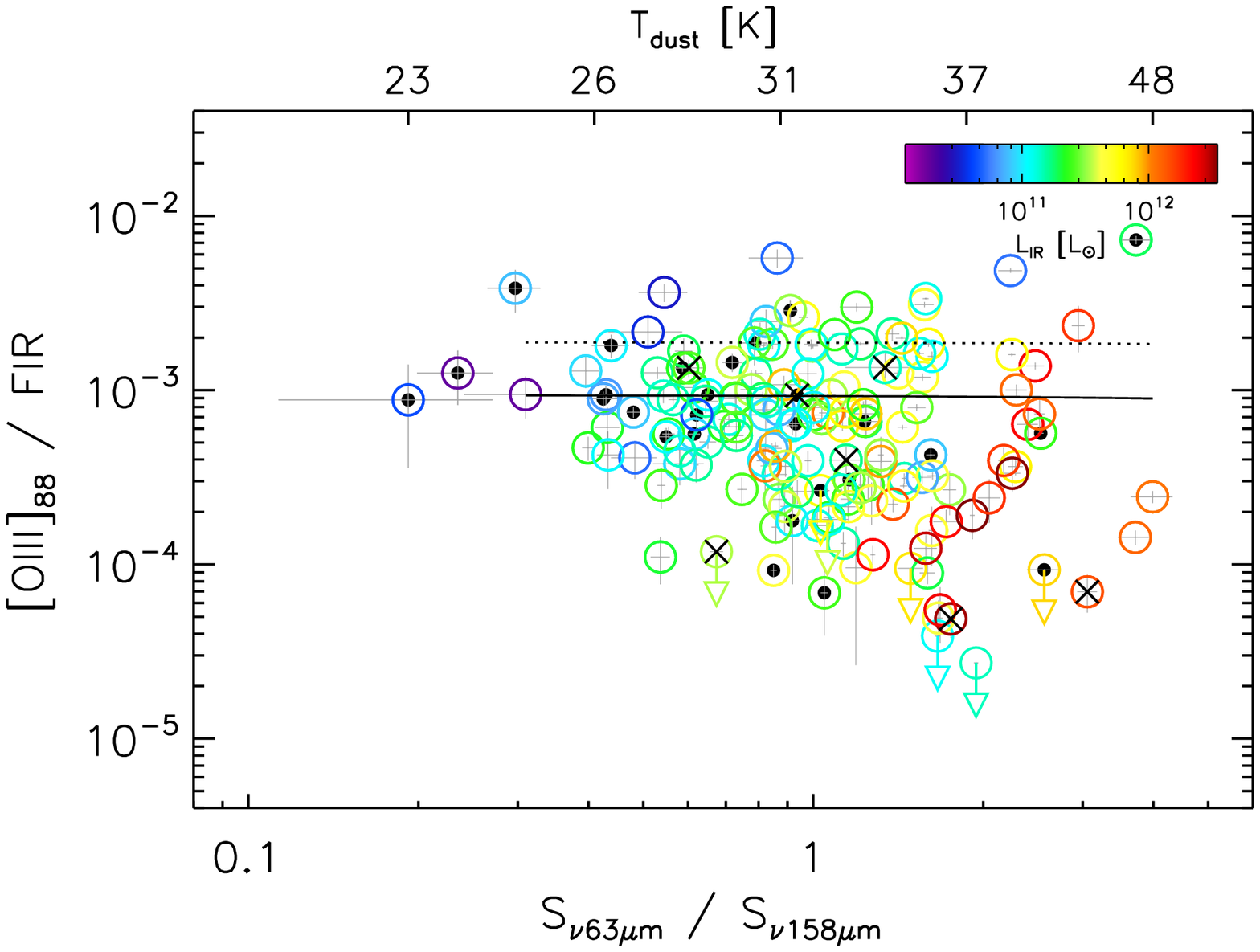}
\plotone{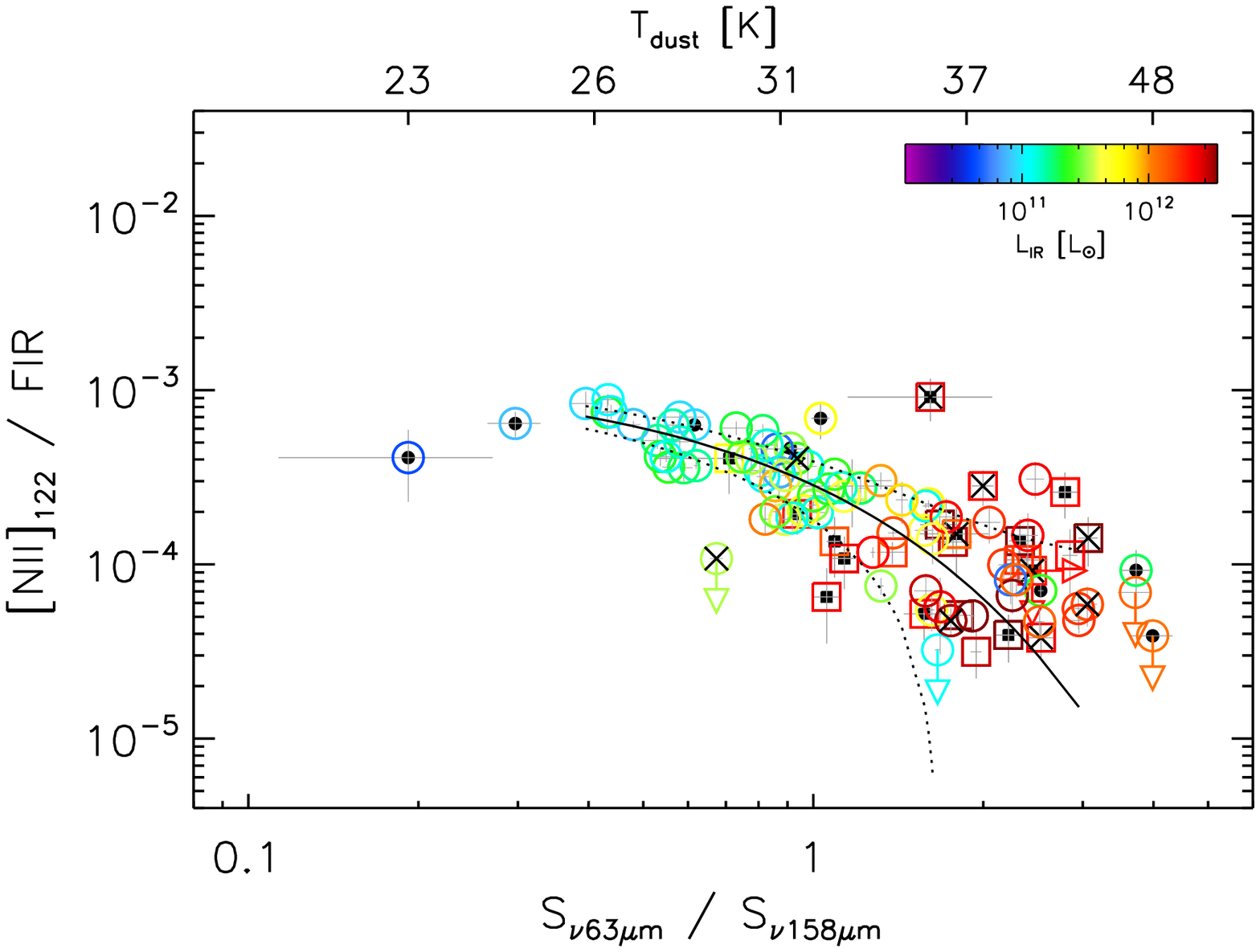}
\plotone{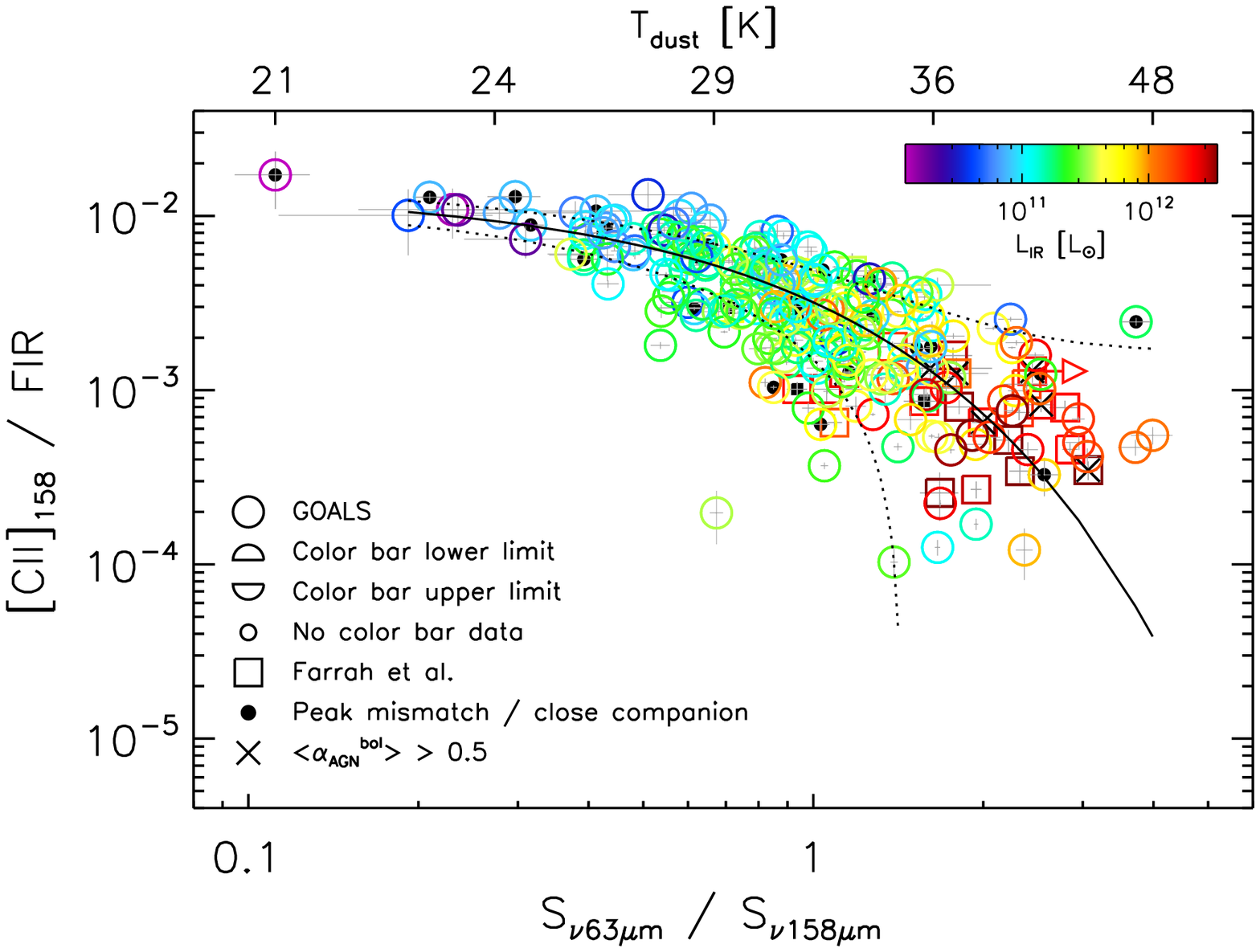}
\plotone{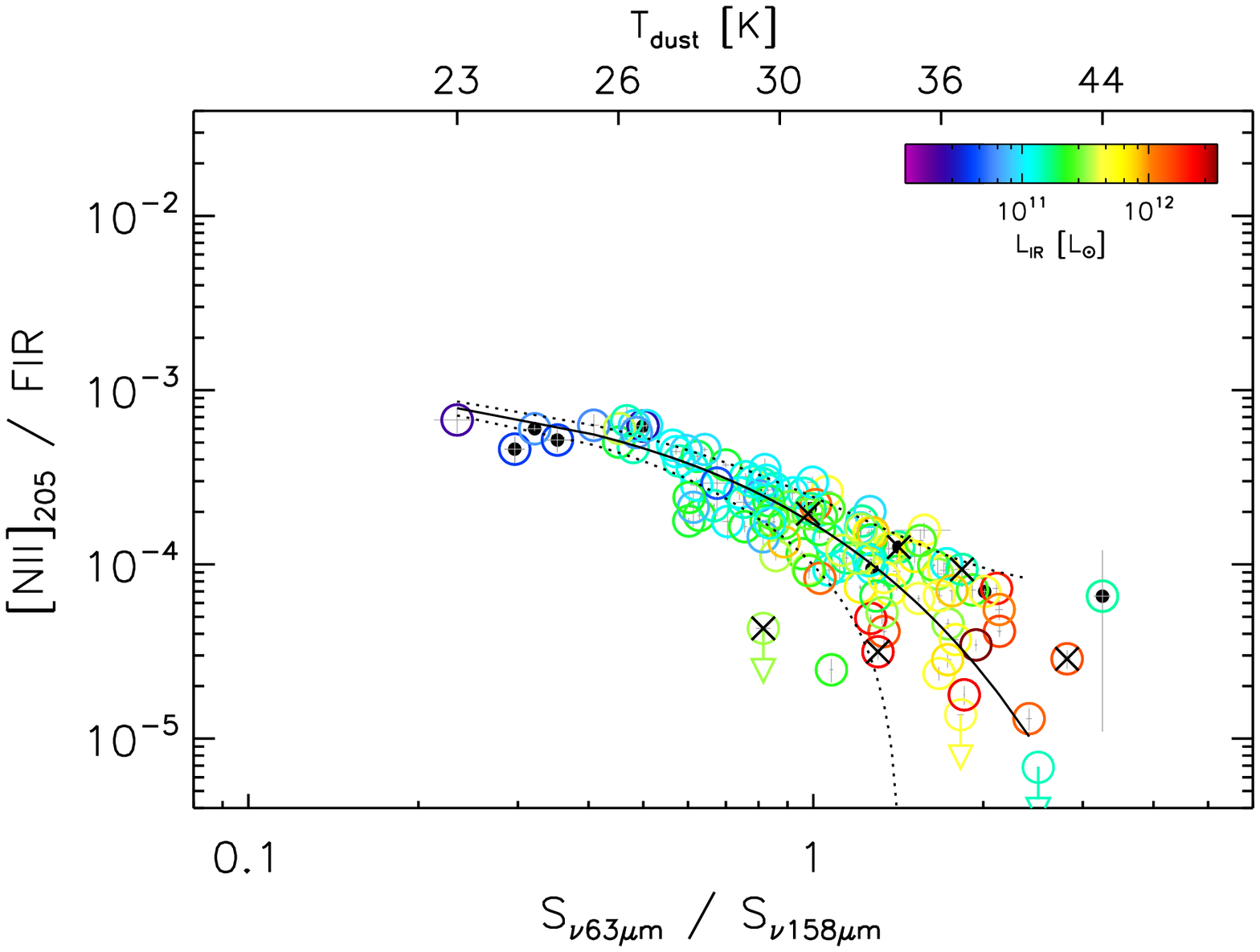}
\vspace{0cm}
\caption{\footnotesize The line deficits of the \OIa, \OIIIb, \NIIa, \CII\, and \NIIb\, emission lines, defined as the line to \FIRwave\, flux ratio as a function of the FIR \Sa/\Sb\, continuum ratio for the entire GOALS sample (open circles; see legend in mid-right panel). The dynamic range of the x- and y-axes is the same for all panels. The data shown here represent the best integrated emission values available for each galaxy and line, as described in section~\ref{ss:dataanalysis}. The top x-axis is the \Tdust\, of a modified black body (with $\beta$\,=\,1.8) that has a \Sa/\Sb\, ratio equal to the values shown in the bottom x-axis. The data are color-coded as a function of the total IR luminosity, \LIRwave, measured within the same aperture used to obtain the line and continuum emissions. Sources marked with black dots are galaxies where there is a mismatch between the location where the line and/or continuum emissions peak (see Table~\ref{t:fluxes}). Galaxies with companions within the aperture used to measure their integrated flux are also marked with black dots. In addition to the GOALS sample we also show the ULIRG sample from \cite{Farrah2013} (open squares), which populate better the high \Tdust\, regime. Galaxies with an AGN contributing more than 50\,\% to their bolometric luminosity, $<$\alphabolAGN$>$\,$\ge$\,0.5 (see section~\ref{ss:bolagnfrac}), are marked with black crosses. The solid lines show fits to the data using equation~\ref{e:linedef}, with the dotted lines representing the 1\,$\sigma$\, dispersion of the data with respect to the best fit. The dispersion is calculated in the same way for all the fits performed in throughout the paper. Neither galaxies marked as black dots nor AGN-dominated sources are used for the fits (here or in any other fit presented throughout the paper).
}\label{f:linedef}
\vspace{1.cm}
\end{figure*}

Using the best galaxy-integrated measurements described in section~\ref{ss:dataanalysis}, we present in Figure~\ref{f:linedef} the \OIasub, \OIIIbsub, \NIIasub, \CIIsub\, and \NIIbsub\, emission line deficits\footnote{The word ``deficit'' refers to the deficiency of a given line flux when compared to the dust continuum emission in a galaxy. This term was historically coined to express the decrease in the line (gas) cooling efficiency with respect to that of the dust in IR luminous galaxies.} --expressed as the line-to-FIR continuum flux ratio-- for the entire GOALS LIRG sample as a function of the $S_{\nu 63\,\mu m}/S_{\nu 158\,\mu m}$ (\Sa/\Sb) continuum flux density ratio, which is a first order tracer of the average dust temperature in galaxies, \Tdust. The dynamic range in both x- and y-axes is the same in all panels to facilitate the comparison. We have fitted the data using a functional form of the type:

\begin{equation}\label{e:linedef}
\begin{split}
L_{\rm line}/\LFIR\,=\,\epsilon_0\,e^{-(S_{63}/S_{158})/\delta}
\end{split}
\end{equation}

\noindent
where $\epsilon_0$ denotes a limiting line/FIR ratio for sources with cold FIR colors and no deficits, and $\delta$ is the \Sa/\Sb\, at which the line/FIR ratio has been reduced by a factor of \textit{e} with respect to $\epsilon_0$. The $\epsilon_0$ parameter can be understood as the nominal cooling efficiency of each line with respect to that of big dust grains, representative of normal star-forming galaxies. The parameters obtained from the fits to the entire GOALS sample can be found in Table~\ref{t:linedeffits} and the fits are displayed in Figure~\ref{f:linedef} as solid black lines, with the associated 1\,$\sigma$\, dispersion around the fit shown as dotted lines. While the choice of this particular functional form is arbitrary, it provides a better description of the trends than a power-law fit, as it is known that the line to FIR ratios do not increase indefinitely at the lower end of the \Tdust\, distribution and IR luminosities \cite[e.g.,][]{Malhotra1997, Malhotra2001, Brauher2008, DeLooze2014, Cormier2015, HC2015}. That is, the line to FIR ratios level off at low \Tdust\, reflecting a cooling efficiency ``ceiling'' (which depends on each line) of the gas in PDRs, \HII\, regions and the diffuse ISM, with respect to the energy dissipated by dust in thermal equilibrium in normal galaxies.

We convert the \Sa/\Sb\, ratios into dust temperatures (see upper x-axis in Figure~\ref{f:linedef}) by assuming that the observed FIR continuum emission is produced by a single-temperature modified black body (mBB) with a fixed emissivity index $\beta$\,=\,1.8 and whose emission is optically thin. This is a reasonable approximation for spectral energy distribution (SED) fits that do not include data at $\lambda$\,$\lesssim$\,60\,$\mu$m. We also provide a practical equation that relates both quantities, \Sa/\Sb\, and \Tdust, using the following approximation:

\begin{equation}\label{e:tdustfircolor}
\begin{split}
T_{\rm dust} = 20.24 + 14.54\,(\Sa/\Sb) - 3.75\,(\Sa/\Sb)^2 \\
+\,0.46\,(\Sa/\Sb)^3
\end{split}
\end{equation}

We note that this equation is only valid for the dust temperatures and \Sa/\Sb\, ratios spanned by the galaxies in the GOALS sample. That is 21\,$\leq$\,\Tdust\,$\leq$\,48\,K or 0.1\,$\leq$\,\Sa/\Sb\,$\leq$\,4. The error in \Tdust\, obtained from this expression is $\lesssim$\,0.5\,K for the dynamic ranges mentioned.

\begin{figure*}[t!]
\vspace{.25cm}
\epsscale{.54}
\plotone{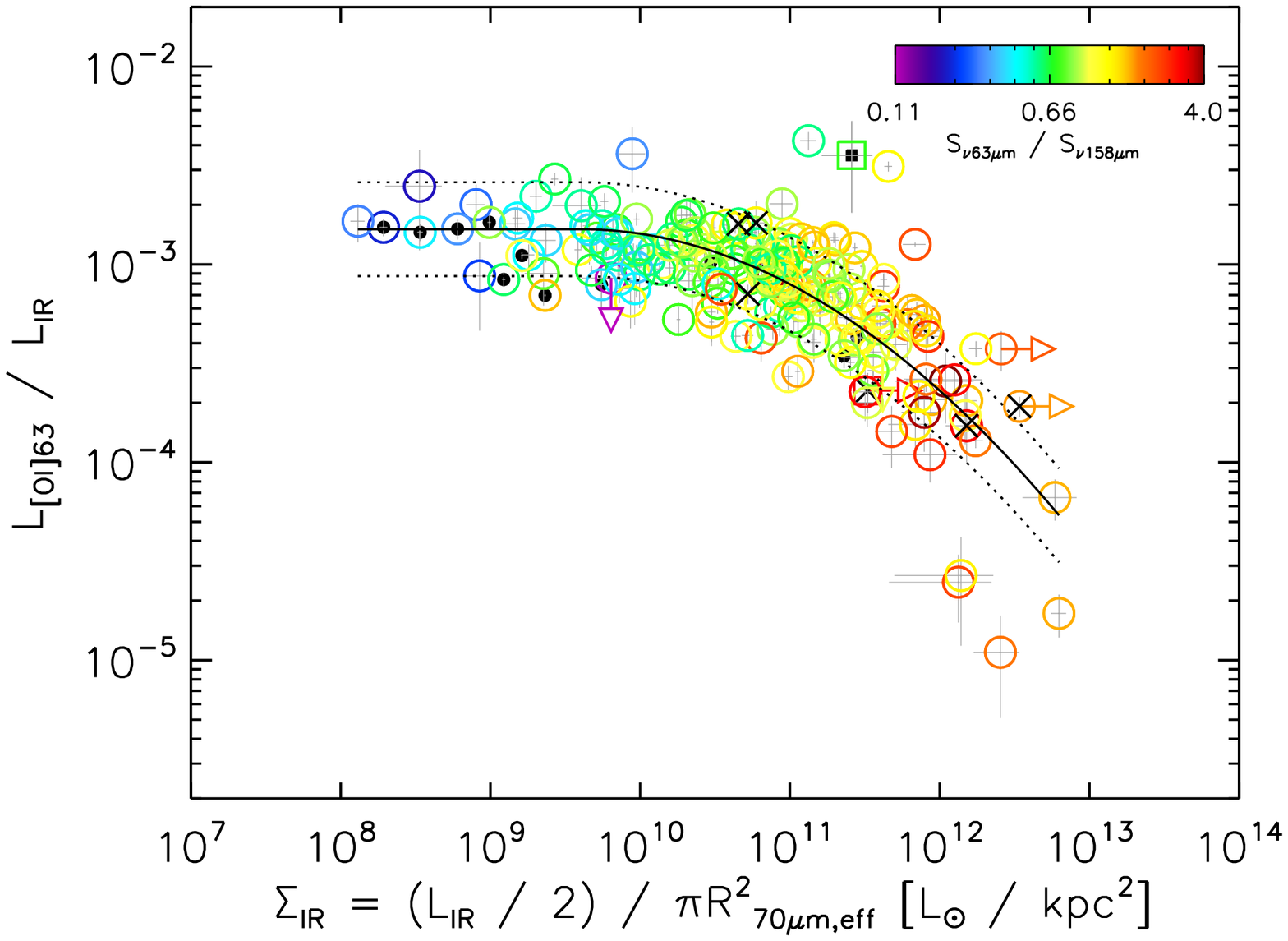}
\plotone{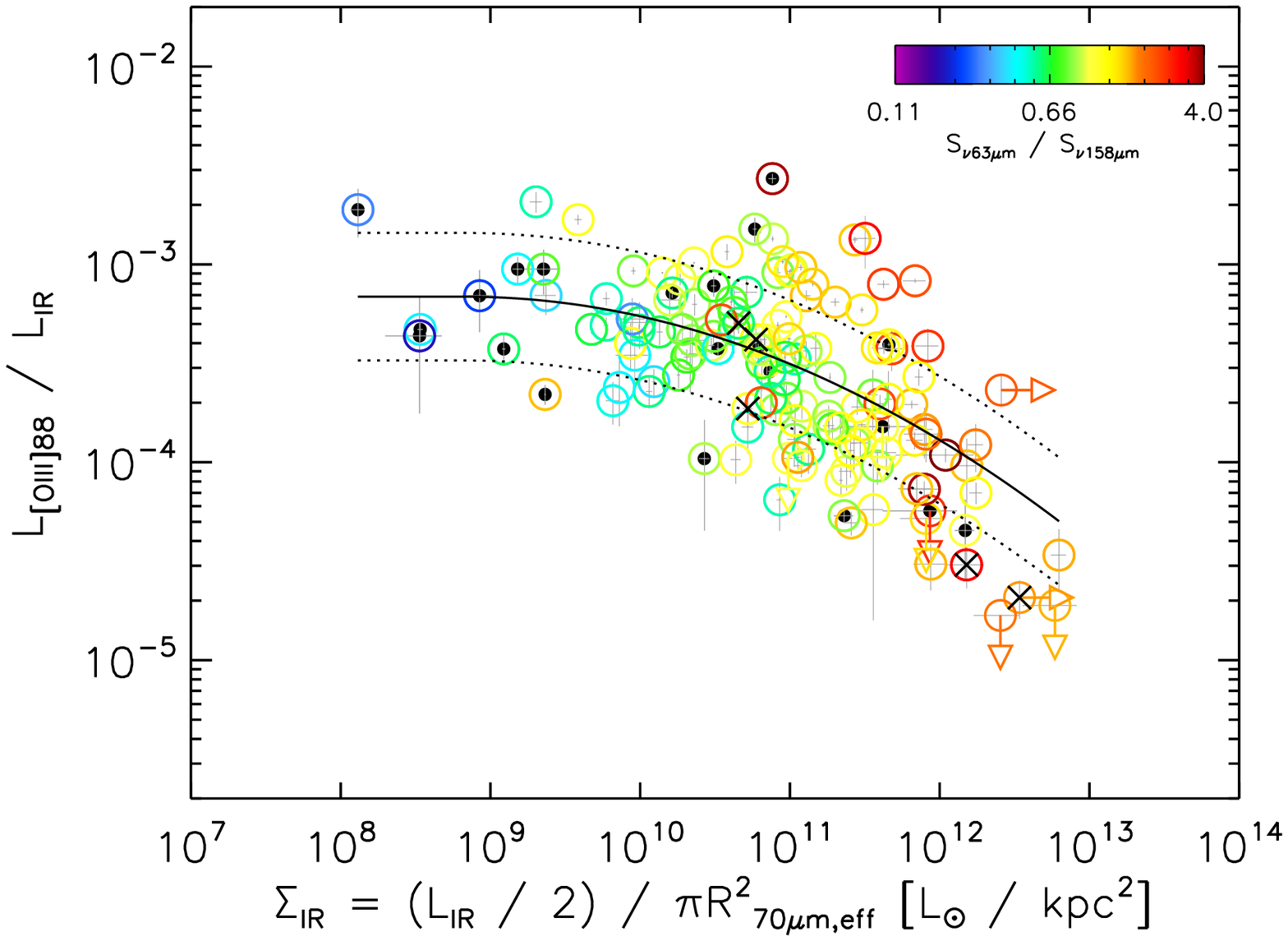}
\plotone{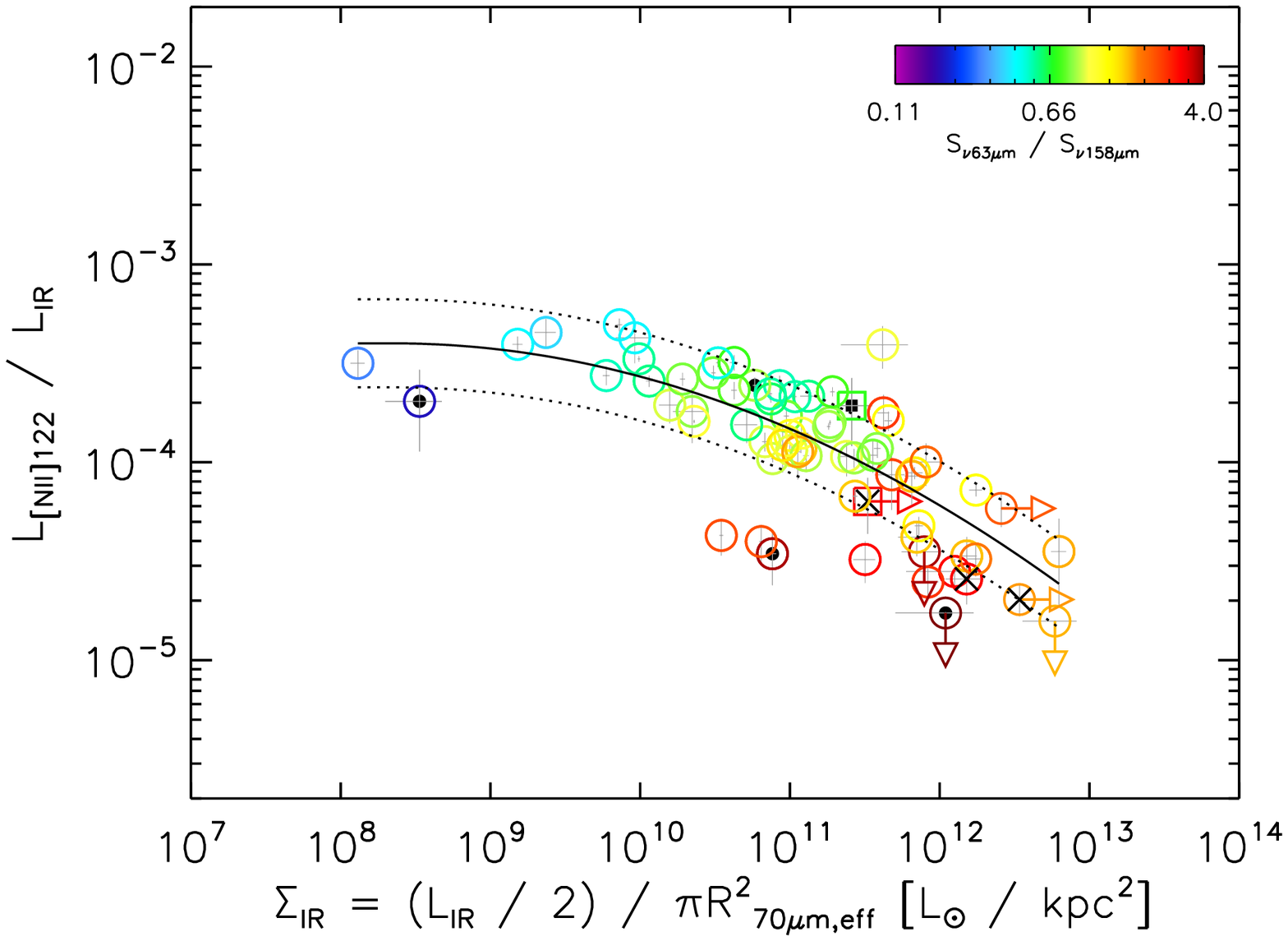}
\plotone{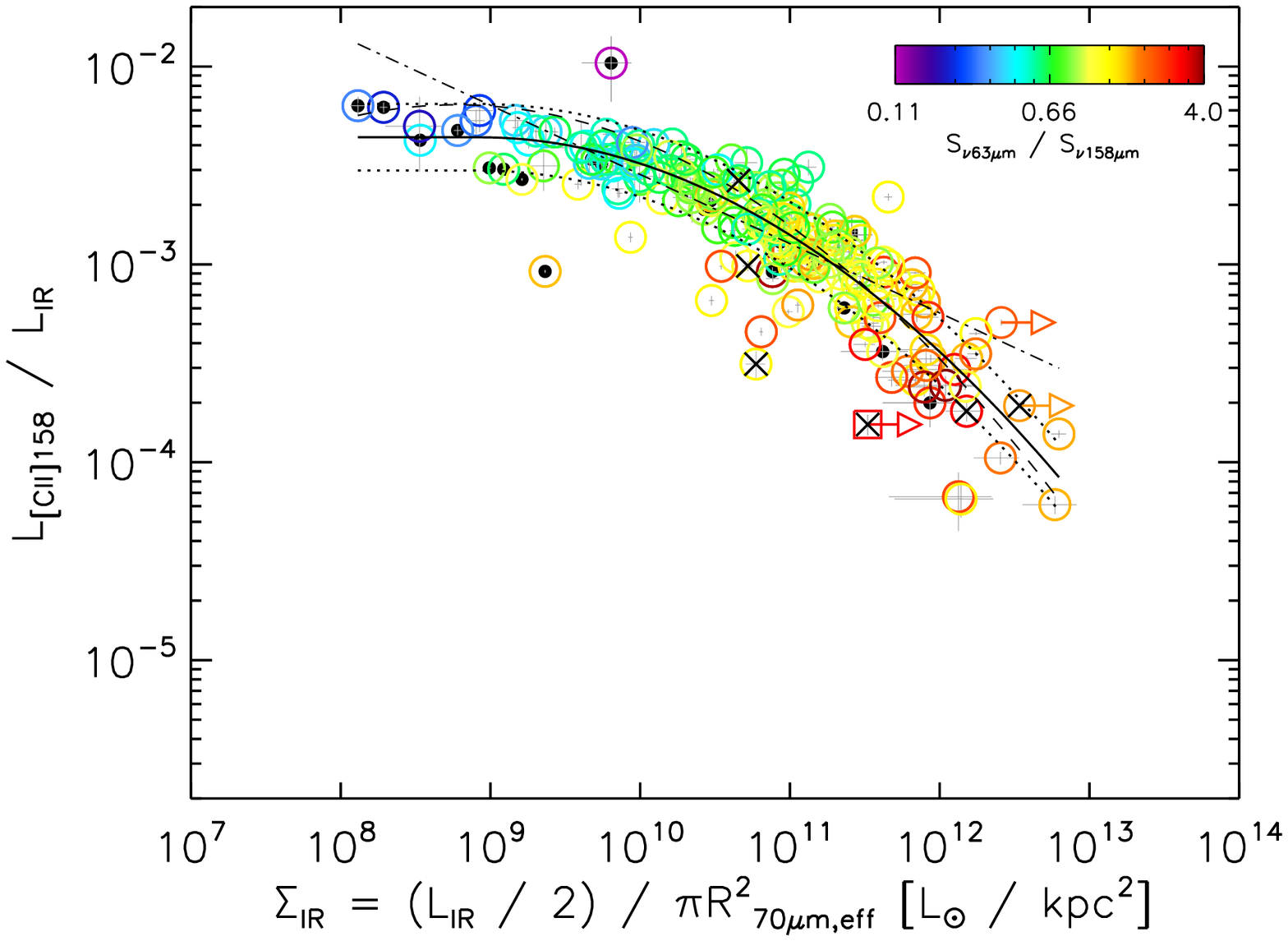}
\plotone{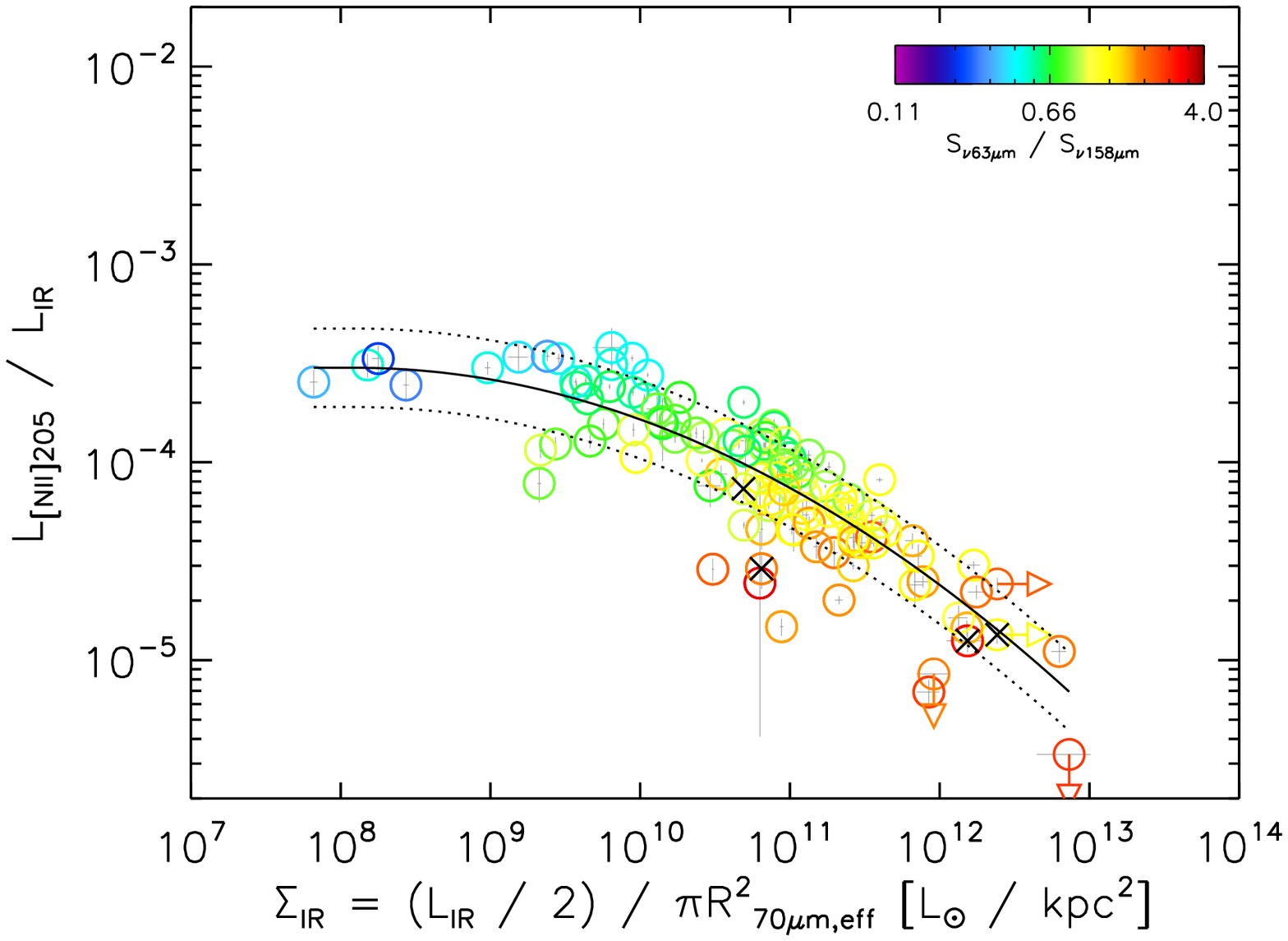}
\vspace{0cm}
\caption{\footnotesize The line deficits of the \OIasub, \OIIIbsub, \NIIasub, \CIIsub\, and \NIIbsub\, emission lines, defined as the line luminosity to \LIRwave\, ratio as a function of the IR luminosity surface density, defined as \SigmaIR\,=\,(\LIR/2)/\Areaeff, for galaxies in GOALS with available measurements of their FIR sizes \citep[taken from][]{Lutz2016} (open circles). Symbols are as in Figure~\ref{f:linedef}. We note that while the \LIR\, used in the y-axis represents the total IR luminosity of the galaxy (calculated in the same aperture as the line luminosity), the value used in the x-axis is the effective luminosity, \LIReff, where \LIReff\,=\LIR/2, since the measured sizes refer to the half-light radii of the sources. The solid lines represent a fit to the data using a second-order polynomial (see equation~\ref{e:linedeflsd}). The best fit parameters are tabulated in Table~\ref{t:linedeflsdfits}. The fits have been set to the maximum value of the respective quadratic equation below the wavelength at which the maximum is reached, such that the ratio remains constant. For reference, in the panel showing the \CIIsub\, deficit we also plot the best fit found by \cite{Lutz2016} using the same parametrization (dashed line; within the errors of our fit), and the best fit originally found by \cite{DS2013} using galaxy sizes measured in the MIR with \textit{Spitzer} and assuming a linear log-log relation (dotted-dashed line).}\label{f:linedeflsd}
\vspace{1.2cm}
\end{figure*}

Figure~\ref{f:linedef} shows that there is a common trend for most lines to show stronger deficits as the average \Tdust\, becomes warmer --including the two \NIIno\, lines, which arise from the ionized medium. The \OIasub\, deficit shows a decline of approximately an order of magnitude and a large scatter. The \NIIasub, \CIIsub\, and \NIIbsub, exhibit stronger deficits, of up to two orders of magnitude, and tighter trends. For the \OIIIbsub\, line, although there may be a deficit at the highest \Sa/\Sb\, values ($\gtrsim$\,2), the exponential fit yields a result that is statistically indistinguishable from a flat trend. Binning the data and obtaining the median values for each bin provides the same result.

A possible interpretation is that the dispersion in the \OIIIbsub/FIR ratio as a function of \Tdust\, may be related to the location where the line and dust emission originate within the star-forming regions. As we argue in sections~\ref{ss:ciidef} and \ref{ss:dustyreg} \citep[see also][]{DS2013}, most of the energy reprocessed by dust may be arising from grains in front of the PDRs, mixed with the ionized gas, and at low optical depths into the molecular cloud. Low ionization lines, that originate from the PDR itself or the outer edge of the \HII\, region (\NIIasub, \NIIbsub\, and \CIIsub) would show less scatter around the mean. On the other hand, lines arising from highly ionized gas within the \HII\, regions, like \OIIIbsub, and lines for which part of the emission may be arising from deeper into the molecular cloud, like \OIasub, would show larger dispersion around the trend average (see $\sigma$/$\epsilon_0$ in Table~\ref{t:linedeffits}). The potential contribution of an AGN could also introduce additional dispersion in the line deficits, specially at high \Sa/\Sb\, \citep{Fischer2014,GA2015} or when the AGN overwhelmingly dominates the MIR emission \citep{DS2014}, but we do not find galaxies with large fractional MIR or bolometric AGN contributions to be distributed with a significantly larger scatter than star-formation dominated sources, based on the \textit{Spitzer}/IRS spectral diagnostics described in section~\ref{ss:bolagnfrac}.

Figure~\ref{f:linedeflsd} presents the same line deficits but as a function of the luminosity surface density, \SigmaIR, defined as the ratio of effective luminosity, \LIReff\,=\,\LIR/2, divided by the effective area of the source that contains half of its luminosity measured at 70\,$\mu$m, \Areaeff\, for galaxies with available FIR size measurements. These correlations are overall tighter than with \Tdust\, (including \OIIIbsub), suggesting a closer physical connection between the cause(s) that give rise to the line deficits, and the concentration of dust-reprocessed energy --or IR ``compactness''-- of LIRGs (see discussion in section~\ref{ss:g0nHSigmaIR}). The scatter in the trends of those lines that have a PDR origin, \OIasub\, and \CIIsub, is especially small and remarkably constant in relative terms at any \SigmaIR. Moreover, the trend is followed by nearly all LIRGs regardless of their FIR color or \Tdust. We have fitted these correlations with a second-order polynomial function:

\begin{equation}\label{e:linedeflsd}
\begin{split}
log(L_{\rm line}/\LIR) = \alpha_0 + \alpha_1\,log\,\SigmaIR + \alpha_2\,(log\,\SigmaIR)^2
\end{split}
\end{equation}

\noindent
The best fit parameters are presented in Table~\ref{t:linedeflsdfits}. Note that in Figure~\ref{f:linedeflsd} the fits have been set to the maximum value of the respective quadratic equation below the wavelength at which the maximum is reached, such that the ratio remains constant.

\begin{deluxetable}{cccc}
\tabletypesize{\scriptsize}
\tablewidth{0pc}
\tablecaption{\scriptsize Best Parameters From The Line Deficit vs. $\Sa/\Sb$ Fits}
\tablehead{\colhead{Line} & \colhead{$\epsilon_0$} & \colhead{$\delta$} & \colhead{1$\sigma$ dispersion} \\
& \colhead{($\times$\,$10^{-3}$)} & \colhead{($e$-fold)} & \colhead{($\times$\,$10^{-3}$)}}
\startdata 
\OIasub  & 4.37\,$\pm$\,0.93 & 1.15\,$\pm$\,0.30 & 1.32 \\
\OIIIbsub & 0.94\,$\pm$\,0.17 & N/A & 0.94 \\
\NIIasub & 1.27\,$\pm$\,0.16 & 0.67\,$\pm$\,0.08 & 0.11 \\
\CIIsub  & 14.0\,$\pm$\,0.9 & 0.68\,$\pm$\,0.04 & 1.69 \\
\NIIbsub & 1.26\,$\pm$\,0.16 & 0.50\,$\pm$\,0.04 & 0.07
\enddata
\tablecomments{\scriptsize The parameters $\epsilon_0$ and $\delta$ correspond to the fits of the line deficits as a function of the \Sa/\Sb\, ratio for the entire GOALS sample using equation~\ref{e:linedef}. The fits are presented in Figure~\ref{f:linedef} (solid lines).}\label{t:linedeffits}
\end{deluxetable}

\begin{deluxetable}{crrrc}
\tabletypesize{\scriptsize}
\tablewidth{0pc}
\tablecaption{\scriptsize Best Parameters From The Line Deficit vs. $\SigmaIR$ Fits}
\tablehead{\colhead{Line} & \colhead{$\alpha_0$} & \colhead{$\alpha_1$} & \colhead{$\alpha_2$} & \colhead{1$\sigma$ dispersion} \\
\colhead{} & \colhead{} & \colhead{} & \colhead{} & \colhead{[dex]}}
\startdata 
\OIasub   & --15.83  &  2.711   &  --0.1412    &  0.24 \\
\OIIIbsub &  --8.78  &  1.272   &  --0.0721    &  0.32 \\
\NIIasub  &  --7.66  &  1.023   &  --0.0613    &  0.22 \\
\CIIsub   & --11.47  &  2.043   &  --0.1145    &  0.17 \\
\NIIbsub  &  --8.24  &  1.165   &  --0.0719    &  0.20 
\enddata
\tablecomments{\scriptsize The parameters correspond to the fits of the line deficits as a function of \SigmaIR\, for galaxies in GOALS with available FIR sizes, using equation~\ref{e:linedeflsd}. The fits are presented in Figure~\ref{f:linedeflsd} (solid lines).}\label{t:linedeflsdfits}
\end{deluxetable}

\section{Discussion}\label{s:discussion}

\subsection{\CIIsub\, Emission from Ionized Gas and PDRs}\label{ss:cii}

Because of the low ionization potential of the Carbon atom (11.26\,eV), the \CIIsub\, emission line can be produced not only in regions of ionized gas (\CIIion) but also in the dense, neutral ISM (\CIIpdr). However, due to the low critical density of the transition when it is collisionally excited by free electrons or protons \cite[$n^{\rm cr,[CII]}_{\rm e}$\,$\simeq$\,44\,\cnmmm, at $T$\,=\,8000\,K;][]{Goldsmith2012}, the \CIIsub\, is rapidly thermalized in mildly dense, ionized environments. Thus, unless the volume filling factor of the diffuse medium is very high (see section~\ref{ss:ciiion}), most of the \CIIsub\, emission --especially in actively star-forming galaxies-- is expected to arise from dense PDRs surrounding young, massive stars \citep{Hollenbach1997}, where the \CIIsub\, is collisionally excited by neutral and molecular hydrogen, with \nhcr$^{\rm ,[CII]}$\,$\simeq$\,3.0\,$\times$\,10$^3$\,\cnmmm\, and \nhhcr$^{\rm ,[CII]}$\,$\simeq$\,6.1\,$\times$\,10$^3$\,\cnmmm,  at $T$\,=\,100\,K \citep{Goldsmith2012}.

\subsubsection{Dense PDRs}\label{ss:ciipdr}

\begin{figure}[t!]
\vspace{.25cm}
\epsscale{1.15}
\plotone{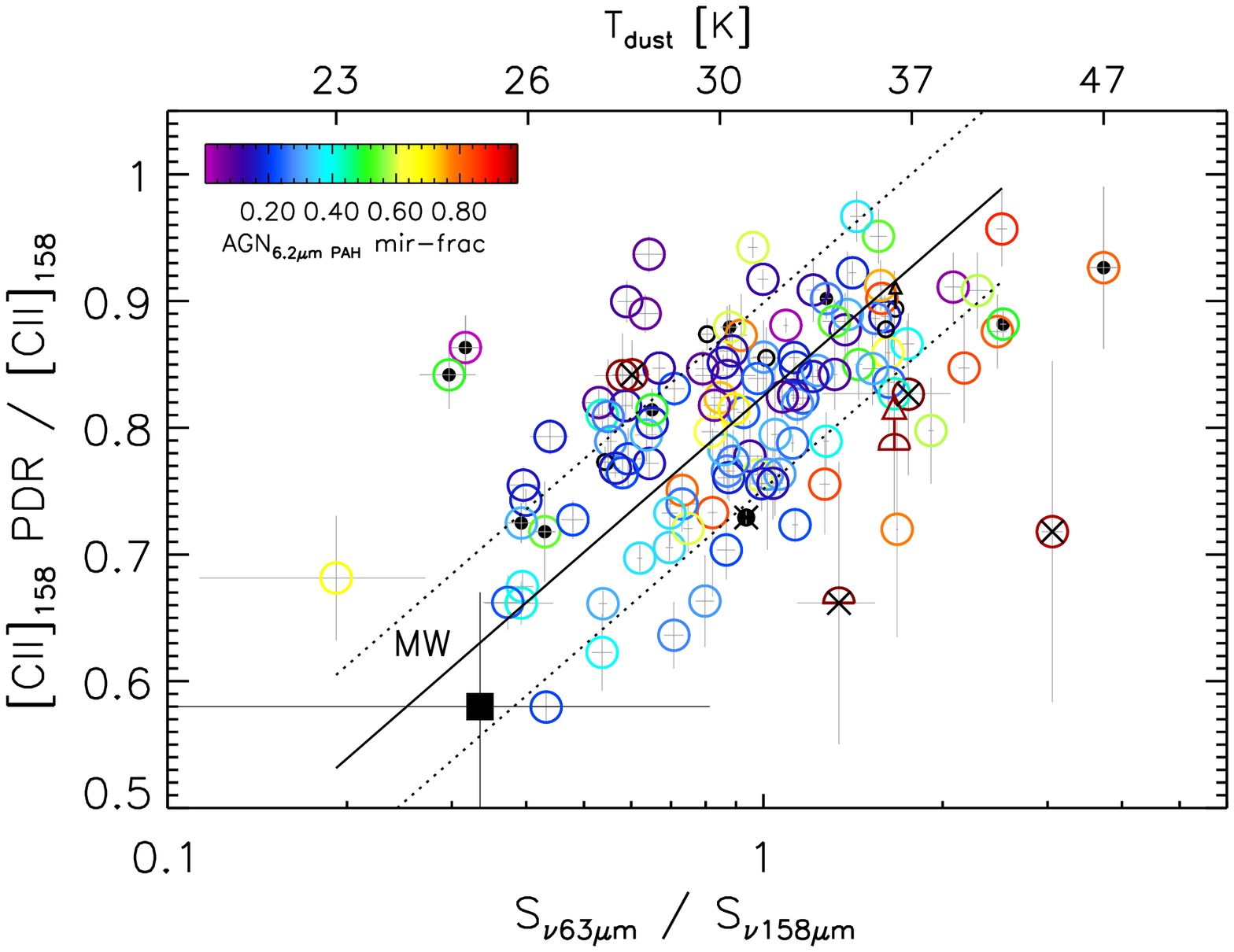}
\vspace{.25cm}
\plotone{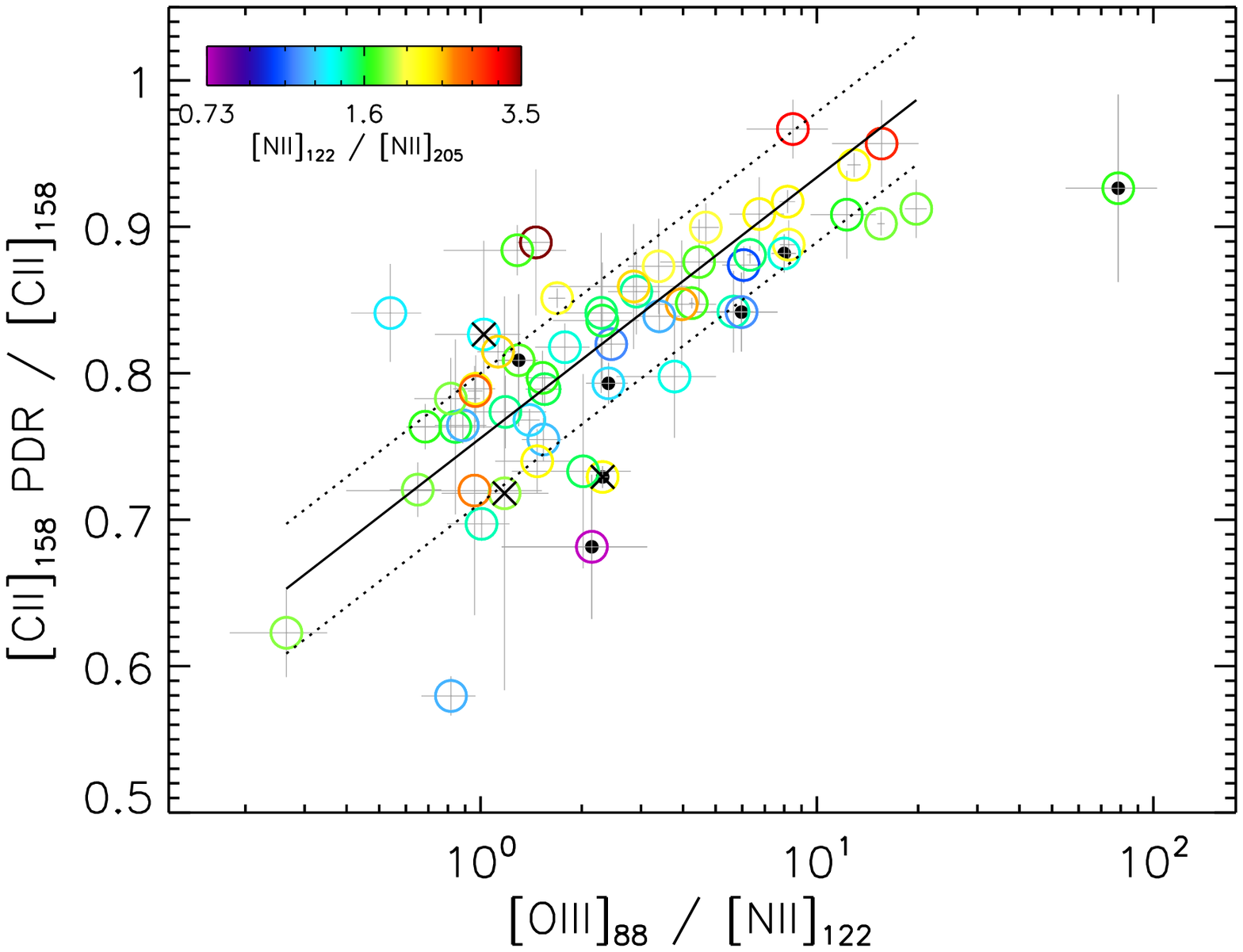}
\vspace{.25cm}
\caption{\footnotesize The fraction of PDR contribution to the total \CIIsub\, emission, $f(\CIIpdr)$, as a function of (1) the FIR \Sa/\Sb\, continuum ratio (top) and (2) the \OIIIbsub/\NIIasub\, emission line ratio (bottom), for the LIRGs in GOALS with available measurements in the relevant lines. Top panel: The position of the Milky Way (MW) is shown for reference as a black square. Bottom panel: LIRGs with measurements of the \NIIasub\, and \NIIbsub\, lines. Galaxies without available data in any of the N$^{+}$ lines are shown as small black open circles. The remaining symbols are as in Figure~\ref{f:linedef}. In the upper panel, galaxies are color-coded as a function of the AGN fractional contribution to their MIR emission, as estimated using the \PAHd\, EW diagnostic (see section~\ref{ss:bolagnfrac}). The solid black line represents a fit to the data (see equation~\ref{e:ciipdr}). In the bottom panel, galaxies are color-coded as a function of their \NIIasub/\NIIbsub\, ratio, showing that $f(\CIIpdr)$ does not depend on \ne\, \citep[c.f.,][]{Accurso2017}.
The correlation (black solid line) is described by the following fit: 
$f(\CIIpdr)$\, = --0.757\,($\pm$\,0.012) + 0.176\,($\pm$\,0.020)\,log\,(\OIIIbsub/\NIIasub), with a dispersion of 0.04.}\label{f:ciipdr}
\vspace{.25cm}
\end{figure}

We can use the \NIIbsub\, line to estimate the amount of \CIIsub\, emission produced in the ionized phase of the ISM, \CIIion\, \citep[e.g.,][]{Oberst2006, Beirao2012, Croxall2012, Kapala2015}. The layer within the Str\"omgren sphere where nitrogen is singly ionized ranges between 29.60 and 14.53\,eV, close to that where \CIIion\, also originates, 24.38 to 13.6\,eV. In addition, given the similar \necr\, and $E_{\rm ul}$/\kB\, 
of both transitions, the \CIIion/\NIIbsub\, ratio is roughly constant and depends weakly on \ne, the intensity of ionizing field, $q$, and the kinetic temperature of the gas, \Tkin.

The photo-ionization models presented in \cite{Oberst2006} predict a \CIIion/\NIIbsub\,$\simeq$\,3\,$\pm$\,0.5 for a range of $n_{\rm e}$ up to $\sim$\,10$^3$\,\cnmmm\, (see section~\ref{ss:ciiion}). For convenience,
we use this constant ratio to estimate \CIIion\, and subtract it from the total \CIIsub\, flux, which yields \CIIpdr\, (=\,\CIIsub--\CIIion). Figure~\ref{f:ciipdr} (upper panel) shows the fraction of \CIIno\, arising from PDRs, $f(\CIIpdr)$\,=\,\CIIpdr/\CIIsub, as a function of \Sa/\Sb\, and \Tdust\, for the entire GOALS sample. In \cite{DS2013} we showed that the presence of an AGN in LIRGs does not play a role in the decreasing of the \CIIsub/FIR\, ratio as a function of \LIR\, or \Tdust. To show that it does not have an affect on $f(\CIIpdr)$ either, we color-code Figure~\ref{f:ciipdr} a function of the fractional contribution of the AGN to the MIR emission of each galaxy, based on the 6.2\,$\mu$m PAH EW diagnostic (see section~\ref{ss:bolagnfrac}). Neither sources with large MIR AGN fractions nor those with the largest bolometric contributions ($<$\alphamirAGN$>$\,$\ge$\,0.5) are found systematically in a different region of the parameter space than star formation dominated galaxies -- or contribute to increase the scatter of the correlation.

A fit to the data provides:

\begin{equation}\label{e:ciipdr}
\begin{split}
f(\CIIpdr) = \CIIpdr/\CIIsub = 0.82\,(\pm\,0.01) \\
+\,0.41\,(\pm\,0.04)\,log\,(\Sa/\Sb)
\end{split}
\end{equation}

\noindent
with a dispersion of 0.07. For reference, the location of the Milky Way is also shown in the figure, assuming a luminosity-weighted average \Tdust\,=\,25\,($\pm$\,5)\,K and considering that PDR emission accounts for the remaining amount of \CIIno\, that is not associated with the ionized medium, which is estimated to be between $\sim$\,1/3 and 1/2 \citep{Goldsmith2015}. While the scatter is large, we can identify a broad trend for galaxies with warmer \Tdust\, to show larger $f(\CIIpdr)$ (black solid line), increasing from $\sim$\,60\,\%, close to the MW value, to nearly 95\,\%. That is, there is a larger PDR contribution to the total \CIIsub\, emission in warmer systems, indicating that, even though the \CIIsub\, line shows a large deficit with respect to the FIR emission in progressively more luminous, warmer galaxies, most of the extra \CIIno\, produced in them originates in dense PDRs.

This scenario is supported by the trend presented in Figure~\ref{f:ciipdr} (bottom panel), which shows also a positive correlation of $f(\CIIpdr)$ with the average hardness of the radiation field seen by the ionized gas, as traced by the \OIIIbsub/\NIIasub\, line ratio (see section~\ref{ss:oiiinii}). Therefore, $f(\CIIpdr)$ increases in environments associated with recent episodes of massive star formation. This is also consistent with \HII\, regions in the LIRG nuclei to be more enshrouded (optically and geometrically thicker; see discussion in section~\ref{ss:g0nHSigmaIR}) than those of evolved star-forming complexes where the stellar winds from massive stars and supernovae have already cleared out most of the dust from the star formation sites \citep{Blitz1980, Larson1981}. That is, galaxies with more evolved \HII\, regions (i.e., posterior to experience a starburst event; or simply having more modest star-formation, like the MW) have more of the \CIIsub\, emission arising from the ionized gas (likely from the low density ISM, see below). We also note that there is no clear trend between $f(\CIIpdr)$ and the electron density of the ionized gas, \ne, as traced by the \NIIasub/\NIIbsub\, ratio (see color code in the figure). We discuss further implications of the trend seen in Figure~\ref{f:ciipdr} (top) in section~\ref{ss:ciidef}.

\subsubsection{Ionized Gas}\label{ss:ciiion}

As shown in Figure~\ref{f:ciipdr}, even though the \CIIpdr\, dominates the total \CIIsub\, emission, the \CIIion\, contribution is not negligible. This ionized component can subsequently originate from both the diffuse (\CIIiondiff) and dense medium (\CIIionhii). \cite{Inami2013} used \textit{Spitzer}/IRS high resolution spectroscopy to probe modestly ionized gas within the \HII\, regions of the LIRG nuclei via the \SIIIa/\SIIIb\, line ratio\footnote{Both lines have \necr\,$\gtrsim$\,2000\,\cnmmm.}, and calculated the average \ne\, for most galaxies to be typically $\sim\,$\,100 to a few hundred \cnmmm, with a median of $\sim$\,300\,\cnmmm. Nearly 30\,\% of the galaxies for which both lines are detected, though, show ratios consistent with the ionized medium being in the low density limit (\ne\,$\lesssim$\,100\,\cnmmm). We note that the layer within the Str\"omgren sphere where Sulfur is doubly ionized is located between 34.79 and 23.34\,eV, while C$^{2+}$ transitions to C$^{+}$ at $h\nu$\,$<$\,24.38\,eV. Considering that the electron density should at least remain constant, if not increase towards the denser PDR region \citep[\ne\, of a few 10$^3$\,\cnmmm\, have been found in \HII\, regions using optical emission lines of singly ionized sulfur;][]{Osterbrock1989}, this means that the \ne\, derived using the \SIIIasub/\SIIIbsub\, ratio likely represents a lower limit to the density of the volume from where \CIIion\, and singly ionized nitrogen emission arises within \HII\, regions. And because the densities derived from the Sulfur lines are significantly larger than the \CIIion\, and \NIIbsub\, critical densities, this implies that the \CIIionhii\, emission has likely been thermalized, therefore suggesting that most \CIIion\, and \NIIbsub\, emission is produced in the diffuse ISM, with a modest contribution from the dense ionized phase -- unless the diffuse medium is extremely thin (low volume density) or its average \Tkin\, very low.

\begin{figure}[t!]
\vspace{.25cm}
\epsscale{1.15}
\plotone{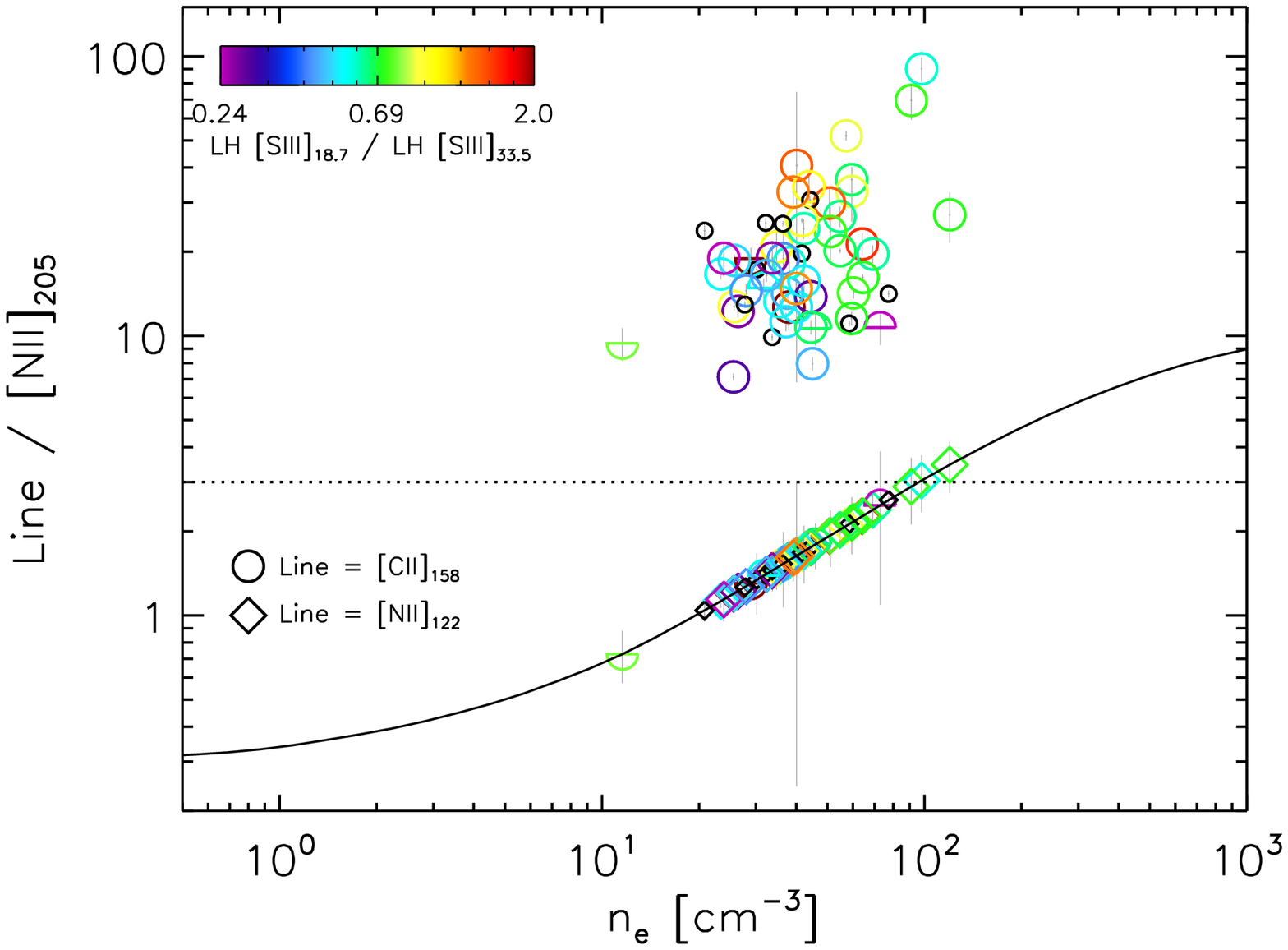}
\vspace{.25cm}
\caption{\footnotesize The \CIIsub/\NIIbsub\, (circles) and \NIIasub/\NIIbsub\, (diamonds) line ratios as a function of the electron density, \ne, for the LIRGs in GOALS with available data on both nitrogen lines, based on the emission line models from \cite{Oberst2006} (solid line). The dotted line represents the typical \CIIsub/\NIIbsub\, ratio for ionized gas that we use to calculate \CIIpdr\, (see text). Galaxies are color-coded as a function of the \SIIIasub/\SIIIbsub\, line ratio measured with long-high (LH) resolution module of \textit{Spitzer}/IRS \citep{Inami2013}, which traces denser (\ne\,$\simeq$\,10$^{3-4.5}$\cnmmm) and higher ionized gas (34.79\,$>$\,h$\nu$\,/\,eV\,$>$\,23.34) than that probed by the nitrogen lines. Galaxies with lower/upper limits in \SIIIasub/\SIIIbsub\, are marked as upward/downward pointing trinangles. Higher \SIIIasub/\SIIIbsub\, ratios imply higher densities. Most LIRGs show \NIIasub/\NIIbsub\, ratios compatible with \ne\,$\simeq$\,10$^{1-2}$\cnmmm, with a median value of 41\,\cnmmm\, and mean of 45\,\cnmmm. The remaining symbols are as in Figure~\ref{f:linedef}.}\label{f:niiratio}
\vspace{.5cm}
\end{figure}

We can also use the \NIIasub/\NIIbsub\, ratio in combination with the models from \cite{Oberst2006} to derive the average electron densities in our LIRG sample. As mentioned above, the region where emission from singly ionized nitrogen atoms originates in an \HII\, region largely overlaps with that of the \CIIionhii\, emission. Figure~\ref{f:niiratio} shows the distribution of \NIIasub/\NIIbsub\, ratios and \ne\, for those galaxies with available measurements of both lines. We find densities between $\sim$\,20 and 100\,\cnmmm, with a median value of 41\,\cnmmm\, and mean of 45\,\cnmmm. These values are very similar to what has been found by other studies of normal and starbursting galaxies. For instance, \cite{Zhao2016b} find \ne\,=\,22\,\cnmmm\, for a subsample of GOALS LIRGs using \textit{ISO} data, and  \cite{HC2016} find \ne\,=\,30\,\cnmmm\, for spatially resolved regions of 21 nearby, normal star-forming galaxies selected from the \textit{Herschel} KINGFISH and Beyond the Peak \textit{Herschel} surveys. In the Milky Way, the average value measured by \cite{Goldsmith2015} with \textit{Herschel}/PACS is 29\,\cnmmm.

As we noted above, the \ne\, values we find based on the nitrogen line ratio are smaller than those inferred from the sulfur lines, suggesting that both \NIIbsub\, and \CIIion\, transitions have been thermalized in the dense \HII\, regions and thus mostly originate from the diffuse ISM. In this case, if we assume a \ne\,$\simeq$\,500\,\cnmmm\, and a \ne\,$\lesssim$\,1\,\cnmmm\, for the dense and diffuse ionized medium respectively, the mean value of the \NIIasub/\NIIbsub\, ratio implies an average volume filling factor, \Vff, of $\lesssim$\,5\,\% for the \HII\, regions with respect to the overall volume of ionized emitting gas. Of course, this is only a very rough estimate, and \Vff\, will vary significantly depending on the compactness of the galaxy.

A number of reasons have been proposed to explain the observed deficit of FIR lines of purely ionized species -- mostly singly and double ionized nitrogen and oxygen (\NIIasub, \NIIbsub, \NIIIsub, \OIIIasub\, and \OIIIbsub). A boost in the average intensity of the radiation field has been suggested by several authors, who have been able to reproduce the low \NIIasub/FIR ratios observed in warm LIRGs using the \textit{Cloudy} spectral synthesis code \citep{Ferland1998} by increasing the ionization parameter from log(\textit{U}) $\sim$\,--2.5 to --1.5 \citep{Voit1992b, Abel2009, GC2011, Fischer2014}. However, the deficits of the higher ionization species, \NIIIsub\, \OIIIasub, and \OIIIbsub\, could not be explained by this effect. In view of the discussion above, our results indicate that a progressive thermalization of those lines with the lowest critical densities could also contribute to the observed deficits. Indeed, we find that the \NIIbsub\, line, which has the lowest \necr\, among our set of ionized lines, presents the largest decline in the \NIIbsub/FIR ratio, of almost two orders of magnitude; followed by the \NIIasub, and \OIIIbsub\, lines, which have increasingly higher \necr.

\subsubsection{Dust Temperature and PDR fraction as Drivers of the \CIIsub\, Deficit}\label{ss:ciidef}

The total luminosity of an optically-thin modified black-body (mBB) increases as a function of its temperature as $T^{4+\beta}$, where $\beta$ is the emissivity index of the emitting material. Thus, an increase of the average \Tdust\, in a LIRG from $\sim$\,25 to $\sim$\,50\,K can boost its \LFIR\, by at least a factor of $\sim$\,50 (for $\beta$\,=\,1.8; \citealt{PC2011})\footnote{Note that the peak emission of a modified black-body around these temperatures is always within the wavelength range used to calculate \LFIR, and therefore it can be approximately scaled using the Stephan-Boltzan law.}. As we argued in \cite{DS2013}, the \CIIsub\, deficit is likely caused by an increase of the amount of dust heated to high temperatures in the ionized region --close to the PDR front-- of dust-bounded star-forming regions \citep[see e.g.,][]{Abel2009, Paladini2012}, where the density of material is expected to increase significantly \citep{Draine2011}. Furthermore, the deficit seems to be restricted to the nuclear region of galaxies, where the starburst is ongoing \citep{DS2014, Smith2016}. Thus, most of the FIR dust continuum emitted at the BB peak would be associated with active, still dust-bounded star-forming regions. On the other hand, and in light of our findings in the previous sections, we expect the \NIIbsub\, and \CIIion\, emission to originate mostly from low density, ``fossil'' \HII\, regions and diffuse ionized gas not associated with OB stars. Then, if the \CIIion\, and FIR emissions truly arise from different ISM phases, the \CIIion/FIR\, ratio should be mostly controlled by the \LFIR\, boosting, due to the increase of \Tdust\, in progressively more active starbursts.

\begin{figure*}[t!]
\vspace{.25cm}
\epsscale{.55}
\plotone{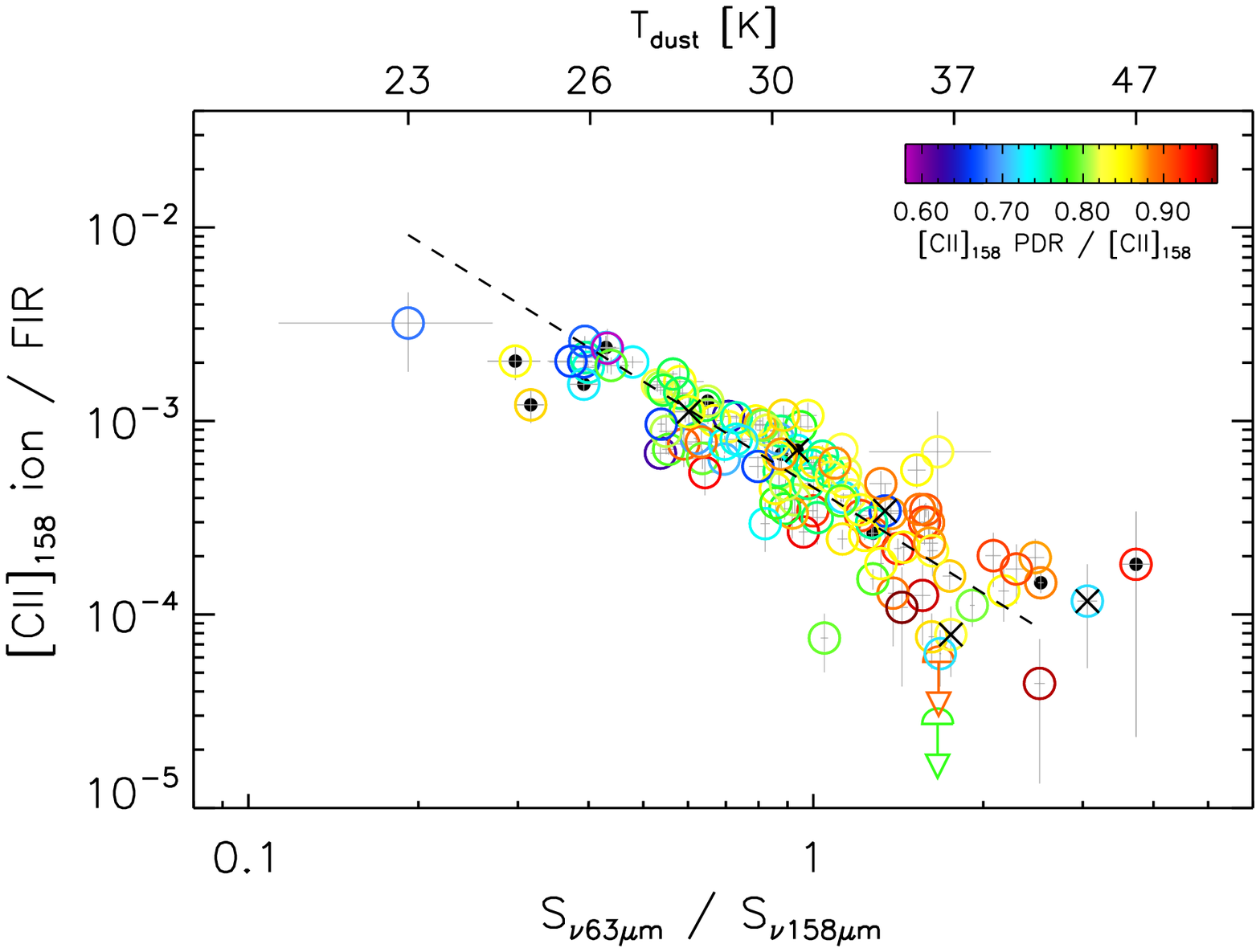}
\hspace{.25cm}
\plotone{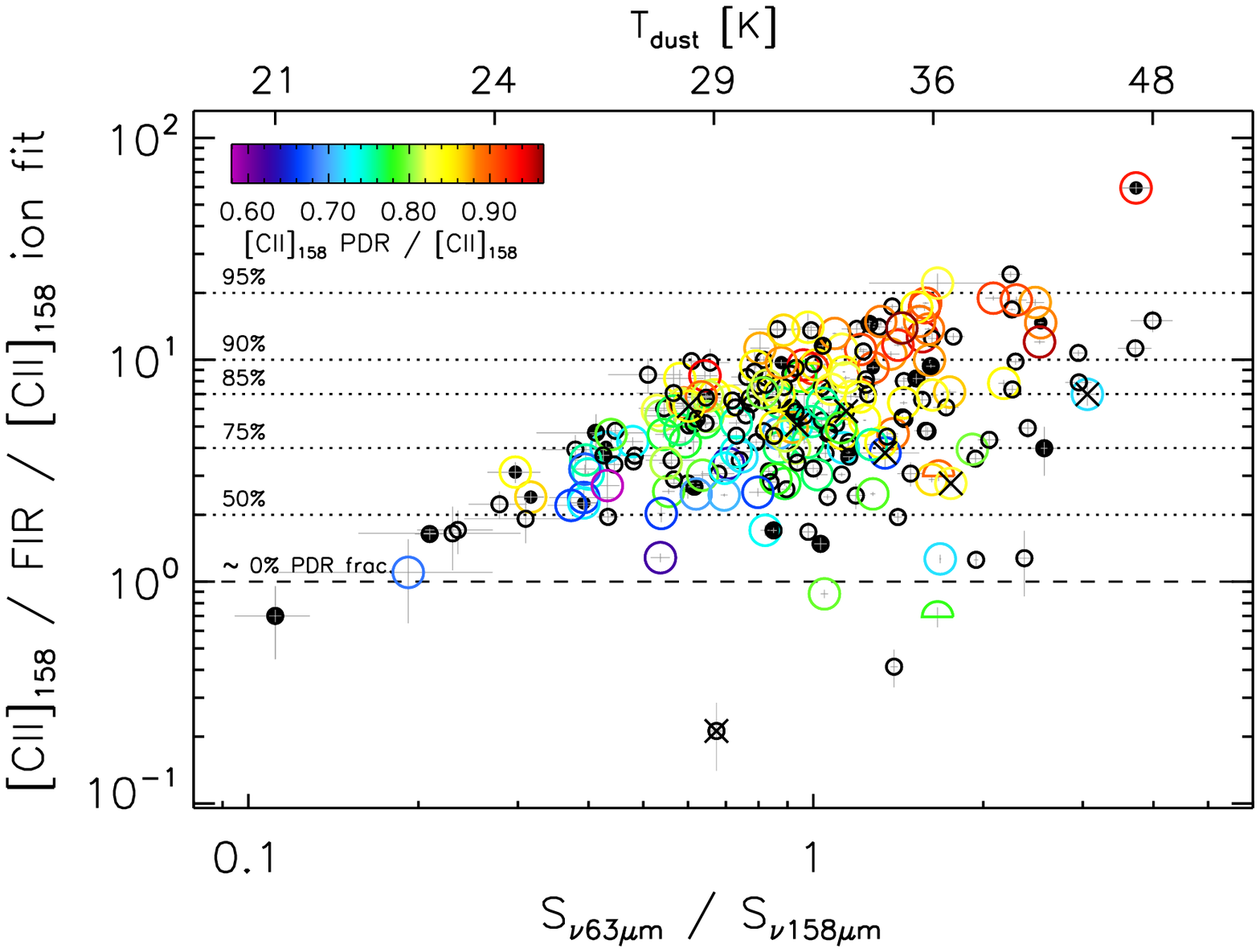}
\epsscale{.80}
\plotone{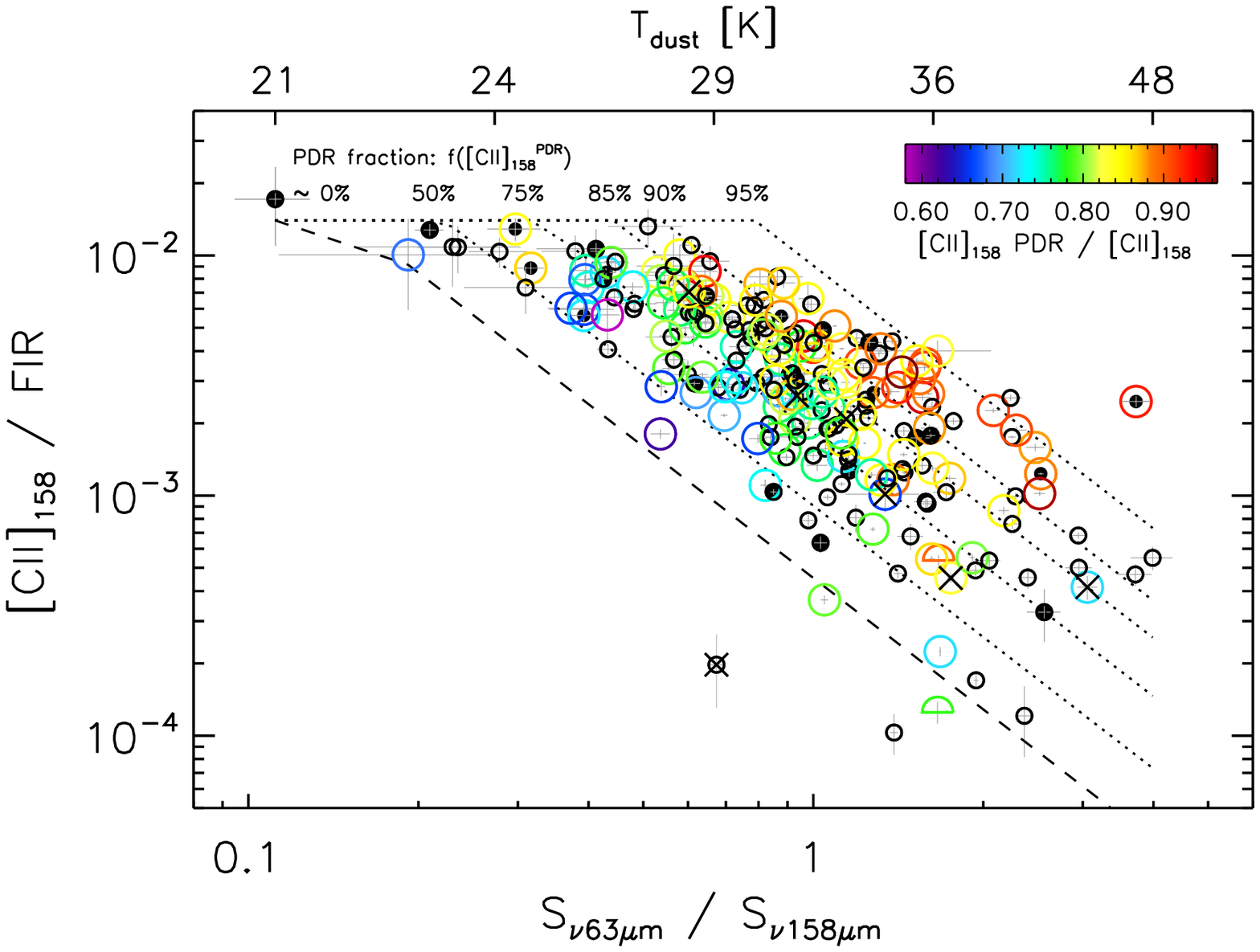}
\vspace{.25cm}
\caption{\footnotesize The quantities shown in the y-axes of all panels, \CIIion/FIR\, (top-left), \CIIsub/FIR/\CIIion-fit (top-right) and \CIIsub/FIR\, (bottom), are presented as a function of the FIR \Sa/\Sb\, continuum ratio for the entire GOALS sample. Top left: \CIIion\, deficit. The dashed black line is a log-log fit to the data where: log(\CIIion/FIR) = --3.343\,($\pm$\,0.021) --1.82\,($\pm$\,0.11)\,$log$\,(\Sa/\Sb), and a dispersion of 0.20\,dex. Top right: \CIIsub\, deficit divided by the fit to the \CIIion/FIR\, ratio, with a diagnostic grid indicating different contributions of the PDR component to the total \CIIsub\, emission, $f(\CIIpdr)$. The dashed line indicates a $\sim$\,0\,\% PDR fraction, set by the fit to the data in the top left panel. Additional increasing contributions of \CIIpdr\,=\,[1, 3, 6, 9, 19]\,$\times$\,\CIIion\, are shown with dotted lines, from $\sim$\,50\,\% to 95\%. Bottom: Same data and grid as in the top-right panel but presented in the more typical form of the \CIIsub\, deficit plot. All panels are color-coded as a function of $f(\CIIpdr)$. Galaxies with lower/upper limits in $f(\CIIpdr)$ are marked as upward/downward pointing triangles. The remaining symbols are as in Figure~\ref{f:linedef}. Small open black cirlces are sources with unavailable $f(\CIIpdr)$ due to lack of measurement of the \NIIbsub\, line.}\label{f:ciipdrgrid}
\vspace{.5cm}
\end{figure*}

The top left panel of Figure~\ref{f:ciipdrgrid} shows \CIIion/FIR\, as a function of \Tdust\, (derived from the \Sa/\Sb\, ratio) as well as a fit in log-log space to the data (dashed line; see the figure caption for the precise equation). Under the assumptions discussed above, this trend indeed suggests that the rise in the average radiation field intensity produced by young, massive stars born during the ongoing starburst is mostly transferred into an increase of \Tdust, which enhances the \CIIsub\, deficit through the boosting of the FIR continuum. If we assume that the FIR emission scales with $\Tdust^{4+\beta}$ (where $\beta$\,=\,1.8), the suppression seen in the \CIIion/FIR\, ratio of the warmest LIRGs can be explained in terms of a factor of almost $\sim$\,30 excess in the IR continuum emission rather than a deficit in the observed line flux. This scaling is also able to recover reasonably well the shape of the trend seen in Figure~\ref{f:ciipdrgrid}. The fit to the data is even better if we consider the BB emission to be optically thick below 100\,$\mu$m -- an assumption frequently made when fitting the SED of heavily dust-obscured sources \citep[e.g.,][]{Wilson2014}.

However, the increase in the dust cooling efficiency\footnote{The term ``dust cooling efficiency'' is used here to refer to the emitted luminosity per dust particle (or per dust mass, assuming that there are no significant variations in the dust grain composition and/or size distribution among our galaxies) due to the increase in temperature.} alone cannot be the only physical parameter involved in the variation of the line-to-FIR ratios because: (1) that would imply that none of the energy produced by massive stars at the core of star-forming regions would be transformed into gas heating, and (2) there is still a significant dispersion of a factor of $\sim$\,3 or larger in the \CIIsub/FIR\, ratio at a given \Tdust\, (see Figure~\ref{f:linedef}). The top right panel of Figure~\ref{f:ciipdrgrid} shows the \CIIsub/FIR\, ratio after correcting the FIR emission for the increase due to hotter \Tdust\, using the fit to the \CIIion/FIR\, ratio described above (top left panel). After this ``IR excess'' has been taken into account, we can see a trend for \CIIsub\, emission to rise as a function of \Tdust, suggesting that part of the heating from the stellar radiation field is indeed transferred to the gas, increasing the cooling carried out through the \CIIsub\, (and other) line(s). Note that the dispersion also increases with \Tdust\, as a result of the PDR contribution to the total \CIIsub\, budget (see color coding in the figure).

Figure~\ref{f:ciipdrgrid} (bottom) conveys the same information but in the form of the classical \CIIsub/FIR\, ratio, which is displayed as a function of \Sa/\Sb\, and \Tdust. Note that the \Sa/\Sb\, ratio shown here is approximately equal to the $S_{\rm 70}$/$S_{\rm 160}$ ratio that can be measured with \textit{Herschel}/PACS or \textit{SOFIA}/FIFI-LS in spatially resolved nearby galaxies, or at high redshifts with a combination of ALMA bands. Regions of similar $f(\CIIpdr)$ contribution are marked with the same line styles as in the top right panel. This figure shows again that the dispersion observed in the \CIIion/FIR\, ratio as a function of \Tdust\, is caused by the different contributions of the PDR component to the total \CIIsub\, emission. This is consistent with the scenario put forward in section~\ref{ss:ciipdr} regarding a larger fraction of a given dust mass (\Mdust) being heated up to higher temperatures by the absorption of ionizing photons in the most compact, luminous starbursts. Because the cooling efficiency of dust increases as \Tdust$^{4+\beta}$, a significant amount of dust present within the \HII\, regions, close to the PDRs \citep[see][]{Draine2011}, could be able to reprocess most of the energy emitted by the stars before it reaches the surrounding neutral/molecular medium. Indeed, the temperatures measured from our data are typical of PDR surfaces (i.e., at \AV\,$\sim$\,0) exposed to moderate to high radiation fields (\G\,$\gtrsim$\,10$^2$\Go; \citealt{Tielens2005}; see section~\ref{ss:pdrmodel}). In addition, a fraction of the energy is transferred via the photo-electric effect into gas heating, as shown in the top right panel, which cools down, increasing the \CIIpdr\, emission.

Recently, using \textit{Herschel}/HIFI, \cite{Goicoechea2015} have studied the variation of the \CIIsub/FIR\, ratio on $\sim$\,1\,pc scales across the Orion molecular cloud 1 (OMC1) and the region surrounding the Trapezium cluster, which is being ionized by the intense UV field of massive O stars. They find that there is a broad correlation between \CIIsub/FIR\, and the column density of dust through the molecular cloud --tighter than with \LFIR/\Mgas\, \citep[see][]{GC2011}--, and conclude that the \CIIsub\,-emitting column relative to the total dust column along each line of sight is responsible for the variations of \CIIsub/FIR\, observed through the cloud. This trend is similar to that found for the nuclear emission of LIRGs, which show a correlation between \CIIsub/FIR\, and (1) the strength of the 9.7$\mu$m silicate absorption feature, \SSi, probing the opacity, and thus the column density and total mass of the MIR-absorbing dust \citep{DS2013}; and (2) the opacity of the free-free emission in \HII\, regions as measured through the spectral index between 1.5 and 6\,GHz continuum emission \citep{BM2017}. However, although these results are in agreement at face value, the argument made for the case of the OMC1 by \cite{Goicoechea2015} --in which the \CIIsub\, deficit would be entirely caused by an increase in the dust column density-- is probably not the complete answer for the case of the kpc-scale, integrated emission of LIRGs. In terms of total \Mdust, the decrease of more than an order of magnitude in \CIIsub/FIR\, would require a similar increase in the amount of dust mass in the galaxy, something that is not seen even in the most extreme ULIRGs, which have similar \Mdust\, as dusty, normal star-forming (MS) galaxies \citep{daCunha2010}. Also, we note that the column density fit used in \cite{Goicoechea2015} assume that there are no dust temperature variations. On the other hand, \cite{Lombardi2014} presented column density and effective dust temperature maps based on NIR, \textit{Herschel} and \textit{Planck} data for the entire Orion complex at a lower angular resolution than \cite{Goicoechea2015}. Considering both works, temperature variations are in the range $\sim$\,15--35\,K, where regions with relatively high temperatures emit significantly higher fluxes, even if the optical depth is substantially lower.

In summary, while it is possible that the column density in LIRGs may indeed increase due to a compactification of the star-forming regions (see end of next section), this interpretation is still compatible with the picture put forward above, where a given amount of dust mass would be accumulated closer to the heating source, thus also increasing \Tdust\, and the corresponding radiative cooling. Disentangling between the two effects is, however, extremely challenging in galaxies where it is not possible to spatially resolve the different temperature components of the dust emission.

\subsection{FIR Oxygen Line Emission}\label{ss:oxygen}

\subsubsection{\OIasub/\CIIpdr\, as a Probe of the Gas Kinetic Temperature}\label{ss:oicii}

The \OIasub\, line is an important coolant of the warm, neutral ISM. A substantial fraction of the \OIasub\, emission is believed to originate in PDRs, mostly at optical depths similar to those of the locations responsible for the C$^+$ emission, \AV\,$\lesssim$\,4 -- although neutral oxygen may exist deeper in the molecular cloud before it combines into CO at \AV\,$\sim$\,10 \citep{Hollenbach1997,Kaufman1999}. While \OIasub\, can be as weak as $\sim$\,20\,\% of the \CIIsub\, luminosity emitted in a normal star-forming galaxy, its relative contribution to the ISM cooling can become increasingly important in the dense and warm environments of extreme starbursting galaxies ($n^{\rm cr,[OI]_{63}}_{\rm H}$\,$\simeq$\,10$^{6}$\,\cnmmm), even surprassing the \CIIsub\, line in the warmest LIRGs \citep{Rosenberg2015}. Using \textit{ISO}/LWS data, \cite{Malhotra2001} found a correlation between the \OIasub/\CIIsub\, ratio and the FIR color of star-forming galaxies. Our \textit{Herschel}/PACS data and \textit{Spitzer}/IRS observations allow us to explore this correlation in depth for the entire GOALS sample; a 4-fold increase in the number of galaxies available relative to previous studies.

\begin{figure}[t!]
\vspace{.5cm}
\epsscale{1.15}
\plotone{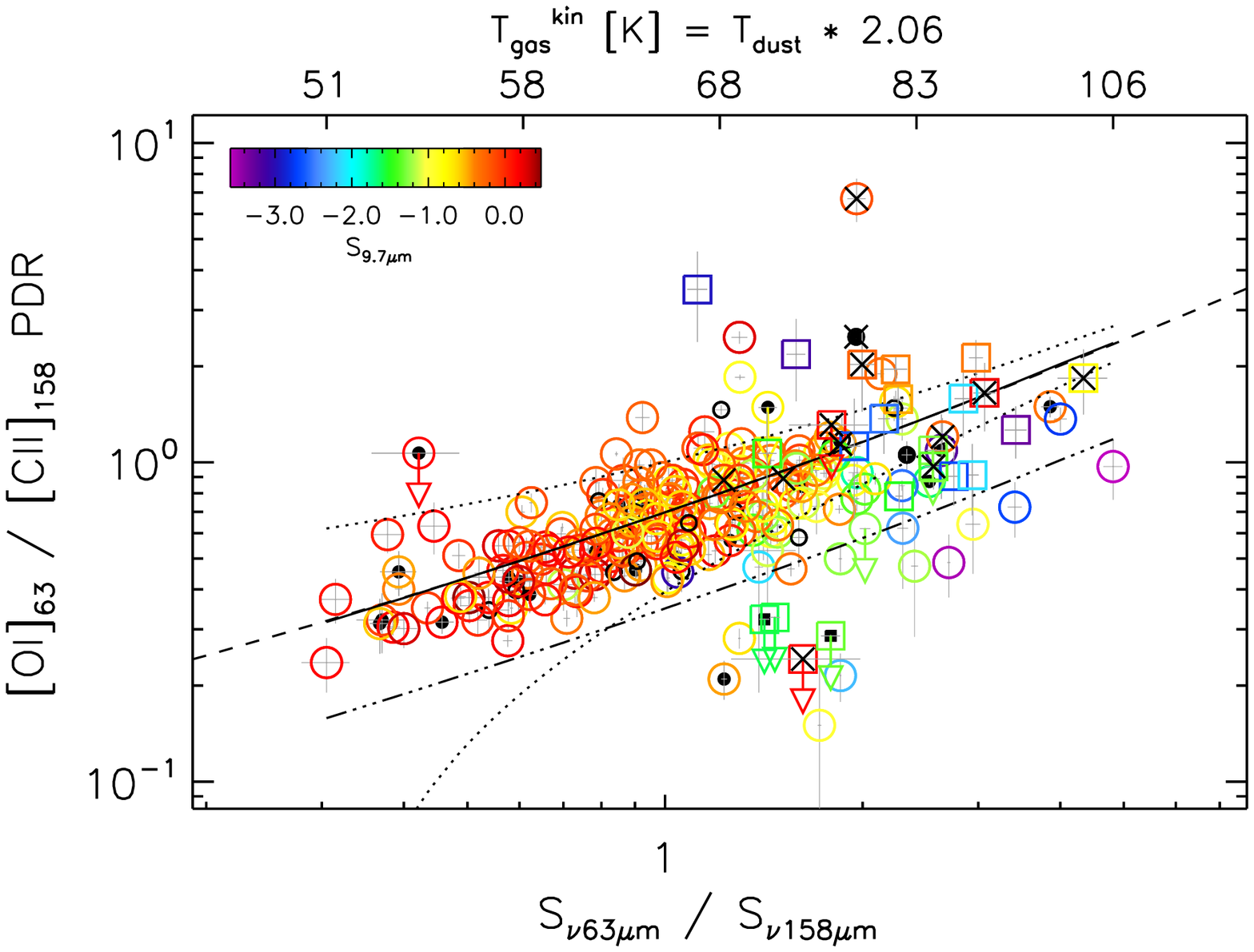}
\vspace{.25cm}
\caption{\footnotesize The \OIasub/\CIIpdr\, ratio as a function of the FIR \Sa/\Sb\, continuum ratio for the entire GOALS sample (open circles) and the ULIRGs from \cite{Farrah2013} (open squares), using the central spaxel aperture to focus on the galaxy nuclei and enhance the \Tdust\, dynamic range. Galaxies are color-coded based on the strenght of the 9.7\,$\mu$m silicate absorption feature, \SSi. The remaining symbols are as in Figure~\ref{f:linedef}. The black line is a fit to the data using the radiative transfer code RADEX. The code calculates the population levels of the fine-strucutre transitions of the O\,{\sc i} and C\,{\sc ii} species as a function of \Tkin\, (see upper y-axis), which we have parametrized to depend linearly on \Tdust\, (see text and equations~\ref{e:oiciipdrtdust} and \ref{e:tdustoiciipdr}). The dotted lines are the 1\,$\sigma$\, dispersion around the trend. The remarkably tight correlation between \OIasub/\CIIpdr\, and \Tdust\, (or \Tkin) only breaks down at low \SSi, albeit only in the most extremely luminous, warm LIRGs and ULIRGs (the dotted-dashed line marks represents 0.5 times the fitted trend, a threshold below which the \OIasub\, emission may be becoming optically thick or be self-absorbed). As can be seen, the majority of galaxies show \Tkin\, lower than the temperatures of the energy levels of the emission lines considered ($E_{\rm ul,[CII]_{158}}$/\kB\,=\,92\,K, $E_{\rm ul,[OI]_{63}}$/\kB\,=\,228\,K).}\label{f:oicii}
\vspace{.5cm}
\end{figure}

Because of the neutral gas phase origin of the oxygen fine-structure emission, we use the \CIIpdr\, component alone to compare it with the \OIasub\, line. For those galaxies without \NIIbsub\, observations we use equation~\ref{e:ciipdr} to estimate their $f(\CIIpdr)$ (capped at 90\,\%) from the \Sa/\Sb\, ratio. Figure~\ref{f:oicii} shows that \OIasub/\CIIpdr\, is tightly correlated with the FIR \Sa/\Sb\, continuum ratio, with a scatter of only 0.13\,dex. We have used the statistical equilibrium radiative transfer code RADEX \citep{vdTak2007}, to model the variation of the \OIasub/\CIIpdr\, ratio as a function of the kinetic temperature of the gas (\Tkin), assuming \Tkin\,=\,$\eta$\,\Tdust. We have further assumed that the excitation is dominated by collisions with neutral hydrogen atoms, a gas density of \nhvol\,=\,10$^{3}$\,\cnmmm\, (see section~\ref{ss:pdrmodel}), and a column density of \nhcol\,=\,10$^{16}$\,\cnmm. In this regime, and for lower volume densities and columns up to \nhcol\,$\simeq$\,10$^{18}$\,\cnmm, both lines are optically thin and also \nhvol\,$<$\,\nhcr. Despite the simplicity of the approach, the RADEX model reproduces the observed trend between \OIasub/\CIIpdr\, and \Tdust\, remarkably well (black solid line in Figure~\ref{f:oicii}; the uncertainty around the trend is 1$\sigma$\,=\,0.30, dotted lines) with a best fit for $\eta$\,=\,2.06\,$\pm$\,0.03. The value of $\eta$ decreases to 2.02 and 1.60 for \nhcol\,=\,10$^{17}$ and 10$^{18}$\,\cnmm\, respectively, since at higher columns the gas does not need to be as warm to produce the same \OIasub/\CIIpdr\, ratios. With any of these scalings we infer gas kinematic temperatures similar or lower than the excitation energy of the \CIIsub\, fine-structure transition, $E_{\rm ul,[CII]_{158}}$/\kB\,=\,92\,K. This is in contrast to suggestions that \CIIsub\, deficit is mainly due to thermal saturation of the \CIIsub\, emission \citep{Munoz2016}.

The trend can be fitted as a function of \Sa/\Sb\, using a second order polynomial function:

\begin{equation}\label{e:oiciipdrtdust}
\begin{split}
log(\OIasub/\CIIpdr) = & -0.16 + 0.71\,log\,\Sa/\Sb \\
& + 0.09\,(log\,\Sa/\Sb)^2
\end{split}
\end{equation}

Inversely, one can obtain \Sa/\Sb\, as a function of the line ratio:

\begin{equation}\label{e:tdustoiciipdr}
\begin{split}
log(\Sa/\Sb) = 0.22 + 1.34\,log\,x\,-\,0.25\,(log\,x)^2
\end{split}
\end{equation}

\noindent
with $x$\,=\,\OIasub/\CIIpdr, and using equation~\ref{e:tdustfircolor} obtain \Tdust\, from \Sa/\Sb, which is interchangeable with \Tkin\, assuming the best scaling factor obtained from the model fit to our data, \Tkin\,=\,2.06\,\Tdust. We caution that these equations assume that the \OIasub\, line is optically thin, and are only valid approximately over the range spanned by our galaxy sample, i.e., 0.25\,$\lesssim$\,\OIasub/\CIIpdr\,$\lesssim$\,3.5 and 28\,$\lesssim$\,\Tdust\,$\lesssim$\,76\,K, or equivalently over the ranges shown in Figure~\ref{f:oicii}.

Some of the warmest, most luminous galaxies seem to deviate from the correlation, showing systematically lower \OIasub/\CIIpdr\, ratios than those predicted by the trend. The most likely explanation is that in these LIRGs the \OIasub\, emission becomes optically thick due to an increase of the \textit{in situ} gas column density and/or foreground absorption by cold gas across the line of sight \citep[e.g.,][]{Poglitsch1996}. This is in agreement with the fact that most galaxies falling below the fitted correlation show large 9.7\,$\mu$m silicate strengths as well, \SSi\,$\lesssim$\,--1.5 (see color-coding in Figure~\ref{f:oicii}), indicating a similar increase of the dust opacity towards the hot, MIR emitting background source. Optically thick \OIasub\, has been already reported by studies with available information of the \OIbsub\, emission line \citep{Malhotra2001,Farrah2013,Rosenberg2015}. Since \OIbsub\, is optically thin, a value of the \OIbsub/\OIasub\, ratio in excess of $\sim$\,0.1 implies some degree of optical thickness in \OIasub.

There are only 12 galaxies in GOALS ($\sim$\,5\% of the sample) displaying \OIasub/\CIIpdr\, ratios lower than half the value of the fit shown in Figure~\ref{f:oicii} (this threshold is denoted by the dotted-dashed line), and five of them show clear absorption in the \OIasub\, line profile (IRASF08572+3915, NGC~4418, Arp~220, IRASF17207--0014 and IRAS17578--0400), in some cases showing P-Cygni or inverse P-Cygni profiles indicating outflowing or inflowing gas \citep{GA2012, Falstad2015, Falstad2017}. Figure~\ref{f:opthickprof} in the Appendix shows the \textit{Herschel}/PACS spectra of these galaxies, where signs of \OIasub\, absorption are visually noticeable. Additionally, a small fraction of galaxies in GOALS with spectra covering the wavelength range of the OH doublet at 119\,$\mu$m, \OHbsub\, (IRAS07251--0248, IRASF15250+3609 and IRASF17207--0014) show signs of strong molecular absorption, which is characteristic of FIR-thick, warm, compact objects \citep{GA2015}. Also, in the ULIRGs sample from \cite{Farrah2013}, 4 out of the 5 galaxies that are below the \OIasub/\CIIpdr\, threshold also present signs of \OHbsub\, absorption. In these highly obscured sources, it is possible that an AGN contributes to the dust heating \citep{Spoon2013}, but it is impossible to determine the exact nature of the buried source due to its extreme optical thickness.

\subsubsection{\OIIIbsub/\OIasub\, and the Ionized/Neutral Gas Filling Factors}\label{ss:oiiioi}

In contrast to the \OIasub\, line, \OIIIbsub\, emission originates from ionized gas with lower densities (\necr\,$\sim$\,500\,\cnmmm). Thus, the two lines clearly probe distinct phases of the ISM, providing a first order approximation to the ratio of volume filling factors, \Vff, between the warm ionized and neutral gas, assuming the lines are optically thin and that the ratio of temperatures between both phases is relatively constant across galaxies. As pointed out by \cite{Cormier2015}, the \OIIIbsub/\OIasub\, ratio is between 2--10 times higher in low metallicity galaxies than in normal star-forming galaxies, which suggest a correspondingly higher \Vff\, of ionized gas with respect to PDR emission in the former.

\begin{figure}[t!]
\vspace{.5cm}
\epsscale{1.15}
\plotone{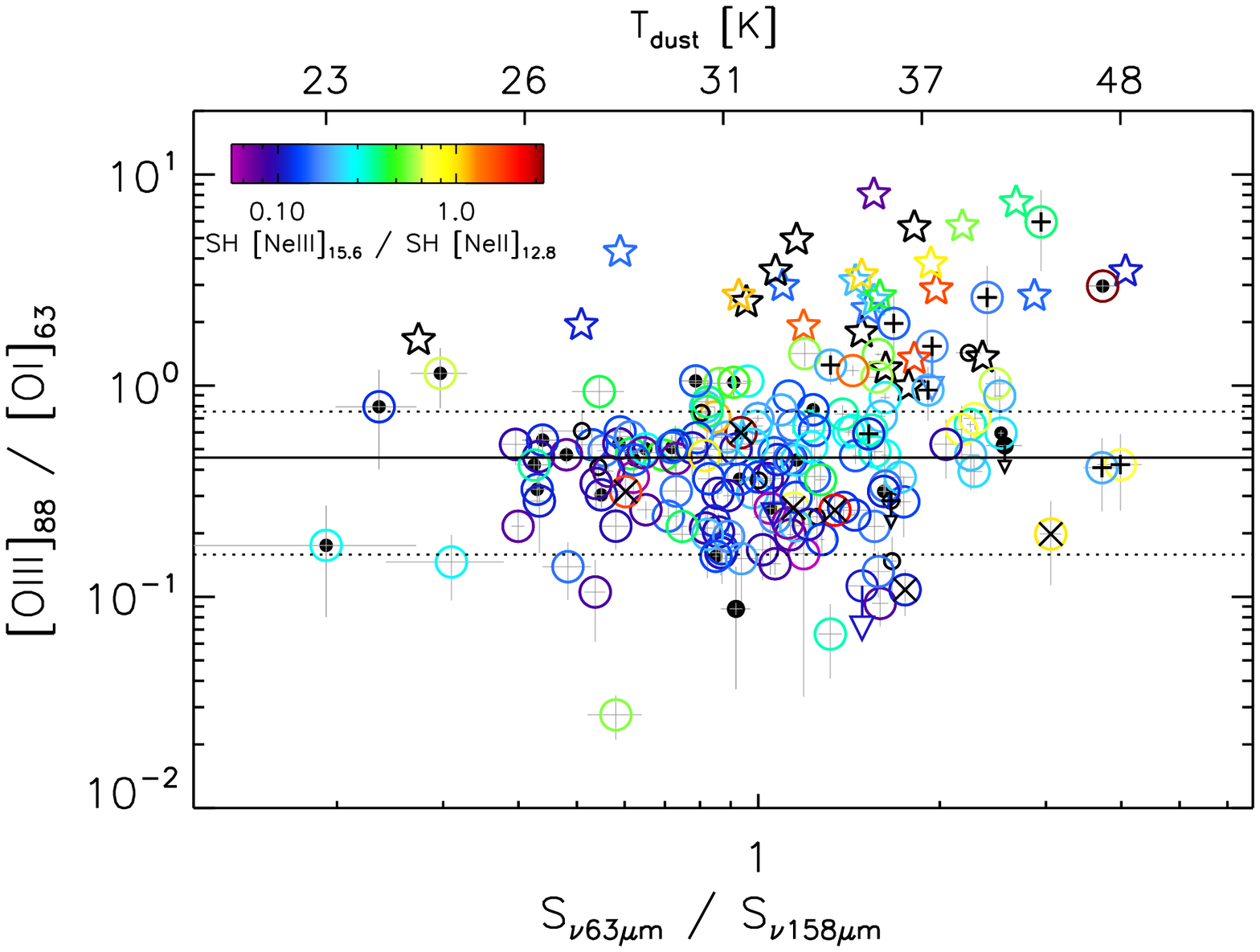}
\vspace{.25cm}
\caption{\footnotesize The \OIIIbsub/\OIasub\, ratio as a function of the FIR \Sa/\Sb\, ratio for the GOALS sample with available measurements of the lines (open circles), and the DGS from \cite{Cormier2015} (open stars). All galaxies are color-coded as a function of the \textit{Spitzer}/IRS short-high (SH) \NeIIIsub/\NeIIsub\, ratio, or as black open symbols when there is no information in any of the neon lines. The solid and dotted lines are the median value for the GOALS sample and the median absolute deviation, respectively. The remaining symbols are as in Figure~\ref{f:linedef}. There is no correlation between \OIIIbsub/\OIasub\, and \Tdust\, for most of the GOALS galaxies. The DGS show higher \OIIIbsub/\OIasub\, and \NeIIIsub/\NeIIsub\, ratios, indicative of larger \Vff\, of ionized gas (with respect to PDRs), and younger starburst and/or more intense radiation fields. A few LIRGs show also high \OIIIbsub/\OIasub\, ratios. However, in most sources this is due to a significant optical thickiness of the \OIasub\, line (see galaxies marked with plus signs in this figure, based on the threshold --dotted-dashed line-- shown in Figure~\ref{f:oicii}).}\label{f:oiiioi}
\vspace{.25cm}
\end{figure}

In Figure~\ref{f:oiiioi} we show \OIIIbsub/\OIasub\, as a function of the FIR \Sa/\Sb\, ratio for the entire GOALS sample as open circles. We also include the low metallicity galaxies from the Dwarf Galaxy Sample (DGS) presented in \cite{Cormier2015}. To obtain the \Sa/\Sb\, ratio for the low metallicity sample, which only have published data of the $S_{\rm 70}$/$S_{\rm 100}$ ratio, we use the same method described in section~\ref{ss:linedef} to obtain \Tdust\, and then interpolate the fitted mBB emission at 63 and 158\,$\mu$m to obtain the \Sa/\Sb\, ratio for each galaxy. Figure~\ref{f:oiiioi} suggests that there is no clear correlation between \OIIIbsub/\OIasub\, and the FIR color in LIRGs. The median \OIIIbsub/\OIasub\, ratio of our sample is 0.46\,$\pm$\,0.30. Low metallicity galaxies populate a region of the parameter space with higher \Tdust\, ($\gtrsim$\,30\,K) and larger \Vff\, of ionized gas ($\gtrsim$\,5\,$\times$) than dusty, metal-rich systems. Figure~\ref{f:oiiioi} also indicates that galaxies with \OIIIbsub/\OIasub\, ratios in excess of $\simeq$\,1 have on average larger \NeIIIsub/\NeIIsub\, emission line ratios as well ($\gtrsim$\,0.5), regardless of the nature of the sample. This is in agreement with a picture in which a larger volume of the ISM is ionized, which is likely caused by a combination of less effective dust cooling (due to a lack of metals) and by the presence of harder ionization fields, rising the gas temperature and the ionization state. In addition, in dusty galaxies the PDR covering factor is likely higher than in normal or low metallicity sources, and thus more energy is captured by neutral and molecular material and cooled via the \OIasub\, line emission.

We note that there are a few LIRGs lying in the same region of the diagnostic diagram as low metallicity galaxies, with \OIIIbsub/\OIasub\,$\geq$\,1. However, most of them have \SSi\,$\lesssim$\,--1.5 and are outliers in the \OIasub/\CIIpdr\, vs. \Sa/\Sb\, correlation (sources below the threshold marked by the dotted-dashed line in Figure~\ref{f:oicii}), indicating a significant optical thickness of the \OIasub\, line, which would artificially boost the oxygen line ratio to higher values (see pluses in Figure~\ref{f:oiiioi}).

\subsubsection{\OIIIbsub/\NIIasub: A Tracer of the Ionization Field Hardness?}\label{ss:oiiinii}

The similar critical densities of the \OIIIbsub\, and \NIIasub\, emission lines (\necr\,$\sim$\,500\,\cnmmm\, and $\sim$\,300\,\cnmmm, respectively) but different ionization potentials ($\sim$\,35.1\,eV and 14.5\,eV, respectively) make their ratio a good tracer of the average hardness of the radiation field in a galaxy, or equivalently, the mean, luminosity-weighted effective temperature of the massive stars born in the starburst, \Teff. Indeed, within the context of star formation, \OIIIbsub\, emission is mostly powered by O stars (\Teff\,$\gtrsim$\,30000\,K \citep[e.g.,][]{Ferkinhoff2011}. Since O stars are short lived, with lifespans of around a few Myr ($\lesssim$\,10\,Myr for an O6 type), \Teff\, rapidly drops with age as the stellar population evolves after a burst of star formation. Alternatively, if a powerful AGN exists, the line flux of highly ionized species can be significantly boosted by its narrow line region. However, we have explored possible correlations between the \OIIIbsub/\NIIasub\, ratio and the AGN fractional contribution to the MIR and/or bolometric luminosity of LIRGs and found no significant trend with any of the diagnostics described in section~\ref{ss:bolagnfrac}, suggesting that these lines originate predominantly from star formation (except maybe in a few particular cases, where the AGN contribution could be significant).

\begin{figure}[t!]
\vspace{.5cm}
\epsscale{1.15}
\plotone{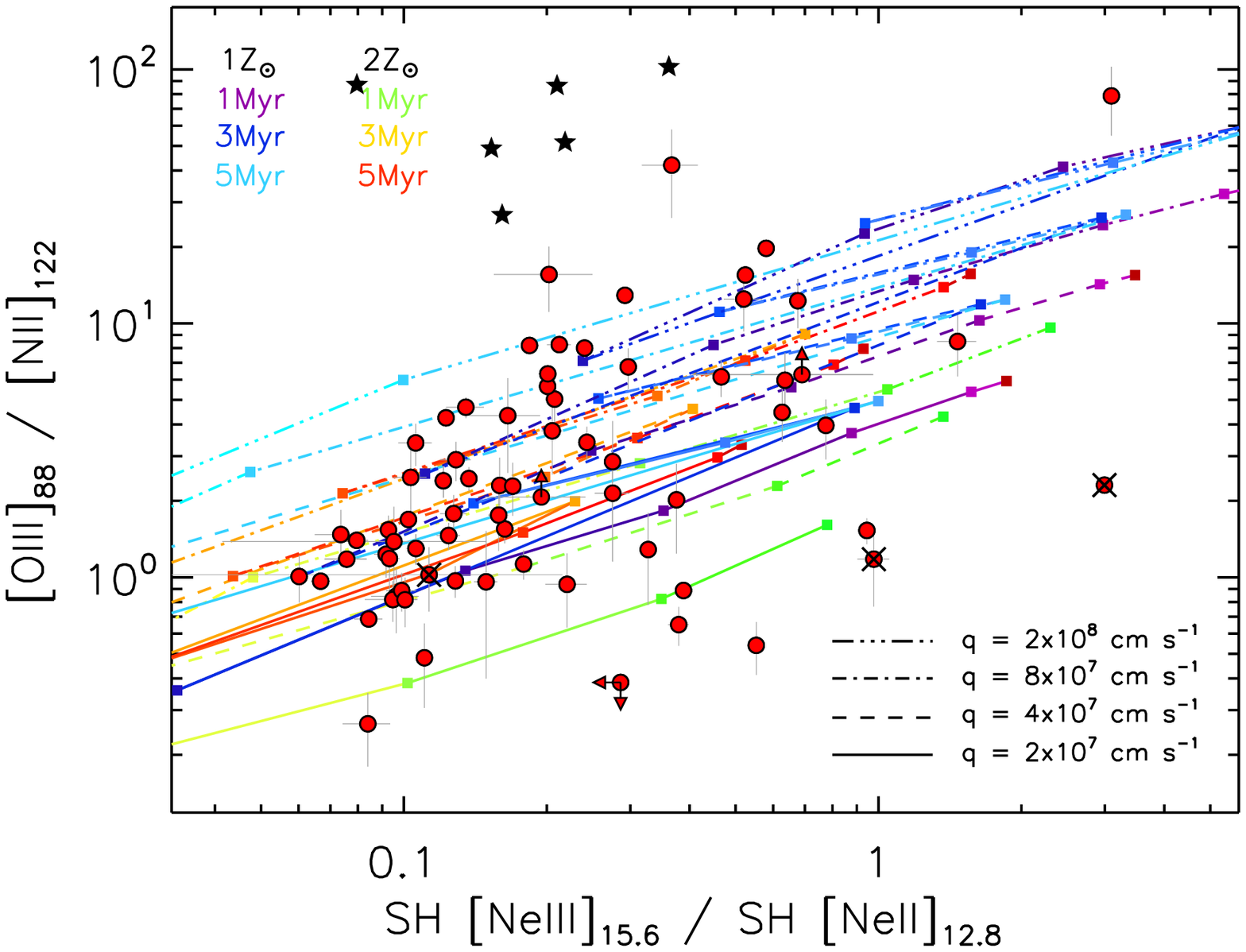}
\vspace{.25cm}
\caption{\footnotesize The \OIIIbsub/\NIIasub\, ratio as a function of the \NeIIIsub/\NeIIsub\, ratio for the GOALS sample (red solid circles). The solid lines represent {\sc mappings iii} models for varying metallicity and age of the starburst (shown in different colors), and ionization parameter, \textit{q} (different line styles). Models with similar ages but different \textit{q} are linked with dotted lines. Most LIRGs are compatible with solar or super-solar metallicity stellar populations $\simeq$\,2--5\,Myr old, except for the ones with the lowest \OIIIbsub/\NIIasub\, ratios, and expand all the range in \textit{q}, from 2\,$\times$\,10$^{7}$ to 2\,$\times$\,10$^{8}$\,\cmns. Galaxies from the DGS from \cite{Cormier2015} are shown as small, black solid stars.}\label{f:oiiinii}
\vspace{.25cm}
\end{figure}

We have used {\sc itera}, the {\sc idl} Tool for Emission-line Ratio Analysis \citep{Groves2008}, which is based on the {\sc Starburst99} stellar evolution synthesis models \citep{Leitherer1999} and the shock and photo-ionization {\sc mappings iii} code \citep{Allen2008}, to explore the properties of the dense ionized gas surrounding newly formed stars as a function of the main physical parameters that describe the emission line nebula, such as \ne, \textit{q}\footnote{\textit{q}\,$\equiv$\,$Q_{0} / 4 \pi R^2 $\ne\,=\,$U c$ where \textit{q} and \textit{U} are different definitions of the ionization parameter, $Q_{0}$ is the number of ionizing Lyman-continuum photons, \textit{R} is the radius of the Str\"omgren sphere, and \ne\, is the electron density in the ionized nebula. Both \textit{q} and \textit{U} are proportional to \G/\nhvol, if a fixed stellar population and size for the star-forming region are assumed.}, metallicity (in units of \Zsun) and the age of the starburst. Figure~\ref{f:oiiinii} shows the \OIIIbsub/\NIIasub\, as a function of \NeIIIsub/\NeIIsub\, for the LIRGs with available \textit{Herschel} and \textit{Spitzer} measurements. For the ITERA models, we have assumed that the starburst has occurred instantaneously (single stellar population models), which is a likely scenario for local LIRGs, and selected the stellar atmospheres from \cite{Levesque2010} with high mass loss. We have adjusted the \OIIIbsub/\NIIasub\, ratio predicted by the models down by 0.18\,dex to roughly account for the difference in oxygen abundances between these models and the values that we use in section~\ref{ss:pdrmodel} to model the PDR emission of the same galaxies with the models from \cite{Kaufman1999} (O/H\,=\,3\,$\times$\,10$^{-4}$). We note that this only affects the y-axis in Figure~\ref{f:oiiinii}. We explore the ITERA models for a range of parameters: \textit{q}\,=\,2--40\,$\times$\,10$^{7}$\,\cmns, starburst ages =\,1--10\,Myr, and metallicities of \Zsun, and 2\,\Zsun.

As we can see, the models reproduce well the range of line ratios observed in our sample, although some of the LIRGs with the lowest \OIIIbsub/\NIIasub\, seem to fall below the model grid. Note that two of these galaxies (IRASF~05189-2524 and NGC~1068) harbor an AGN with $<$\alphabolAGN$>$\,$\ge$\,0.5 (flagged with crosses in Figure~\ref{f:oiiinii}). Sources with \OIIIbsub/\NIIasub\,$\lesssim$\,1 can only be reproduced by models with 2\,\Zsun, starburst ages of $\simeq$\,1\,Myr and the lowest ionization parameter \textit{q}\,=\,2\,$\times$\,10$^{7}$\,\cmns\, (solid green line). We note however, that this region of the parameter space could be reached at slightly higher metallicities than 2\,\Zsun, but such models are not currently available in {\sc itera}. The remaining bulk of the sample is well described by models with solar or super-solar metallicity and starburst ages ranging from $\sim$\,2 to 5\,Myr, in agreement with \cite{Inami2013}. The starburst age is only slightly degenerate with \textit{q}, and variations in the \OIIIbsub/\NIIasub\, ratio can be mostly attributed to a change in the ionization parameter (regardless of \textit{Z}, except for ages $\lesssim$\,2\,Myr). Particularly, values of \OIIIbsub/\NIIasub\,$\gtrsim$\,5 (or lower at \NeIIIsub/\NeIIsub\,$\lesssim$\,0.5), cannot be reproduced by any model of any age unless \textit{q}\,$>$\,4\,$\times$\,10$^{7}$\,\cmns, and $\gtrsim$\,8\,$\times$\,10$^{7}$\,\cmns\, for \OIIIbsub/\NIIasub\, close to $\simeq$\,10.

We note that even though higher \OIIIbsub/\NIIasub\, can be reached at sub-solar metallicities and could explain the location of the sources in the DGS studied by \cite{Cormier2015} (filled stars in Figure~\ref{f:oiiinii}), this is not likely the reason for the increase in the line ratio observed in dusty systems like LIRGs, but rather an increase in \textit{q}. Thus, in dusty star-forming galaxies the \OIIIbsub/\NIIasub\, ratio probes mostly the ionizing parameter \textit{q} for starbursts of a few Myr (i.e., when these emission lines are actually detected), with \OIIIbsub/\NIIasub\, scaling roughly linearly from $\simeq$\,1 to 10 for values of \textit{q} ranging from 2\,$\times$\,10$^{7}$ to 2\,$\times$\,10$^{8}$\,\cmns. This relation can be refined if a measurement of the \NeIIIsub/\NeIIsub\, ratio is also available.

\subsection{PDR Modeling}\label{ss:pdrmodel}

The \CIIsub\, and \OIasub\, fine-structure lines are two of the main coolants of the neutral ISM \citep{Malhotra1997,Rosenberg2015} and therefore they are the most useful to constrain its physical conditions. We have used the {\sc PDR Toolkit} (PDRT) wrapper \citep{Pound2008} to derive the main parameters that govern the PDR emission: the intensity of the UV inter-stellar radiation field (ISRF) that impinges the PDR surface, \G, measured in units of the local Galactic value (\Go\,=\,1.6\,$\times$\,10$^{-3}$\,erg\,s$^{-1}$\,cm$^{-2}$; \citealt{Habing1968}), and the volume density of the neutral gas, \nhvol. The PDRT is based on the models developed by \cite{Kaufman1999,Kaufman2006}, which calculate the temperature structure and line emission from a PDR, based on the formalism for a one-dimensional semi-infinite slab \citep{Tielens1985}. We assume the case in which the PDRs are illuminated only on one side. This choice is the most reasonable for starburst galaxies, especially LIRGs, where most of the newly-formed massive stars are likely still embedded in their parent molecular clouds \citep[e.g.,][]{DS2007}. That is, while a generic molecular cloud in the disk of a LIRG may be seeing an isotropic ISRF from older stellar populations, most of the line flux will be overwhelmingly dominated by the energy cooled in warm PDRs associated with the most recent burst. Strictly, these models can only be compared to individual star-forming regions. However, they can also provide useful information regarding the typical conditions of the PDRs in a given galaxy when only spatially integrated measurements are available.

\begin{figure*}[t!]
\vspace{.5cm}
\epsscale{.32}
\hspace{-.5cm}
\plotone{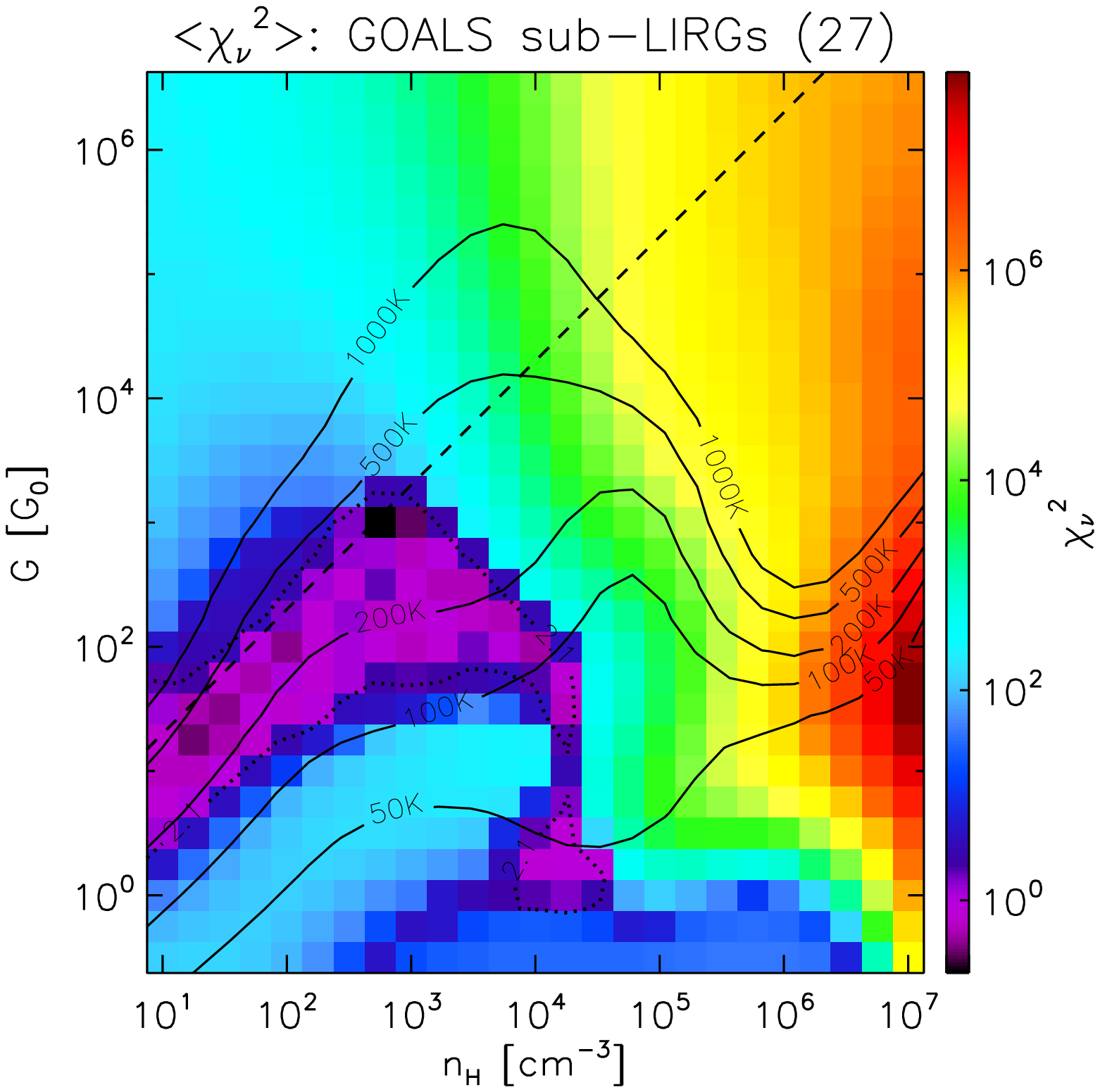}
\hspace{.85cm}
\plotone{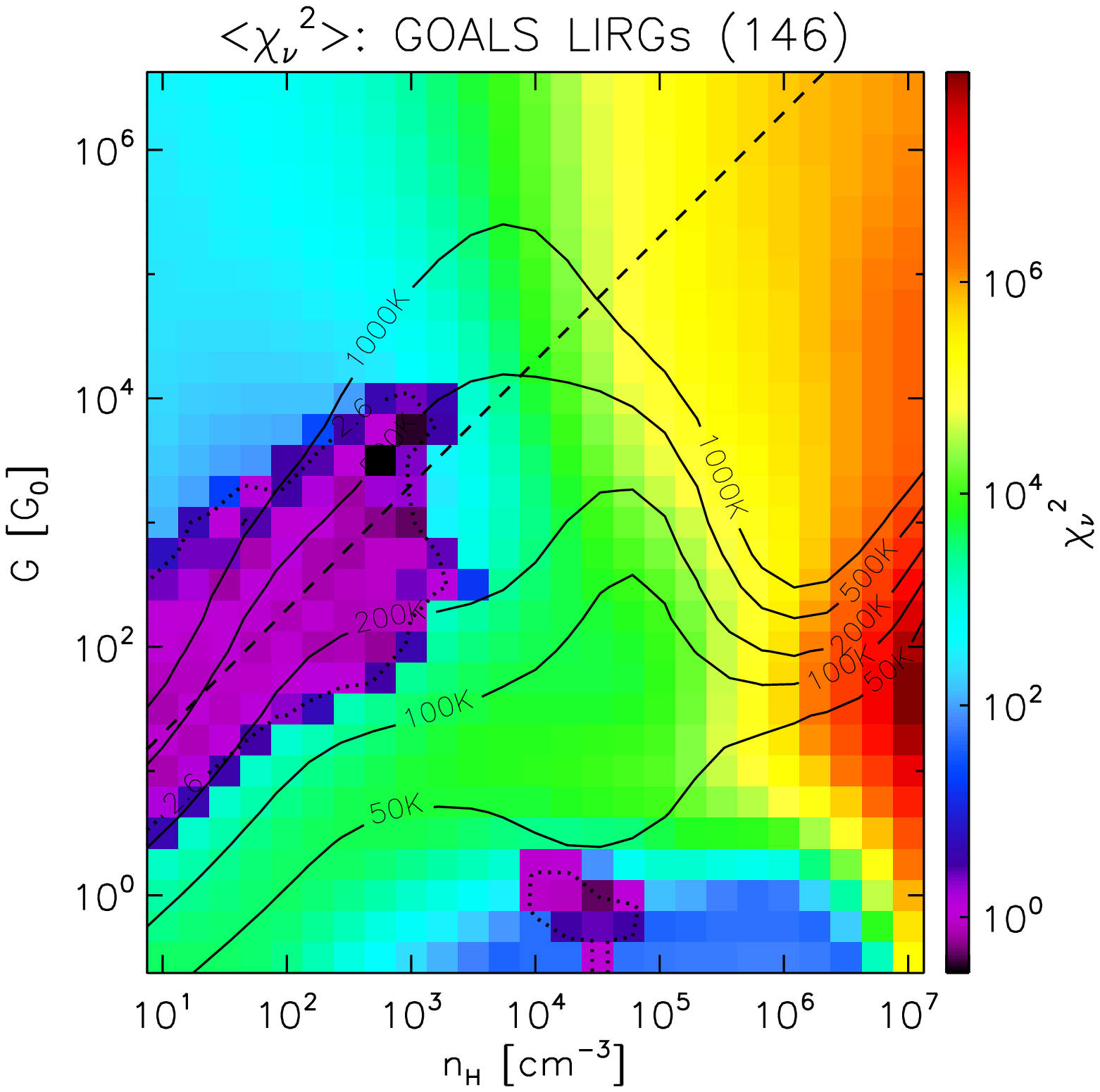}
\hspace{.85cm}
\plotone{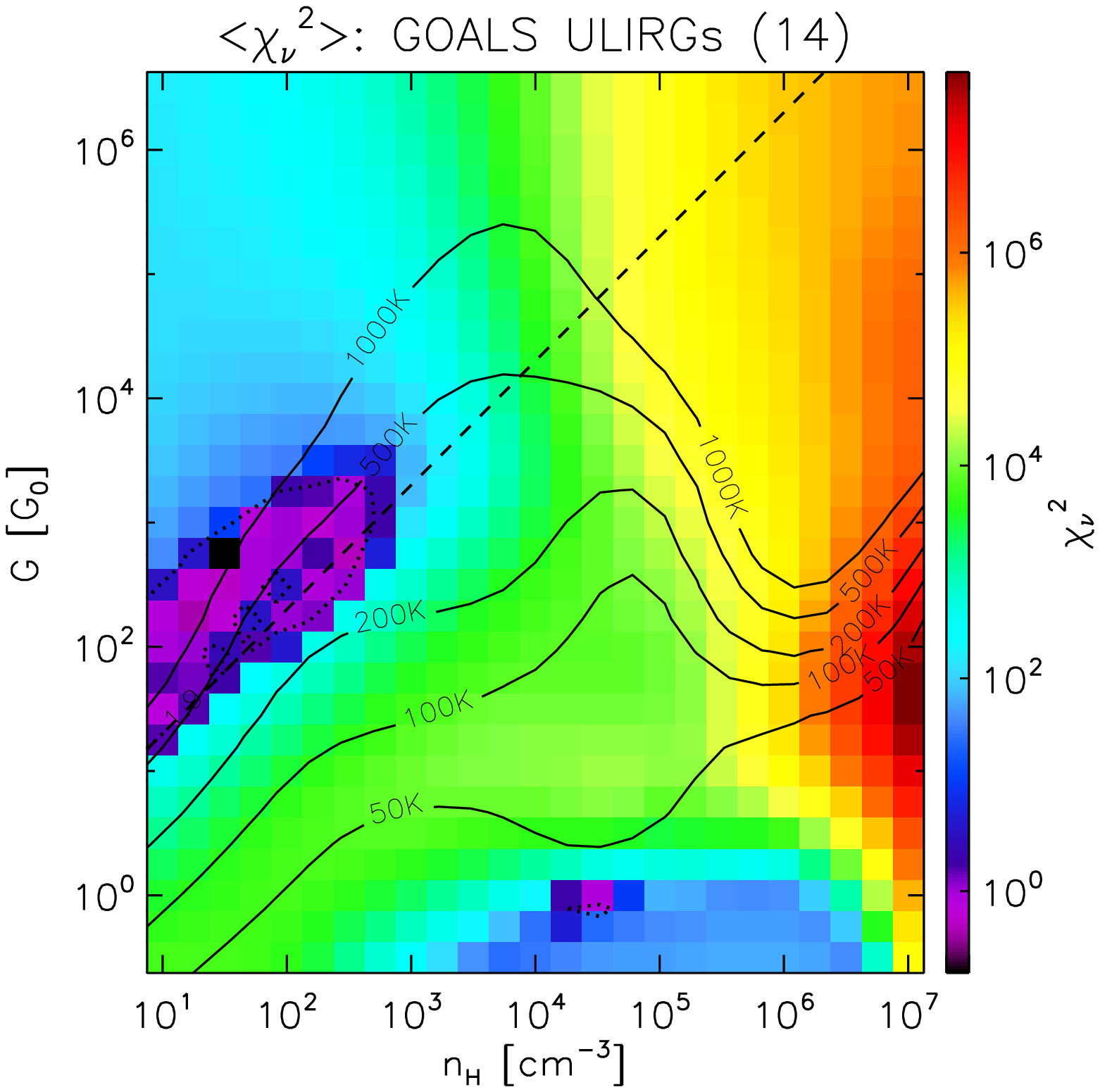}
\vspace{1.cm}
\caption{\footnotesize Results from the PDRT model fitting for three subsamples of galaxies from GOALS: sub-LIRGs (left), LIRGs (center), and ULIRGs (right). The figures show the average probability distribution for each subsample, $<$\chinusq$>$, as a function of \G\, and \nhvol. See text for details about what galaxies were used in the modeling, as well as how the \chinusq\, maps of the galaxies were averaged. The combined best-fit values are the regions of the parameter space colored as purple. The input data for the models are the measured \CIIpdr, \OIasub\, and FIR fluxes of each galaxy. The dashed line denotes a \G/\nhvol\,=\,2\,\cmmm, the threshold at which $v_{\rm drift}$\,$\simeq$\,$v_{\rm turb}$ \citep{Kaufman1999}. The solid lines indicate the temperature of the PDR surface, \Tpdr, for a given combination of \G\, and \nhvol\, as derived from the model.}\label{f:GOnH}
\vspace{.5cm}
\end{figure*}

We use the integrated \CIIpdr, \OIasub\, and FIR fluxes as input to the PDRT models, excluding sources with limits in any of these quantities. Note that the \CIIno\, fluxes used here are those obtained after subtracting the ionized gas contribution from the line emission\footnote{To obtain the \CIIpdr\, flux for galaxies without a measurement of the \NIIbsub\, line we use equation~\ref{e:ciipdr} to estimate $f(\CIIpdr)$ (capped at 90\,\%) from the FIR \Sa/\Sb\, continuum ratio. The errors are fully propagated based on the dispersion of the correlation.} (see section~\ref{ss:ciipdr}); a critical step needed to obtain consistent results from the PDR analysis. The data of each galaxy are compared, via \chisq\, minimization, to the (\CIIpdr+\OIasub)/FIR and \CIIpdr/\OIasub\, ratios predicted by the PDR model for each combination of parameters, --0.5\,$\leq$\,log\,(\G/\Go)\,$\leq$\,6.5 and 1\,$\leq$\,log\,(\nhvol/\cnmmm)\,$\leq$\,7, in 0.25\,dex steps. The flux uncertainties employed to calculate the \chisq\, include a 11\,\% error due to the absolute photometric uncertainty, added in quadrature to the measured error in the actual spectra. To obtain the map of the probability distribution for each \G\, and \nhvol\, combination averaged over entire GOALS sample we combine the \chinusq\, maps of each galaxy using a weighting of $e^{-\chi_{\nu [G_{\rm 0},n_{\rm H}]}^{2}/2}$.

We present the results in Figure~\ref{f:GOnH} separately for sub-LIRGs (\LIR\,$<$\,10$^{11}$\,\Lsun) (left panel), LIRGs (central panel) and ULIRGs (right panel). We exclude from this analysis those galaxies whose bolometric luminosity is dominated by an AGN ($<$\alphabolAGN$>$\,$\ge$\,0.5), sources where there is a mismatch between the spatial location of the lines and/or continuum emission peaks, galaxies with companions within the aperture used to measure their fluxes, and sources where the \OIasub\, line shows signs of self-absorption (see section~\ref{ss:oicii}). We do not attempt to correct for the latter effect since the models already account for the possible optical thickness of the line within the PDR, and self-absorption due to intervening cold material along the line(s) of sight is completely unconstrained. Observations at significantly higher spectral resolution than our PACS data, such as those obtained for the Milky-Way with \textit{Herschel}/HIFI, would be needed in order to correct for this effect \citep{Pineda2013,Gerin2015,Langer2016}.

As we can see, the location of the main cloud of best fitting parameters (purple color, \chinusq\,$\sim$\,1) show that most star-forming LIRGs have average ISRF intensities in the range of $\sim$\,10$^{1-3.5}$\,\Go\, and gas densities \nhvol\,$\sim$\,1--10$^3$\,\cnmmm. In this regime of the parameter space of relatively low \G\, and \nhvol\, where the \CIIno\, line dominates the cooling (\OIasub/\CIIpdr\,$\lesssim$\,1), both parameters are degenerate\footnote{This is more clear when looking at the parameter space of the \chinusq\, map of an individual galaxy. In Figure~\ref{f:GOnH} we show the averaged map, $<$\chinusq$>$, which is biased towards the bottom (best value) of the \chinusq\, distribution for each individual source due to the weighting. Thus the overall $<$\chinusq$>$ map is effectively the accumulation of the best \chinusq\, values for the entire galaxy sample.}, and the relevant quantity setting the observed line ratio is \G/\nhvol\, \citep[see also][]{Malhotra2001}, which is approximately proportional to the temperature of the gas. This is because the numerator and the denominator of the (\CIIpdr+\OIasub)/FIR ratio mainly depend on the \nhvol\, and \G, respectively. In turn, the \OIasub/\CIIpdr\, ratio is sensitive to \G, but also particularly to the gas temperature as we showed in section~\ref{ss:oicii} (\Tkin\,$\propto$\,\Tdust).

Because in a one-side illuminated cloud the (optically thin) FIR intensity radiated back is \IFIR\,$\simeq$\,2\,$\times$\,\G\,/4$\pi$ [erg\,s$^{-1}$\,\cnmm\,sr$^{-1}$] \citep{Kaufman1999}, the integrated FIR luminosity produced by a galaxy containing such typical PDR regions is:

\begin{equation}\label{e:beamff}
L_{\rm FIR}\,\approx\,2\,G\,\pi\,(D/2)^2\,\phi_{\rm A}
\end{equation}

where $D$ is the angular diameter of the overall starburst, and \Aff\, is the beam (area) filling factor of the PDR-emitting regions within it. We note that there is a secondary region of the parameter space, around \G\,$\simeq$\,1\,\Go\, and \nhvol\,$\simeq$\,10$^{4-5}$\,\cnmmm, where low \chinusq\, values are also found. While this range of parameters is equally probable based on the input data, such a low \G\, is rather unlikely. Even assuming \Aff\,=\,1 (but see section~\ref{ss:g0nHSigmaIR}), a \G\,=\,1\,\Go\, implies that in order to generate the total \LFIR\, emitted by a LIRG, a few 10$^{11}$\,\Lsun, the size of the starbursting region would have to be $D$\,$\gtrsim$\,10\,kpc (increasing as $\phi_{\rm A}^{-1/2}$). While this might be possible for a few objects, the typical sizes of LIRGs measured in the MIR and FIR are on average significantly smaller than that, with diameters of just a few kpc, or even below 1\,kpc in the case of ULIRGs \citep{DS2010b, Rujopakarn2011, Lutz2016, Chu2017}.
Thus, we only consider solutions with \G\,$>$\,10$^{0.5}$\,\Go.

The best fitted \G\, values obtained through the PDR modeling are several orders of magnitude lower than 10$^6$\,\Go, which implies that grain charging should not be a major cause of the reduced photoelectric heating efficiency of the gas (see \citealt{Kaufman1999} for details). In other words, PAH molecules should remain mostly neutral on average, even in the warmest LIRGs. This is in agreement with results based on \textit{ISO} data showing that the \CIIsub\, to \PAHdsub\, (or overall 5--10\,$\mu$m wavelength range) emission ratio remains roughly constant in normal star-forming galaxies and ULIRGs \citep[][respectively]{Luhman2003, Helou2001}. We note that in those works the ionized gas component was not subtracted from the total \CIIsub\, emission when compared to the \PAHdsub\, (see section \ref{ss:ciipdr}) and while AGN identification was performed, its contribution to the IR or bolometric power output of the galaxies was not taken into account. Both issues probably contribute to the large dispersion seen in the \CIIsub/\PAHdsub\, ratio across each sample ($\sim$\,0.2\,dex). The values of our best fitted \G\, and the implication of having mostly a neutral PAH population, are also in agreement with the fact that the \PAHcsub/\PAHasub\, ratio --a tracer of the ionization state of the PAH molecules \citep{Draine2001}-- does not vary significantly in LIRGs \citep[$\lesssim$\,30\%;][]{Stierwalt2014}.

Interestingly, within the context of the PDR models used here, the \G/\nhvol\, ratios of around one third of our LIRGs (see section~\ref{ss:g0nHSigmaIR}) are larger than the threshold (\G/\nhvol\,$\simeq$\,2\,\cmmm) at which the radiation pressure on the dust starts driving grains through the gas at velocities greater than the average turbulent velocity of the gas assumed by the model ($v_{\rm drift}$\,$\simeq$\,$v_{\rm turb}$\,=\,1.5\,\kms). This suggests that galaxies with \G/\nhvol\, above this critical value (dotted line in Figure~\ref{f:GOnH}) could have PDRs that may be dynamically unstable \citep[][see also \citealt{Draine2011} for a discussion regarding dust drift velocities within \HII\, regions]{Kaufman1999}.

The purple regions in the left panel of Figure~\ref{f:GOnH} show that sub-LIRGs have \G\, values spanning the lower end of the parameter space of LIRGs (see center panel), but have similar gas densities. On the other hand, the most luminous galaxies in GOALS (ULIRGs, \LIR\,$\geq$\,10$^{12}$\,\Lsun; right panel) have the highest \G\, values, $\simeq$\,10\,$\times$ the average of the sample, but also the same range of gas densities, \nhvol\,$\sim$\,1--10$^3$\,\cnmmm. Therefore, there is a progression for \G\, to increase with \LIR, but not an equivalent increase in the PDR gas density, as it would be expected from more compact environments associated with merger-driven star-formation. The GOALS ULIRG sample agrees reasonably well with the results obtained for the local ULIRGs in the HERUS sample studied by \cite{Farrah2013}, who obtained similar \G\, values, between $\simeq$\,3\,$\times$\,10$^{2-3}$\,\Go, albeit with an upper limit for the derived gas densities, \nhvol\,$\lesssim$\,2\,$\times$\,10$^{2}$\,\cnmmm, a factor of a few lower than our estimates.

In section~\ref{ss:oicii} we estimated \Tkin\,$\simeq$\,50-100\,K, significantly lower than the PDR surface temperatures favored by the models, \Tpdr\,$\simeq$\,150--1000\,K (see contours in Figure~\ref{f:GOnH}). A possible explanation for this discrepancy is that most of the line emission arises from a region deep in the cloud -- far removed from the warm PDR surface. In this case, thermal saturation of the \CIIsub\, line would occur at higher \G/\nhvol\, than expected in the models, closer to the upper limit that we measure in the most luminous sources, around \G/\nhvol\,$\simeq$\,10\,\cmmm\, and \Tkin\,$\simeq$\,100\,K\,$\approx$\,$\Delta$E/\kB\,=\,92\,K. We note, however, that despite the discrepancy in absolute values there is a broad correlation between \Tpdr\, and \Tkin.

\subsection{Probing the ISM Structure Using \G/\nhvol:\\ A Characteristic Break in \SigmaIR}\label{ss:g0nHSigmaIR}

\begin{figure}[t!]
\vspace{0.5cm}
\epsscale{1.15}
\plotone{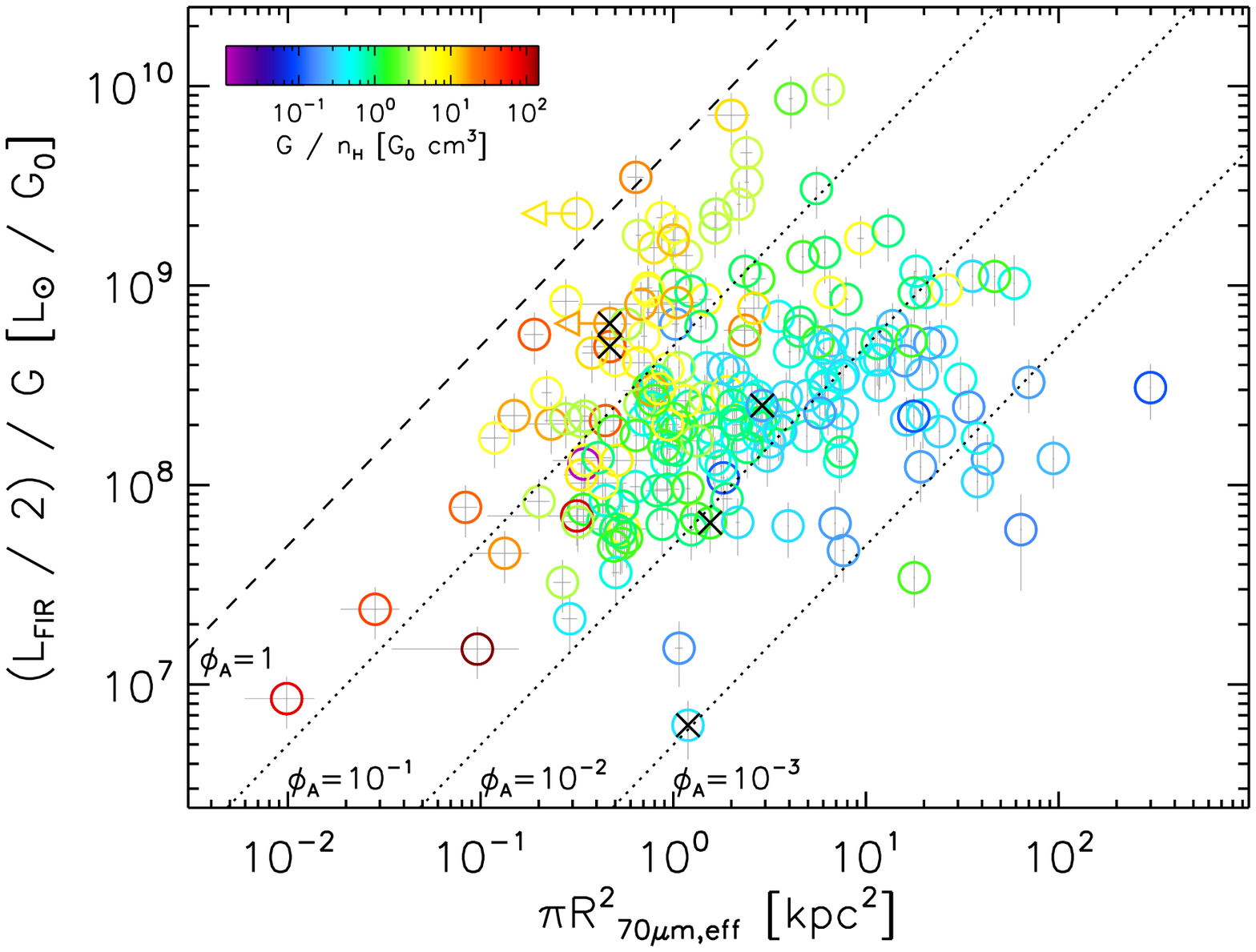}
\caption{\footnotesize. The \LFIReff/\G\, ratio (where \LFIReff\,=\,\LFIR/2) as a function of the effective (half-light) area of the LIRGs in GOALS. The sizes have been taken from \cite{Lutz2016}. Galaxies are color-coded as a function of the \G/\nhvol\, ratio. Dashed and dotted lines indicate different values of the area filling factor of the PDR emitting regions within the galaxy, \Aff, as derived from equation~\ref{e:beamff}.}\label{f:beamff}
\vspace{.25cm}
\end{figure}

If most of the IR luminosity of a galaxy is produced in PDRs, the local ISRF, traced by \G, and the galaxy-integrated \LFIR\, are tied, via equation~\ref{e:beamff}, through the overall size of the starburst, \textit{D}, and the area filling factor of the PDR-emitting star-forming regions within the source, \Aff\footnote{We note that area filling factors larger than unity are possible if multiple PDRs are stacked along the same line of sight.}. This means that if we know the actual size of the starburst in a galaxy, we can estimate \Aff. Figure~\ref{f:beamff} shows the ratio of \LFIReff\, to \G\, as a function of the effective physical area of our galaxies measured at 70\,$\mu$m, where the starburst emission peaks. Within the uncertainties, we find that all LIRGs fall below the dashed line marking \Aff\,=\,1. However, \Aff\, can be as small as $\simeq$\,10$^{-3}$ for some of the largest galaxies.

\begin{figure*}[t!]
\vspace{.5cm}
\epsscale{.8}
\plotone{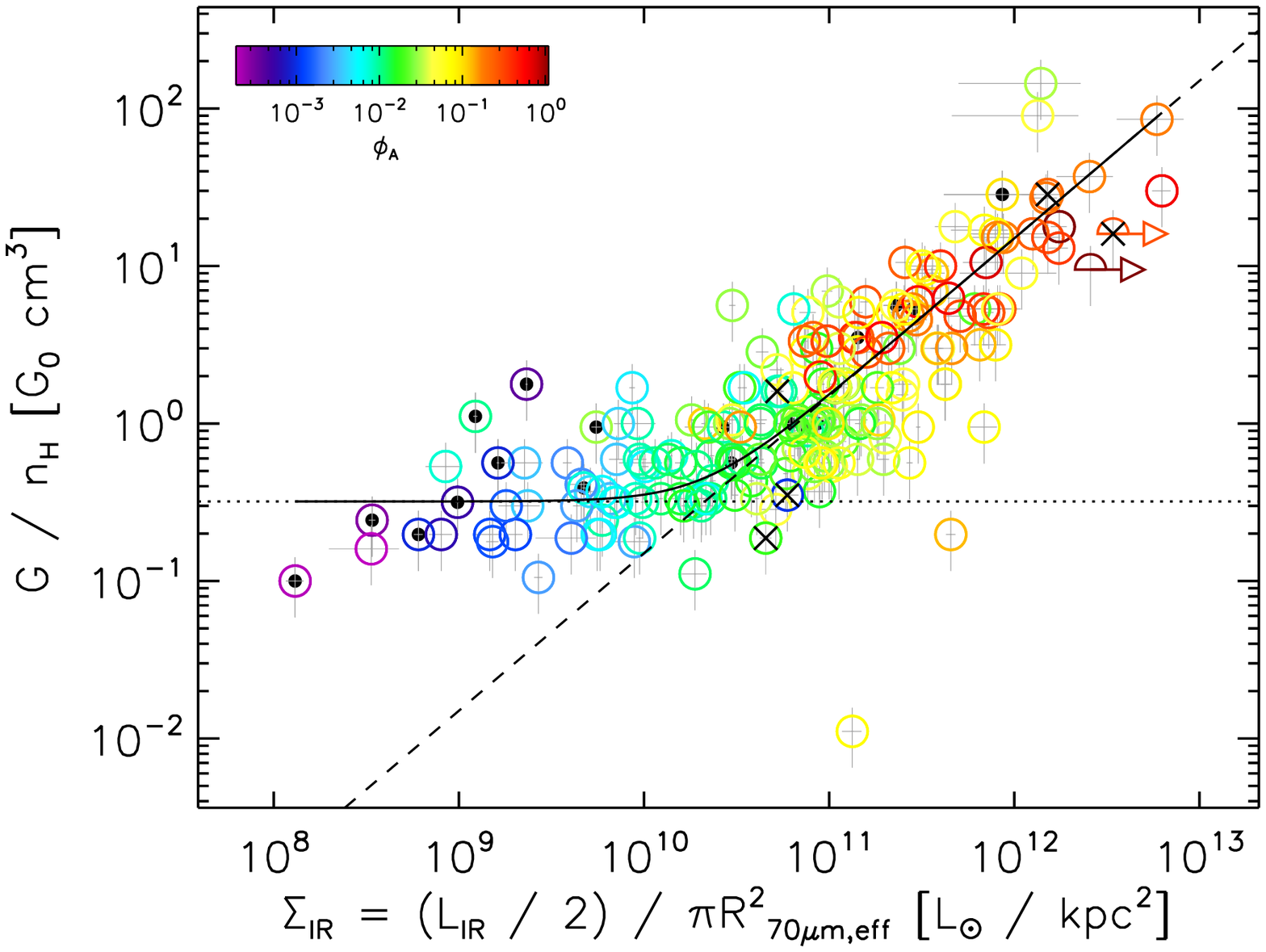}
\caption{\footnotesize The \G/\nhvol\, ratio as a function of the effective \SigmaIR\, for the GOALS sample (where \SigmaIR\,=\,\LIReff/$\pi$R$^2_{\rm eff,70\mu m}$ and \LIReff\,=\,\LIR/2). The galaxies are color-coded as a function of the PDR area filling factor, \Aff. Below \SigmaIRstar\,$\simeq$\,5\,$\times$\,10$^{10}$\,\lsd, \G/\nhvol\, remains constant, ranging between $\simeq$\,0.2--0.6\,\cmmm\, (the dotted line shows the median value, 0.32\,$\pm$\,0.21\,\Go\,\cmmm), suggesting that the boosting in \SigmaIR\, is driven by a higher \Aff, i.e., a larger number of star-forming regions per unit area. Above that threshold, \G/\nhvol\, increases linearly with \SigmaIR\, (dashed line; where \G/\nhvol\,[\Go\,\cmmm]\,=\,1.5\,$\times$\,10$^{-11}$\,$\times$\,\SigmaIR\,[\lsd]) indicating a change in the physical properties of the PDRs. At a fixed \SigmaIR\, the scatter in \G/\nhvol\, is mostly due to variations in \nhvol, with lower \nhvol\, galaxies displaying higher \G/\nhvol\, ratios.}\label{f:G0nHlsd}
\vspace{.25cm}
\end{figure*}

In Figure~\ref{f:G0nHlsd} we show the \G/\nhvol\, ratio as a function of the luminosity surface density for our LIRG sample. We can see that there is a clear correlation between both quantities, but only above \SigmaIRstar\,$\simeq$\,5\,$\times$\,10$^{10}$\,\lsd\, \citep[see also][]{Elbaz2011}. Below this threshold, an increase in \SigmaIR\, is not followed by an increase of \G/\nhvol. This implies that while the number density of star-forming regions within LIRGs in this dynamic range is progressively larger, the average PDR properties do not vary significantly. This is clear from the color-coding of the data, which indicates the \Aff\, of each galaxy. Increasing \SigmaIR\, from $\sim$\,10$^{9}$ to $\sim$\,5\,$\times$\,10$^{10}$\,\lsd\, can be accomplished by an equal increase in \Aff, from 10$^{-3}$ to a few 10$^{-2}$, with \G/\nhvol\, remaining at a relatively constant value of $\simeq$\,0.32\,\Go\,\cmmm\, (see dotted line). Indeed, \G/\nhvol\, ratios of the order of unity or lower are also found in the nuclei and extended disks of normal, nearby star-forming galaxies, like M~82 or NGC~891 \citep[e.g.,][]{Contursi2013, Hughes2015}. Above \SigmaIRstar, even though \Aff\, still keeps increasing, the upturn in \SigmaIR\, is mostly driven by a rise in \G/\nhvol, pointing to a change in the physical conditions of the PDRs towards more intense ISRFs for a given gas density. In fact, most of the increase in \G/\nhvol\, at high \SigmaIR\, is due to an increase in \G, while variations in \nhvol\, are the source of scatter in the trend at a any \SigmaIR. For reference, the solid line in Figure~\ref{f:G0nHlsd} shows a linear dependence (not a fit) of the form: \G/\nhvol\,[\Go\,\cmmm]\,=\,1.5\,$\times$\,10$^{-11}$\,$\times$\,\SigmaIR\,[\lsd].

\subsection{Physical Interpretation:\\ Young, Compact, Dusty Star-Forming Regions}\label{ss:dustyreg}

It has been proposed that the deficit of PDR (\CIIno, \OIno) line emission with respect to that of the IR continuum could be understood in terms of the geometry of the star-forming regions. This essentially represents the change of the ratio between the PDR the surface area and volume of the dust-emitting region -- a ratio that would decrease linearly as the star-forming region grows in size. This would be further accentuated as the starburst evolves and \HII\, regions begin to overlap, sharing a common PDR envelope, creating effectively a giant, single star-forming region -- a ``giant Orion''. This picture does not seem consistent with our results for two reasons: (1) the areal filling factor of PDR emission we measure in the majority of LIRGs is much less than unity as might be expected from a single, large \HII\, region filling the central starburst, and (2) we estimate a volume filling factor of dense ionized gas of $\lesssim$\,5\%, implying that it mostly originates from compact \HII\, regions.

The connection between \SigmaIR\, and \G/\nhvol\, shown in Figure~\ref{f:G0nHlsd} relates a kpc-scale property of LIRGs to the physics of the individual PDRs. In other words, the fact that both quantities scale with each other above \SigmaIRstar\, indicates that in this regime, the luminosity surface density of the entire starburst increases with the average energy per particle density at the surface of PDRs inside the few pc-scale star-forming regions. A joint, progressive increase of \SigmaIR\, and \G/\nhvol\, can be accomplished by increasing the fraction of younger, more massive stars per star-forming region\footnote{Here we do not imply a change in the initial mass function, but rather refer to younger star-forming regions in which massive stars, with orders of magnitude higher light-to-mass ratios than solar mass stars, have still not disappeared \citep[ages less than a few Myr;][]{Inami2013}.}. Indeed, the presence of a larger number of younger stars with harder ionization fields would provide the necessary boost in \G\, and \LIR\, to produce the observed correlation seen in Figure~\ref{f:G0nHlsd}, above \SigmaIRstar. Besides, the spatial distribution of gas and dust within these regions must be also different. Very young, less-developed star-forming regions have not had time to expel their dusty envelopes and thus have a larger fraction of the surrounding material closer to the newly born stars. At the same time, their radial density structure peaks at the position of the ionization front, close to the PDR, in contrast to dustless \HII\, regions where the density profile is rather constant \citep{Draine2011}. Note that in this case the volume and column density of gas and dust may remain constant, as only a redistribution of the intervening material along the radial direction is needed. However, \Tdust, and thus \G, must increase since dust grains are overall closer to the heating source.

Our results indicate that this dusty phase, in which we are able to see a significant number of star-forming regions that are likely matter-bounded and still embedded in their molecular cocoon \citep[e.g.,][]{Kawamura2009, Miura2012, Whitmore2014}, 
seems to be associated primarily with the period when the overall starburst region is also most compact (above \SigmaIRstar).
A mechanism often proposed to produce this compaction of the gas and dust in galaxies are major mergers \citep{Sanders1988a, Hopkins2008a}.
While there is not a clear sequence where galaxies with increasingly larger \SigmaIR\, and \G/\nhvol\, are found to be in progressively more advanced interacting stages, we find that most ($\sim$\,80\,\%) late-stage mergers \citep{Stierwalt2013} have \SigmaIR\,$>$\,5\,$\times$\,10$^{10}$\,\lsd, and lie along the slope in Figure~\ref{f:G0nHlsd}. This lack of a simple correlation with merger stage is consistent with the idea that star-forming clumps and stellar clusters can form and be quickly destroyed by powerful stellar feedback in gas-rich starbursts and mergers on relatively short (10--20 Myr) timescales \citep{Whitmore2014, Oklopcic2017, Linden2017}, significantly shorter than the timescale of the entire merging process (over several hundred Myr). Thus, rather than a single or small number of giant, long-lived, and relatively normal \HII\, regions, it may be more accurate to think of the nuclear starburst in a LIRG as a collection of dense, dusty and young star-forming regions that are individually short-lived but constantly replenished, like ``fireworks'' triggered by a galactic merger.

\section{Conclusions}\label{s:summary}

We obtained \textit{Herschel}/PACS spectroscopy of the main FIR fine-structure emission lines for a sample of $\sim$\,240 galaxies in GOALS, a complete flux-limited and luminosity-limited sample of all 60$\,\mu$m-selected LIRGs systems detected in the nearby universe (\textit{z}\,$<$\,0.09).
We combined these observations with \textit{Herschel}/SPIRE and MIR \textit{Spitzer}/IRS spectroscopic data to derive the main physical characteristics of the neutral and ionized gas in dense PDRs and \HII\, regions and provide a comprehensive view of the heating and cooling of the ISM as a function of galaxy-integrated properties. We have found the following results:

\begin{itemize}

\item The FIR lines with the most pronounced ``deficits'' (greatest decline of the line flux with respect to FIR continuum emission) as a function of \Sa/\Sb, or equivalently \Tdust, are the \CIIsub, \NIIasub\, and \NIIbsub\, emission lines, which show a decrease of a factor $\sim$\,50 in the warmest LIRGs. The \OIasub\, line displays a smaller deficit (a factor of $\sim$\,10). The \OIIIbsub\, line shows no deficit with \Tdust\, but rather an increasing scatter at high \Sa/\Sb. However, \textit{all} lines show an increasing deficit with rising infrared luminosity surface density, \SigmaIR.

\item We use the \CIIsub/\NIIbsub\, ratio to derive the contribution of photo-dissociation regions (PDRs) to the total \CIIsub\, emission, \CIIpdr, as well as its fractional contribution, $f(\CIIpdr)$\,=\,\CIIpdr/\CIIsub. We find that $f(\CIIpdr)$ broadly increases as a function of \Sa/\Sb\, from $\sim$\,60\,\%, a value typical of Milky Way-like, to up to 95\,\% in the warmest LIRGs. $f(\CIIpdr)$ also increases as a function of the hardness of the ionizing radiation field as traced by the \OIIIbsub/\NIIasub\, ratio.

\item We find that the scatter in the \CIIsub-to-FIR\, ratio as a function of \Sa/\Sb\, is correlated with $f(\CIIpdr)$. LIRGs with higher $f(\CIIpdr)$ show progressively smaller \CIIsub\, deficits at a given \Tdust, indicating that while warmer LIRGs do in fact have larger \CIIsub\, deficits than normal star-forming galaxies, those with higher $f(\CIIpdr)$ show less pronounced ones.

\item Most LIRGs show \NIIasub/\NIIbsub\, ratios compatible with electron densities of the ionized gas in the range \ne\,$\simeq$\,10$^{1-2}$\cnmmm, with a median of 41\,\cnmmm. Given the higher \ne\, values obtained from the MIR sulfur line ratios, this suggests that most of the \NIIbsub\, emission (and \CIIion) is thermalized in the \HII\, regions and thus mostly originate from the diffuse ionized gas phase. Assuming a simple, two-component ISM model with dense (\ne\,$\simeq$\,500\,\cnmmm) and diffuse (\ne\,$\lesssim$\,1\,\cnmmm) ionized gas regions, the measured \NIIasub/\NIIbsub\, ratio implies an average volume filling factor \Vff\,$\leq$\,5\,\% for dense \HII\, regions with respect to the overall volume of ionized emitting gas in the majority of LIRGs.

\item We model the \OIasub/\CIIpdr\, ratio and find that it is well correlated with the kinetic temperature of the gas in PDRs when the \OIasub\, emission is optically thin. Galaxies with \OIasub/\CIIpdr\, ratios lower than expected by the model tend to display strong absorption of the 9.7\,$\mu$m silicate feature, indicating large dust opacities and suggesting that \OIasub\, may be optically thick or self-absorbed in about 5\,\% of the sample.

\item We find that the \OIIIbsub/\NIIasub\, ratio is a good tracer of the ionization parameter, \textit{q}, which ranges from 2\,$\times$\,10$^7$ to 2\,$\times$\,10$^8$\,\cmns\, proportionally with \OIIIbsub/\NIIasub\,$\sim$\,1--10. Moreover, most LIRGs are compatible with solar or super-solar metallicities and starburst ages $\simeq$\,2--5\,Myr old, in agreement with previous results.

\item We derive the intensity of the UV radiation field heating the PDRs, \G, and the volume density of their neutral gas, \nhvol. We find values in the range \G\,$\sim$\,10$^{1-3.5}$\,\Go, increasing with \LIR, and \nhvol\,$\sim$\,1--10$^3$\cnmmm\, regardless of the luminosity of galaxies. The \G/\nhvol, ratio ranges between $\simeq$\,0.1 and 50\,\Go\,\cmmm\, for the entire sample, with ULIRGs populating the upper end of the distribution at \G/\nhvol\,$\gtrsim$\,2\,\Go\,\cmmm. We use the scaling relation between \G\, and \LIR\, to estimate the area filling factor of the PDRs within the overall area of the starburst region, \Aff, and find \Aff\,$\simeq$\,1--10$^{-3}$.

\item The \G/\nhvol\, ratio shows a distinct transition at \SigmaIRstar\,$\simeq$\,5\,$\times$\,10$^{10}$\,\lsd, with the ratio being constant below this value, and increasing linearly with \SigmaIR\, above it. We suggest that below this threshold, the increase in \SigmaIR\, is purely driven by an increase of the number density of star-forming regions in the galaxy ($\propto$\,\Aff), without changing the average PDR physical conditions. Above \SigmaIRstar, \G/\nhvol\, also starts to increase, signaling a departure from the typical PDR properties of normal star-forming galaxies towards more intense/harder radiation fields and compact geometries typical of starbursting sources.

\end{itemize}

We note that the correlation between \SigmaIR\, and \G/\nhvol\, links a macroscopic property of galaxies with the local, small-scale physics of their PDR regions, and indicates that there is a threshold above which the average radiation field intensity per particle density in \textit{individual} star-forming regions scales linearly with the overall energy density of the \textit{entire} starburst region. We discuss the possibility that the observed correlation above \SigmaIRstar\, arises not only from a boost in the radiation field intensity provided by young massive stars but also due to finding these star-forming regions in an earlier, dustier phase of their evolution. While we do not find a clear trend between the position of galaxies along the \G/\nhvol\, vs. \SigmaIR\, correlation and the interaction stage of the systems, we find that most late-stage mergers ($\sim$\,80\%) are above \SigmaIRstar\, and show enhanced \G/\nhvol\, ratios, probably reflecting the fast, ``fireworks''-like process of star-cluster formation and disruption during the merger.

\section*{Acknowledgments}

We thank the referee for her/his useful comments. T.D-S. would like to thank M. Wolfire, J. Pineda, C. Ferkinhoff, D. Brisbin and N. Scoville for stimulating discussions about PDR physics and models. T.D-S. acknowledges support from ALMA-CONICYT project 31130005 and FONDECYT regular project 1151239. G.C.P. was supported by a FONDECYT Postdoctoral Fellowship (No. 3150361). N.L. acknowledges support from the NSFC grant No. 11673028. K.I. acknowledges support by the Spanish MINECO under grant AYA2016-76012-C3-1-P and MDM-2014-0369 of ICCUB (Unidad de Excelencia 'Mar\'ia de Maeztu'). This work was carried out in part at the Jet Propulsion Laboratory, which is operated for NASA by the California Institute of Technology. This work is based on observations made with the \textit{Herschel Space Observatory}, an European Space Agency Cornerstone Mission with science instruments provided by European-led Principal Investigator consortia and significant participation from NASA. The \textit{Spitzer Space Telescope} is operated by the Jet Propulsion Laboratory, California Institute of Technology, under NASA contract 1407. This research has made use of the NASA/IPAC Extragalactic Database (NED), which is operated by the Jet Propulsion Laboratory, California Institute of Technology, under contract with the National Aeronautics and Space Administration, and of NASA's Astrophysics Data System (ADS) abstract service. T.D-S. wants to thank the NASA Herschel Science Center (NHSC), and in particular D. Shupe, for providing access to their computer cluster with which most of the \textit{Herschel}/PACS data-sets were processed. Part of this work was carried out at the Aspen Center for Physics, which is supported by the National Science Foundation grant PHY-1066293.


\appendix

\begin{figure*}[t!]
\vspace{.5cm}
\epsscale{1.}
\plotone{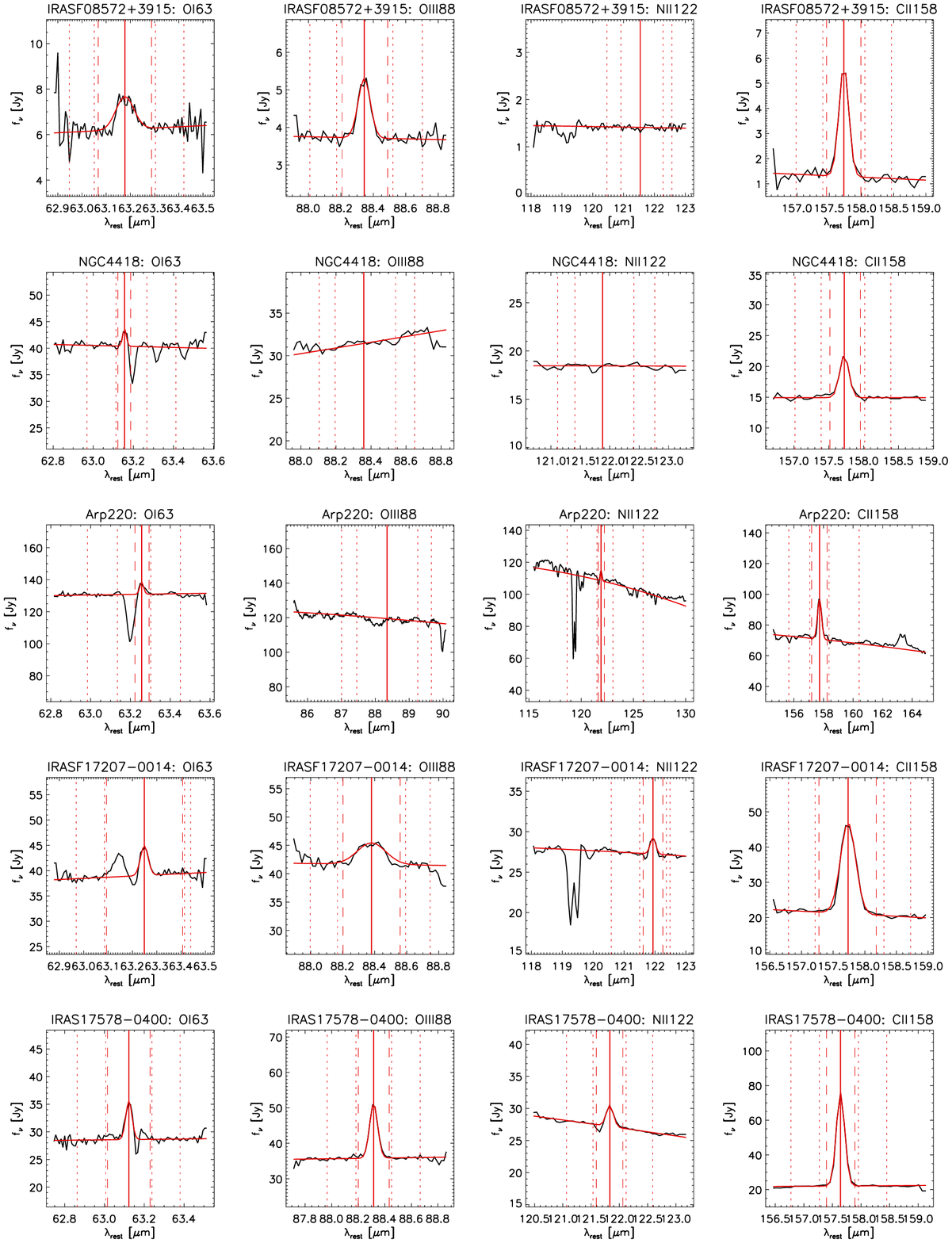}
\vspace{.25cm}
\caption{\footnotesize The \textit{Herschel}/PACS spectra (black line) of the 5 galaxies with \OIasub/\CIIpdr\, ratios lower than half the value of the fit to the correlation in Figure~\ref{f:oicii} (dotted-dashed line) that show clear signs of absorption in their \OIasub\, line profiles: IRASF08572+3915, NGC~4418, Arp~220, IRASF17207--0014 and IRAS17578--0400. The continuum emission was estimated using the wavelength ranges delimited by the two pairs of vertical dotted lines at the sides of the emission line (see section~\ref{s:datared} for details). The Gaussian fit to the line uses the wavelength range delimited by the two vertical dashed lines, which are also employed for integrating the line flux.}\label{f:opthickprof}
\vspace{.5cm}
\end{figure*}

\end{document}